\documentclass[twoside,12pt]{article}
\usepackage{epsfig,bbold,bm}
\usepackage{verbatim}
\def\Journal#1#2#3#4{{#1} {#2} (#4) #3}

\def\NPA{{\em Nucl. Phys.} A}

\def\NPB{{\em Nucl. Phys.} B}

\def\PLB{{\em Phys. Lett.} B}

\def\PRL{\em Phys. Rev. Lett.}

\def\PREP{\em Phys. Rep.}

\def\PRD{{\em Phys. Rev.} D}
\def\PRC{{\em Phys. Rev.} C}

\def\ZPA{{\em Z. Phys.} A}

\def\RMP{{\em Rev. Mod. Phys.}}

\def\INT{{\em Int. J. Mod. Phys.} E}
\def\ChinC{{\em Chin. Phys.} C}
\def\AIPCP{{\em AIP Conf. Proc.}}
\def\PTP{{\em Prog. Theor. Phys.}}

\def\JPG{{\em J. Phys.} G}

\def\r{\vec r}
\def\R{\vec R}
\def\p{\vec p}

\def\k{\vec k}
\def\vq{\vec q\,}
\def\El{\vec E}
\def\B{\vec B}
\def\D{\vec D}
\def\vsigma{\vec{\sigma}}

\newcommand{\be}{\begin{equation}}
\newcommand{\beq}{\begin{equation}}
\newcommand{\ee}{\end{equation}}
\newcommand{\eeq}{\end{equation}}
\newcommand{\bea}{\begin{eqnarray}}
\newcommand{\eea}{\end{eqnarray}}
\newcommand{\nn}{\nonumber}

\newcommand{\bfm}[1]{\mbox{\boldmath$#1$}}

\def\Dsl{\,\raise.15ex\hbox{/}\mkern-13.5mu D}
\def\dsl{\,\raise.15ex\hbox{/}\mkern-10.5mu \partial}

\def\d{$d{\hat{\sigma}}$}

\def\d{\lambda}

\def\bar#1{\overline{#1}}

\def\frac#1#2{{\textstyle{#1\over #2}}} %puts a small fraction
\def\dfrac#1#2{{\displaystyle{#1\over #2}}} %puts a small fraction
\def\Im{\mathop{\rm Im}}
\def\Re{\mathop{\rm Re}}

\def\ltap{\ \raise.3ex\hbox{$<$\kern-.75em\lower1ex\hbox{$\sim$}}\ }
\def\gtap{\ \raise.3ex\hbox{$>$\kern-.75em\lower1ex\hbox{$\sim$}}\ }
\def\gl{\ \raise.5ex\hbox{$>$}\kern-.8em\lower.5ex\hbox{$<$}\ }
\def\roughly#1{\raise.3ex\hbox{$#1$\kern-.75em\lower1ex\hbox{$\sim$}}}
\def\ltap{\ \raise.3ex\hbox{$<$\kern-.75em\lower1ex\hbox{$\sim$}}\ }
\def\gtap{\ \raise.3ex\hbox{$>$\kern-.75em\lower1ex\hbox{$\sim$}}\ }
\def\gl{\ \raise.5ex\hbox{$>$}\kern-.8em\lower.5ex\hbox{$<$}\ }
\def\roughly#1{\raise.3ex\hbox{$#1$\kern-.75em\lower1ex\hbox{$\sim$}}}
\relax
\def\Dsl{\,\raise.15ex \hbox{/}\mkern-13.5mu D}
\def\vsl{\,\raise.15ex \hbox{/}\mkern-10.5mu v}
\def\frac#1#2{{\textstyle{#1 \over #2}}}

\def\[{\left[}
\def\]{\right]}
\def\({\left(}
\def\){\right)}

\newcommand{\mi}{\mathrm{i}} %math i

\newcommand{\dfd}[3]{\hspace{-0.4em}\ensuremath{\frac{\mathrm{d}^{#1}#3}{(2\pi)^{#2}}}\,}

\newcommand{\eqn}[1]{Eq.~(\ref{#1})}

\newcommand{\tab}[1]{Tab.~\ref{#1}}

\def\Kbar{\overline{K}}
\def\K0bar{\overline{K^0}}

\def\qbar{\bar{q}}
\def\Qbar{\overline{Q}}
\def\Kbar{\overline{K}}
\def\Dbar{\overline{D}}

\def\vr{\vec{r}}

\newcommand{\bge}{\begin{equation}}
\newcommand{\ene}{\end{equation}}
\newcommand{\bg}{\begin{eqnarray}}
\newcommand{\en}{\end{eqnarray}}

\def\bge{\begin{equation}}
\def\ene{\end{equation}}
\def\bg{\begin{eqnarray}}
\def\en{\end{eqnarray}}
\def\nn{\nonumber}

\def\qbar{{\bar{q}}}

\def\d0bar{{\bar{D}^0}}
\def\Dbar{{\bar{D}}}

\def\vr{\vec{r}}

\begin{document}

\title{ 
\vspace{-1cm} 
\begin{flushright}
    {\large ADP-17-25/T1031, \hspace{2ex} LFTC-17-4/4 \vspace{1cm}}
\end{flushright}
Nuclear-bound quarkonia and heavy-flavor hadrons}
\author{G.\ Krein,$^{1}$ A. W.\ Thomas,$^2$ K.\ Tsushima$^3$ \\
\\
\noindent
$^1$Instituto de F\'{\i}sica Te\'orica, Universidade Estadual Paulista, \\
Rua Dr. Bento Teobaldo Ferraz, 271 - Bloco II, 01140-070 S\~ao Paulo, SP, Brazil \\
\noindent
$^2$ARC Centre of Excellence for Particle Physics at the Terascale and CSSM, \\
School of Physical Sciences, Department of Physics, The University of Adelaide,  \\ 
Adelaide SA 5005, Australia \\
\noindent
$^3$Laborat\'orio de F\'{\i}sica Te\'orica e Computacional, Universidade Cruzeiro do Sul, 
\\ Rua Galv\~ao Bueno, 868, Liberdade 01506-000, S\~ao Paulo, SP, Brazil}
\maketitle
\begin{abstract}
In our quest to win a deeper understanding of how QCD actually works, the study of the 
binding of heavy quarkonia and heavy-flavor hadrons to atomic nuclei offers enormous promise.
Modern experimental facilities such as FAIR, Jefferson Lab at 12 GeV and J-PARC offer 
exciting new experimental opportunities to study such systems. 
These experimental advances are complemented by new 
theoretical approaches and predictions, which will both 
guide these experimental efforts and be informed and improved by them. This review  
will outline the main theoretical approaches, beginning with QCD itself, summarize 
recent theoretical predictions and relate them both to past experiments and those from 
which we may expect results in the near future.
\end{abstract}
%\eject
%\tableofcontents

\eject
\tableofcontents

% % % % % % % % % % % % % % % % % % % % % %
%
%\input{sec_intro.tex}
%
%
%%%%%%%%%%%%%%%%%%%%%%%%%%%%%%%%%%%%%%%%%%%%%
\section{Introduction}
\label{sec:into}
%%%%%%%%%%%%%%%%%%%%%%%%%%%%%%%%%%%%%%%%%%%%%

There is now overwhelming evidence that Quantum Chromodynamics (QCD) is indeed the fundamental 
theory of the strong interaction and yet we are very far from 
actually understanding how it works. 
Certainly lattice QCD has had remarkable success with ground state hadrons, mesons with 
a simple quark-anti-quark pair and baryons with three valence quarks~\cite{Durr:2008zz}. 
Yet beyond that our ignorance 
is profound. After decades of speculation about, and searches for, colorless states with 
more than the minimal number of valence quarks we still have no idea whether such states 
exist or indeed whether QCD predicts them or not. 

In the list of particles that have come and gone at various 
times we think of dibaryons~\cite{Jaffe:1976yi,Mulders:1982da,Clement:2016vnl}, 
pentaquarks~\cite{DeVita:2006aaq,Aaij:2015tga} and exotic mesons~\cite{Maiani:2017mpo}.
Certainly, in the study of excited baryons it has recently 
become clear that the $\Lambda(1405)$ is an anti-kaon-nucleon bound 
state~\cite{Hall:2014uca} and that the Roper resonance is almost certainly 
dynamically generated through meson-baryon coupled channel dynamics~\cite{Liu:2016uzk},
although there is no consensus on this~\cite{Segovia:2015hra}.      
There is now a zoo of excited states of the $c\overline{c}$ system that must involve 
more than a simple quark-anti-quark pair. Yet there is no consensus as to whether these 
are molecular states, threshold effects or genuine exotic states bound by gluonic 
forces~\cite{Lebed:2016hpi}. Until we can find definitive answers to such issues, it is 
not possible to classify our understanding of QCD as more than superficial. 

The study of the interactions of quarkonia with atomic nuclei is an extremely promising avenue 
for exploring such issues. Because of the Zweig rule, 
the mean fields generated by light meson exchange, which provide a 
natural explanation of the binding of atomic nuclei, cannot bind a $c\overline{c}$ 
or $b\overline{b}$ pair and {\em if}~such states are indeed bound to nuclei 
one must look to other mechanisms, including gluon exchange. 

Thus the key issues in this field are firstly whether quarkonia are indeed bound to nuclei, 
secondly by how much and what are the properties of such states and thirdly how can 
one quantitatively understand these observations in terms of QCD. This work 
aims to summarize the status of the theoretical and experimental work in this exciting area.
It is complementary in some aspects to the recent review by A. Hosaka et al. in 
Ref.~\cite{Hosaka:2016ypm}; while the main focus of the latter is on properties
of heavy hadrons in nuclear matter and few-body systems, the present review focuses on 
nuclear binding phenomena of quarkonia and heavy-flavor hadrons that are of direct 
relevance for the experiments planned at existing and forthcoming facilities that include
FAIR, Jefferson Lab at 12 GeV and J-PARC.

The review begins with a reminder, in section 2, 
of QCD and especially its adaptation to heavy quark systems, 
with an emphasis on non-relativistic QCD, NRQCD, which has proven very successful in 
dealing with heavy quark systems. This is followed by a discussion of potential NRQCD, 
pNRQCD, which combines NRQCD with effective field theory to yield a Hamiltonian describing the 
gluon-mediated interactions of quarkonia with other colorless  systems.

Section 3 is an outline of the quark-meson coupling (QMC) model. This model starts at 
the level of the quark structure of hadrons, focusing on the interactions between them 
mediated by meson exchange, including the change of the internal structure of a hadron 
immersed in a nuclear medium implied by the self-consistent solution of the field equations. 
The QMC model serves as a natural way to calculate the non-gluonic mechanisms which 
may contribute to the binding of quarkonia. 

Section 4 deals with the ideal case of the interactions of the $J/\Psi$ with nuclei. 
In terms of QCD this may be viewed as a small color dipole immersed in a nuclear 
medium and so goes to the heart of the issue elaborated earlier, namely how QCD actually 
works in systems with more than the smallest number of valence quarks. While the 
$J/\Psi$ presents particular challenges because of its larger mass, any nuclear levels 
should be quite narrow and this should make their experimental identification, once formed, 
quite straightforward.

Although in the context of this review the strange quark is not particularly heavy, because the 
$\phi$ meson is almost entirely $s\overline{s}$, with little light quark content, it serves as a 
promising way to access the physics at the heart of this review. In section 5 we outline the 
many experiments which have already given some hints of binding of the $\phi$ to atomic 
nuclei as well as the more modern attempts in preparation. This experimental work is 
carefully placed in the context of modern theoretical expectations.

Although they are not strictly quarkonia, there has recently been quite a bit of 
experimental interest in the possible binding of $\eta$, $\eta^\prime$ and $\omega$ 
mesons and we devote section 6 to this topic. Section 7 deals with the natural 
extension of heavy flavored hadrons, $D$, $\overline{D}$ mesons and hypernuclei, 
namely bound states of $D$, $\overline{D}$ mesons as well as  
$\Lambda_c^+$ and $\Lambda_b$ baryons to nuclei. 
Finally, section 8 contains some concluding remarks.
%
%
%
%\input{sec_QCD.tex}
%
% % % % % % % % % % % % % % % % % % % % % 
\section{Quantum chromodynamics}
\label{sec:QCD}

In this section we collect different pieces of common knowledge on QCD
most relevant for the present review. Without aiming at 
being exhaustive or self-contained, the basic facts regarding QCD which we have chosen 
to cover in this section are directly related to the material discussed in the rest 
of this review. We refer the reader to the indicated references throughout the text
for a more complete and comprehensive account of the subjects discussed.

\subsection{QCD Lagrangian, quark masses and $\Lambda_{\rm QCD}$}
\label{subQCD-Lag}

We write the Lagrangian density of QCD as 
\bea
{\cal L}_{\rm QCD} &=& {\cal L}^{(0)}_q + {\cal L}^{(0)}_Q + {\cal L}^{(0)}_g \nn \\[0.25true cm]
&=& {\cal L}_q + {\cal L}_Q + {\cal L}_g + {\cal L}_{\rm c.t.},
\label{L-QCD}
\eea
wherein the terms with or without the superscript $(0)$ have the same structure and are given 
as follows. ${\cal L}_q$ contains the light-flavor quark Dirac fields $q=(u,d,s)$, ${\cal L}_Q$ 
the heavy-flavor quark Dirac fields $Q=(c,b,t)$, 
\beq
{\cal L}_q = {\overline q}(x)\left(i\Dsl - m_q\right)q(x), \hspace{1.0cm} {\cal L}_Q 
= {\overline Q}(x)\left(i\Dsl - m_Q\right)Q(x),
\label{LqQ}
\eeq
and ${\cal L}_g$ contains the pure-gluon part
\beq
{\cal L}_g = - \dfrac{1}{4} G^a_{\mu\nu}(x) G^{a \mu\nu}(x).  
\eeq
In Eq.~(\ref{LqQ}), $m_q$ and $m_Q$ are the light and heavy quark mass matrices
\be
m_q = \left( 
\begin{array}{ccc} 
m_u & 0 & 0 \\
0 & m_d & 0 \\
0 & 0 & m_s 
\end{array}\right), \hspace{1.0cm}
m_Q = \left( 
\begin{array}{ccc} 
m_c & 0 & 0 \\
0 & m_b & 0 \\
0 & 0 & m_t 
\end{array}\right).
\label{mq&mQ}
\ee
In addition, $\Dsl = \gamma^\mu D_\mu$ and $G^a_{\mu\nu}$, with $a=1,\dots,8$, are
respectively the SU(3) color covariant derivative and gluon field strength tensor,
\be
D_\mu = \partial_\mu + ig \, A_\mu(x), \hspace{1.0cm}
G^a_{\mu\nu}(x) = \partial_\mu A^a_\nu(x) - \partial_\nu A^a_\mu(x) - g f^{abc} A^b_\mu(x) A^c_\nu(x),
\label{D&G}
\ee 
where $A_\mu(x) = T^a A^a_\mu(x)$, $f^{abc}$ are the structure constants of the color SU(3) group, 
defined through the commutator $[T^a,T^b] = i f^{abc} T^b$, with $T^a = \lambda^a/2$ and normalized as 
$T^aT^a = 4/3$, where $\lambda^a$ are the SU(3) Gell-Mann matrices. Finally, ${\cal L}_{\rm c.t.}$ 
refers to the counterterm Lagrangian density that is required for renormalization; it contains terms
of the same structure as those in the other three terms in the second line
of Eq.~(\ref{L-QCD}). The relation between a field with and without a superscript $(0)$ is
through {\em wave-function renormalization} factors as $\phi^{(0)}(x) = Z^{1/2}_\phi\phi(x)$; 
the renormalization procedure determines the relation between $(m_q, m_Q, g)$ and 
$(m^{(0)}_q, m^{(0)}_Q, g^{(0)})$. $\phi^{(0)}$ is a {\em bare} field
and $\phi$ is a {\em renormalized} field, and likewise for the mass and coupling parameters. 
The Lagrangian in Eq.~(\ref{L-QCD}) is not yet complete as gauge fixing terms are necessary 
for implementing the calculations; we have omitted them because they are not needed for the 
present discussions.

The separation into light and heavy quark sectors of QCD is dictated by the values
of the quark masses. Since free quarks are not observed, their masses cannot be 
measured directly. When one refers to a value for the mass of a quark, it refers to a 
particular theoretical framework used to define it. For an observable particle, the position 
of the pole of the full propagator of the particle is taken as the definition of its 
mass{\textemdash}this is known as the pole mass. In QCD, the pole position of the quark propagator 
can be defined unambiguously in the context of perturbation theory only. In that context, the  
pole mass is infrared finite and gauge independent to all orders in perturbation 
theory~\cite{Kronfeld:1998di}. 

Renormalization in a perturbative calculation in a field theory like QCD is based on subtraction 
schemes to handle ultraviolet divergences. Invariably, renormalization schemes introduce an arbitrary mass 
scale, usually denoted by~$\mu$. The counterterm Lagrangian density ${\cal L}_{\rm c.t.}$ in Eq.~(\ref{L-QCD}) 
is treated as part of the interaction and is used to cancel the ultraviolet divergences; 
$\mu$ is a momentum at which one chooses to cancel the ultraviolet divergences and hence it its 
called the subtraction or renormalization scale. In~this way, the values of the quark masses 
$m_q$ and $m_Q$ and the coupling constant $\alpha_s \equiv g^2/4\pi$ depend on the scheme used to 
render amplitudes finite; they become {\em running} quantities, in that $m_q = m_q(\mu)$, $m_Q = m_Q(\mu)$, 
$\alpha_s = \alpha_s(\mu)$, with the running given by renormalization group (RG) equations. 
The most used renormalization scheme is the so-called $\overline{\rm MS}$ scheme~\cite{Collins:2011zzd}; 
in that scheme, the RG equation for a running quark mass $m_f(\mu)$, where $f=(q,Q)$, is given 
by (we present the one-loop contributions only)~\cite{PDG}
\beq
\mu^2 \dfrac{d m_f(\mu)}{d\mu^2} =  \left[ - \dfrac{\alpha_s(\mu)}{\pi} 
+ {\cal O}(\alpha^2_s)\right] m_f(\mu),
\label{RG-m}
\eeq
while the RG equation for $\alpha_s(\mu)$ is
\beq
\mu^2 \dfrac{d \alpha_s(\mu)}{d\mu^2} = \beta(\alpha_s) 
= - b_0 \, \alpha^2_s + {\cal O}(\alpha^3_s),
\label{RG-alphas}
\eeq
where $b_0$ is the one-loop beta-function coefficient, given in terms of the number of flavors 
$N_f$~as 
\beq
b_0 = \dfrac{33 - 2 N_f}{12\pi}, 
\label{b_0}
\eeq
where $N_f$ is the number of light flavors with masses less than the scale $\mu$. Complications arise 
when heavy flavors need to be taken into account in the loops~\cite{Collins:2011zzd} but they are of
no concern at the level of the present discussion. The essence of RG equations is that $\mu$ is an 
extraneous parameter to the theory and physical 
quantities must be $\mu$-independent; when changing $\mu$, the running masses and coupling must be 
readjusted in compensation to leave observables invariant.  These RG equations are commonly derived
by using the fact that the bare parameters of the theory are $\mu$-independent. An intuitive meaning 
of $\mu$ is given in an enlightened discussion in Ref.~\cite{Collins:2011zzd} in the context of a 
deep inelastic scattering process: for large values of $\mu$, $\mu$ plays the role of a cutoff on the 
transverse momentum in a loop integral contributing to such a process. In general, $\mu$ is chosen to 
be of the same size as the typical momentum transfer involved in the process under study, and $\alpha_s(\mu)$ 
at that scale gives the effective strength of the strong interaction in the process. 

For $\alpha_s(\mu)$ sufficiently small, one can obtain its explicit $\mu$ dependence
by integrating Eq.~(\ref{RG-alphas}) from $\mu_0$ to $\mu$:
\beq
\dfrac{1}{\alpha_s(\mu)} = \dfrac{1}{\alpha_s(\mu_0)} + b_0 \ln \left(\dfrac{\mu^2}{\mu^2_0}\right),
\eeq
which can be rewritten in terms of a $\mu$-independent mass scale $\Lambda_{\rm QCD}$ as
\beq
\dfrac{1}{\alpha_s(\mu)}  -  b_0 \ln \left(\dfrac{\mu^2}{\Lambda^2_{\rm QCD}}\right) 
= \dfrac{1}{\alpha_s(\mu_0)} - b_0 \ln \left(\dfrac{\mu^2_0}{\Lambda^2_{\rm QCD}}\right).
\eeq
This, on the other hand, means that 
\beq
\Lambda_{\rm QCD} = \mu \, e^{-1/[2b_0 \alpha_s(\mu)]},
\label{Lambda-QCD1}
\eeq
or, equivalently
\beq
\alpha_s(\mu) = \dfrac{1}{b_0 \ln (\mu^2/\Lambda^2_{\rm QCD})} 
= \dfrac{12\pi}{(33 - 2N_f)\, \ln(\mu^2/\Lambda^2_{\rm QCD})}.
\label{RG-alphas-sol}
\eeq
As $\mu$ increases, $\alpha_s(\mu)$ decreases, QCD becomes weakly coupled and perturbation 
is applicable. This feature is the renowned property of {\em asymptotic freedom}~\cite{{Gross:1973id}, 
{Politzer:1973fx}}. Therefore, processes amenable to perturbation theory are those involving large momentum 
transfers, or short distances. On the other hand, processes 
involving small momentum transfers are not amenable to perturbation theory because the effective 
coupling becomes large. From Eq.~(\ref{RG-alphas-sol}) one has that for $\mu \sim \Lambda_{\rm QCD}$,
the coupling increases substantially. Although Eq.~(\ref{RG-alphas-sol}) is not valid at such values 
of $\mu$, extractions of $\alpha_s$ from different experimental sources show that it is a growing 
function of $\mu$, varying typically from $\alpha_s \simeq 0.08$ at $\mu = 1000~{\rm GeV}$ to 
$\alpha_s \simeq 0.4$ at $\mu = 2~{\rm GeV}$~\cite{PDG}. 

The appearance of a mass scale in the theory due to the renormalization has deep consequences in the 
way a symmetry of the classical QCD Lagrangian is realized when taking into account quantum fluctuations.
In general, when a symmetry of the classical Lagrangian of a theory is not realized in its quantum version,
one says that there exists an {\em anomaly} in the theory. One particular anomaly that is of importance for 
the subject of quarkonia in nuclei, as we will discuss shortly, is related to scale invariance 
of (massless) QCD, known as the anomaly in the trace of the energy-momentum tensor 
(or, simply the {\em trace anomaly})~\cite{Chanowitz:1972vd,
Chanowitz:1972da,Crewther:1972kn, Freedman:1974gs,Collins:1976yq,Voloshin:1980zf, Novikov:1980fa}. 

The mass scale $\Lambda_{\rm QCD}$ is a parameter of QCD; it is the value at which the theory becomes
strongly coupled and nonperturbative. Its precise value depends on the renormalization scheme, on the number of
active flavors, on the perturbative order at which the $\beta$ function is evaluated and also on approximations 
used to solve the RG equation. That is, the determination of $\Lambda_{\rm QCD}$ from data involves 
the use of an $\alpha_s(\mu)$ obtained from the solution of the RG equation in some renormalization 
scheme and calculated up to a certain number of loops; Equation~(\ref{RG-alphas}), for example,  is derived 
in the $\overline{\rm MS}$ scheme and a calculation that retains the first term only in that equation 
is a one-loop calculation. The present (central) values in the $\overline{MS}$ scheme for
different flavor numbers $N_f$ are~\cite{PDG}
\bea
\Lambda_{\rm QCD} = \left\{ \begin{array}{r}332 \;\;{\rm for}\;\;N_f = 3\\ 292 \hspace{1.5cm} = 4\\ 
210 \hspace{1.5cm} = 5 \\ 89 \hspace{1.5cm} = 6\end{array} \right. 
\label{Lambda-QCD2}
\eea
Reference~\cite{Deur:2016tte} is a very recent review of the present theoretical and empirical knowledge 
for $\alpha_s$. The review discusses several issues 
related to different definitions of $\alpha_s$ and $\Lambda_{\rm QCD}$, their scheme dependence and
extraction of their values from experimental data. We strongly recommend this reference to the reader 
interested in more details on these fundamental parameters of QCD. 

Knowing $\alpha_s(\mu)$, one can solve the RG equation for the quark masses $m_f(\mu)$. The PDG~\cite{PDG} 
quotes values of the masses of the light $(u,d,s)$ quarks 
in the $\overline{\rm MS}$ scheme at the renormalization scale $\mu = 2$~GeV, and those of the
heavy quarks $(c,b,t)$ as pole masses or as $\overline{\rm MS}$ masses evaluated at a scale equal 
to the mass, i.e. $m_Q(\mu=m_Q)$. The pole mass $m_f$ and the $\overline{\rm MS}$ masses $m_f(\mu)$ 
evaluated at $\mu = m_f$ are related~by
\beq
m_f = m_f(m_f) \left[ 1 + \dfrac{4\alpha_s(m_f)}{3\pi} + {\cal O}(\alpha^2_s) \right].
\label{pole-MS}
\eeq
Table~\ref{tab:quarks} presents the values of $\overline{\rm MS}$ masses collected by
the PDG~\cite{PDG}{\textemdash}for completeness, the electric charges and flavor quantum numbers
(isospin I and third component of isospin I$_3$, strangeness~S, charm~C, bottom~B and top~T) of the 
quarks are also shown in the table. For the bottom quark, using the $\overline{\rm MS}$ mass 
given in the table leads to $m_b = 1.18\,\overline{m}_b(\overline{m}_b)$ = 4.93~GeV for the
bottom pole mass, when using Eq.~(\ref{pole-MS}) evaluated up to $\alpha^3_s$~\cite{PDG}.

\begin{table}[htb]
\caption{\label{tab:quarks} Quark masses (in GeV). The values of quark masses are central values 
(i.e. without uncertainties) from the PDG~\cite{PDG}{\textemdash}they are $\overline{\rm MS}$
masses, $m_q = m_q(\mu = 2$~GeV) for the light quarks $q=(u,d,s)$, and $m_Q = m_Q(\mu=m_Q)$ 
for the heavy quarks $Q=(c,b,t)$, where $\mu$ is the renormalization scale. Also shown are the 
electric charges (in units of the electron charge magnitude $e$) and flavor quantum numbers of the quarks.}
\begin{center}
\begin{tabular}{lc|c|c|c|c|c}
\hline 
{} & {} & {} & {} & {} & {} & {} \\[-0.45true cm]
Quark & $u$ & $d$  & $s$  & $c$ & $b$  & $t$ \\
{} & {} & {} & {} & {} & {} & {} \\[-0.45true cm]
\hline %[-0.25true cm]
{} & {} & {} & {} & {} & {} & {} \\[-0.15true cm]
Mass   & 0.0022 & 0.0047 & 0.096 & 1.27 & 4.18 & 160  \\[0.2true cm]
\hline
{} & {} & {} & {} & {} & {} & {} \\[-0.15true cm]
Charge & $+ \frac{2}{3}$ & $- \frac{1}{3}$  & $+\frac{2}{3}$   & $+\frac{2}{3}$   & $+\frac{2}{3}$   
& $+\frac{2}{3}$   \\[0.2true cm]
\hline
{} & {} & {} & {} & {} & {} & {} \\[-0.15true cm]
Flavor & I=$\frac{1}{2},$ I$_3$=$+\frac{1}{2}$  & I=$\frac{1}{2}$ , I$_3$=$-\frac{1}{2}$ & S=$-1$ 
& C=$+1$ & B=$-1$ & T=$+1$ \\[0.10true cm]
\hline 
\end{tabular}
\end{center}
\end{table}

Inspection of Tab.~\ref{tab:quarks} reveals that the $(u,d,s)$ quarks are clearly much lighter than 
the $(c,b,t)$ quarks. Although the $s$ quark is much heavier than the $(u,d)$ quarks, still, it is 
an order of magnitude lighter than the lightest of the heavy quarks, the $c$ quark. This large 
difference in the masses of light and heavy quarks has dramatic consequences in the dynamics  
governing the internal structure of quark-gluon bound states{\textemdash}the 
hadrons{\textemdash}containing different combinations of quarks of different 
flavors. In~QCD, a hadron is described by a state in Hilbert space, the state being characterized
by a set of quantum numbers, such as spin, parity, flavor, etc. Hadron properties can be studied 
by analysing correlation functions (also known as propagators, or Green functions) of interpolators. Interpolators
are gauge-invariant products of quark and gluon field operators having the quantum numbers of the 
hadron of interest. The mass spectrum of a hadron, for example, can be extracted by searching for isolated poles
in such a correlation function. Let $h$ denote a light-flavor hadron. Even though the interpolators 
for $h$ contain gluon and light-flavor quark fields only, the heavy quark fields do contribute to the $h$ 
correlation functions through quantum fluctuations.  However, in view of their large mass, heavy quarks 
play a minor role in the properties of~$h$ in view of the {\em decoupling theorem}~\cite{Appelquist:1974tg},
which in its essence asserts that the contributions of a large mass in quantum fluctuations are suppressed 
when they appear in a finite amplitude, and when they appear in a divergent amplitude they can be eliminated 
through the renormalization of the parameters of the theory. 

\subsection{Symmetries and anomalies}
\label{sub:symm}

Consider the hypothetical situation in which the light quarks are massless, $m_q = 0$, and the heavy 
quarks are infinitely heavy, $m_Q = \infty$. The heavy quarks do not contribute at all to the structure 
of the light-flavor hadron~$h$. In this hypothetical situation, the only dimensionful parameters in the theory are 
$\Lambda_{\rm QCD}$ and the mass $m_h$ and a typical size $r_h$ of $h$ would be
\beq
m_h = a_h \, \Lambda_{\rm QCD}, \hspace{1.0cm}
r_h = b_h \, \Lambda^{-1}_{\rm QCD},
\label{m_h-r_h}
\eeq
where $a_h$ and $b_h$ are dimensionless numbers. Different hadrons have different values for 
$a_h$ and $b_h$. The value of $m_h$ is deeply connected with the breaking of the scale invariance 
of the QCD Lagrangian due to the renormalization. Let us consider a scale transformation of the 
spacetime coordinates $x^\mu$ by a dimensionless parameter $\lambda$
\be
x^\mu \rightarrow x^{\prime\mu} = \lambda \, x^\mu,
\ee
and the associated transformations of the quark $q(x)$ and gluon fields $A_\mu(x)$
\be
q(x) \rightarrow q^\prime(x) = \lambda^{3/2} \, q(\lambda x), \hspace{1.0cm} 
A_\mu(x) \rightarrow A^\prime_\mu(x) =  \lambda \, A_\mu(\lambda x).
\ee
Under this scale transformation, while the the classical QCD (massless) Lagrangian would change as
\be
{\cal L}_{\rm QCD}(x) \rightarrow {\cal L}^\prime_{\rm QCD}(x)  = \lambda^4\, {\cal L}_{\rm QCD}(\lambda x) ,
\ee 
the action remains unchanged:
\be
S_{\rm QCD} = \int d^4x \, {\cal L}_{\rm QCD}(x) \rightarrow \int d^4x \, \lambda^4 \, {\cal L}_{\rm QCD}(\lambda x)
= \int d^4x' \, {\cal L}_{\rm QCD}(x') = S_{\rm QCD}.
\ee
The associated Noether current, the dilatational current, is given by
\be
J^{\mu}_{\rm dilat}(x) = T^{\mu\nu}(x) \, x_\nu ,
\ee
where $T^{\mu \nu}(x)$ is the energy-momentum tensor, which is a conserved quantity, 
$\partial_ \mu T^{\mu \nu}(x) = 0$. The divergence of $J^{\mu}_{\rm dilat}(x)$ is the trace
of the energy-momentum tensor:
\be
\partial_\mu J^{\mu}_{\rm dilat}(x) = T^{\mu}_{\mu}(x) = 0,
\label{trace-mass}
\ee
where the vanishing of $T^\mu_\mu(x)$ follows from the classical equations of motion of the quark
and gluon fields. On the other hand, since the expectation value of $T^\mu_\mu$ in the state $|h\rangle$ 
of a hadron at rest gives the mass of this state, 
\be
\langle h|T^\mu_\mu|h\rangle = m_h, 
\ee
where we are using a nonrelativistic normalization for the hadron states $|h\rangle$, the result 
in Eq.~(\ref{trace-mass}) would suggest a vanishing mass for all hadrons. 

A very simple, heuristic way to see how the situation changes due to quantum effects 
is as follows~\cite{Donoghue:1992dd}. If one considers that the coupling $g$ that appears in the 
Lagrangian is scale dependent, $g = g(\mu)$, the massless action is not scale invariant anymore.
First, let us rescale the gluon field by $g$ as $\overline{A}^a_\mu \equiv g\,A^a_\mu$, the massless 
Lagrangian becomes
\be
{\cal L}_{\rm QCD} = - \dfrac{1}{4g^2} \overline{G}^a_{\mu\nu}(x) \overline{G}^{a \mu\nu}(x) 
+ {\overline q}(x)\left[i \gamma^\mu (\partial_\mu + i \, {\overline A}_\mu(x)) \right]q(x),
\ee
so that one sees that all dependence on $g$ appears as an overall factor in the pure-gluon term.
Next, let us consider an infinitesimal scale transformation, $\mu \rightarrow \lambda \mu$ with  
$\lambda = 1 + \delta \lambda$, so that
\be
\delta S_{\rm QCD}  =  \delta \, \left( - \dfrac{1}{4\pi \alpha_s} 
\dfrac{1}{4} \int d^4x \,\overline{G}^a_{\mu\nu}(x) \overline{G}^{a \mu\nu}(x) \right) 
= - \dfrac{2\beta(\alpha_s)}{\alpha_s} S_{\rm QCD} \, \delta \lambda .
\ee
From Noether's theorem, one then has that the trace of the energy momentum tensor
does not vanish, which is the statement of the trace anomaly~\cite{Chanowitz:1972vd,
Chanowitz:1972da,Crewther:1972kn, Freedman:1974gs,Collins:1976yq,Voloshin:1980zf, Novikov:1980fa}: 
\bea
T^\mu_\mu(x) &=& \dfrac{2\beta(\alpha_s)}{\alpha_s} \, \dfrac{1}{4} G^a_{\mu\nu}(x) G^{a \mu \nu}(x)
= - \dfrac{1}{2}\, b_0 \alpha_s \, G^a_{\mu\nu}(x) G^{a \mu \nu}(x)
\nn \\
&=& - \dfrac{9}{32\pi^2} \, g^2 G^a_{\mu\nu}(x) G^{a \mu \nu}(x),  
\eea
where we used the $\beta-$function at lowest order in perturbation theory. A more sophisticated proof 
of this result can be found in Ref.~\cite{Smilga:2001ck}. Therefore,
\beq
m_h = - \dfrac{9}{32\pi^2} \,\langle h|g^2G^a_{\mu\nu} G^{a \mu \nu}|h\rangle.
\label{mass-anomaly}
\eeq
That is, the entire mass of a hadron is due to gluons.

When $m_q \neq 0$, the trace of the energy-momentum tensor receives contributions from the mass 
term in the light-quark part of the QCD Lagrangian in Eq.~(\ref{LqQ}) and $m_h$ is given by
\beq
m_h = \dfrac{\beta(\alpha_s)}{2\alpha_s} \, \langle h|G^a_{\mu\nu} G^{a \mu \nu}|h\rangle
+ \langle h|\overline{q}\,m_q\,q|h\rangle.
\label{mh-trace-anom}
\eeq
The interesting point of this result is that the matrix element 
$\langle h|g^2G^a_{\mu\nu} G^{a \mu \nu}|h\rangle$ contributes to the amplitude of 
quarkonium-hadron scattering at threshold, as we discuss further ahead in this review. 
This means that one can probe the distribution of mass inside a hadron 
via close to threshold quarkonium-hadron scattering~\cite{Kharzeev:1995ij}. Also,
the combined use of the trace anomaly and a deep-inelastic momentum sum rule allows
to derive a separation of the nucleon mass into contributions of the quark and gluon kinetic 
and potential energies, quark masses, and the trace anomaly~\cite{Ji:1994av}.  

For $m_q \neq 0$, the parameters $a_h$ and $b_h$ 
receive contributions from $m_q$. Since $m_q/\Lambda_{\rm QCD} \ll 1$, one can in principle calculate 
corrections to $a_h$ and $b_h$ by treating the mass term $m_q\, \overline{q}(x)q(x)$ in the QCD Lagrangian 
in Eq.~(\ref{L-QCD}) as a perturbation and the results would be expressed in the form of (integer and 
noninteger) powers of $m_q/\Lambda_{\rm QCD}$ and of $\ln (m_q/\Lambda_{\rm QCD})$. There is, however, 
a more practical way of implementing a calculation of this kind by using an EFT employing hadron degrees 
of freedom. When endowed with an appropriate power counting scheme, the EFT provides a powerful way of
calculating quark-mass corrections to observables. The EFT is based on the fact that in 
the $m_q \rightarrow~0$ limit, QCD acquires an $SU(3)_R \times SU(3)_L$ chiral symmetry that is dynamically 
broken by the strong QCD interactions to an $SU(3)_V$ flavor symmetry. Specifically, the light-quark part 
of the QCD Lagrangian in Eq.~(\ref{LqQ}) can be rewritten in terms of the right-handed and left-handed quark
field operators $q_R(x) = 1/2 (1+\gamma_5)q(x)$ and $q_L(x) = 1/2 (1-\gamma_5)q(x)$ as (we follow closely
the presentation in Ref.~\cite{Manohar:2000dt}):
\be
{\cal L}_q = {\overline q}(x) \, i\Dsl \, q(x) 
= {\overline q}_R(x)\,i\Dsl\, q_R(x) + {\overline q}_L(x)\,i\Dsl\,q_L(x),
\label{L-RL}
\ee
which clearly is invariant under the transformations
\be
q_R(x) \rightarrow q'_R = R \, q_R(x), \hspace{1.0cm} q_L(x) \rightarrow q'_L = L \, q_L(x), 
\label{LR-transf}
\ee
with $R \in SU_R(3)$ and $L \in SU_L(3)$. The dynamical breaking 
of the $SU(3)_R \times SU(3)_L$ is commonly discussed in terms of the nonzero value of 
the vacuum expectation value (v.e.v.) of quark bilinears  
$\overline{q}^f(x) q^{f'}(x) = \overline{q}^f_R q^{f'}_L(x) + \overline{q}^{f}_L q^{f'}_R(x)$:  
\be
\langle \overline{q}^f_R(x) q^{f'}_L(x)\rangle = B \; \delta^{f f'},
\label{condensate}
\ee
where $B \sim \Lambda^3_{\rm QCD}$ and $f$ and $f'$ are flavor indices. The v.e.v. 
$\langle \overline{q}^f(x) q^{f}(x)\rangle$ is known as the quark condensate. In perturbation theory, 
it is trivially zero (in the massless limit) and its nonzero value is driven by nonperturbative gluon 
dynamics. There is a profound interrelationship between the trace anomaly and 
dynamical chiral symmetry breaking, in that the smallness of the pion mass would be difficult to 
reconcile with Eq.~(\ref{mh-trace-anom}) were it not for the dynamical breaking of a
symmetry, in the present case, chiral symmetry{\textemdash}for a very recent discussion on this, 
see Ref.~\cite{Roberts:2016mhh}. A nonzero value of the condensate implies in a dynamically generated
quark mass and thereof hadron masses can be understood as coming from massive quarks, as in the 
pre-QCD quark models. Our understanding on how this exactly happens and its impact on hadron observables
has evolved tremendously during the last decades through numerical simulations of QCD discretized 
on a space-time Euclidean lattice~\cite{Kronfeld:2012ym} and also from studies in the 
continuum~\cite{{Cloet:2013jya},{Eichmann:2016yit}}.  

Under the transformations in Eq.~(\ref{LR-transf}), this v.e.v. 
transforms as $\langle \overline{q}'^{f}_R q'^{f'}_L\rangle = B\, (LR^\dag)_{f'f}$. Since for $L=R$ 
the v.e.v. is unchanged, the $SU(3)_R \times SU(3)_L$ symmetry of the Lagrangian is broken to its 
diagonal subgroup $SU(3)_V$ and one identifies the eight pseudoscalar mesons $\pi^0$, $\pi^{\pm}$, 
$K^0$, ${\bar K}^0$, $K^{\pm}$, and $\eta$ with the (pseudo-)Goldstone bosons with the 
fluctuations of $\overline{q}_Rq_L$. The fields corresponding to these particles can be described 
by a special unitary $3 \,\times \, 3$ matrix $U(x)$ 
\be
U = e^{i \Phi/f}\, \hspace{1.0cm}
\Phi = \left( \begin{array}{ccc}
\pi^0 + \eta/\sqrt{3} & \sqrt{2} \, \pi^+      & \sqrt{2} \, K^+  \\[0.25true cm] 
\sqrt{2} \,\pi^-      & -\pi^0 + \eta/\sqrt{3} &  \sqrt{2} \, K^0 \\[0.25true cm] 
\sqrt{2} \, K^-       &  \overline{K}^0        &  - 2 \eta/\sqrt{3}
\end{array}
\right),
\label{GB-fields}
\ee
where $f$ is the leptonic decay constant of the Goldstone bosons (with this normalization, the 
experimental value of the pion decay constant is $f_\pi = 93$~MeV). This matrix 
transforms under $SU(3)_R \times SU(3)_L$ as
\be
U(x) \rightarrow L\,U(x) \, R^\dag.
\ee
An effective Lagrangian density respecting $SU(3)_R \times SU(3)_L$ can be constructed by
using as building blocks products of $U(x)$ and its derivative $\partial_\mu U(x)$. The 
leading order Lagrangian in the number of derivatives is
\be
{\cal L}_{\rm GB} = \dfrac{f^2}{4} {\rm Tr}\left[\partial_\mu U(x) \; \partial^\mu U^\dagger(x)\right].
\label{L-GB}
\ee 
The factors $1/f$ in Eq.~(\ref{GB-fields}) and $f^2/4$ in the Lagrangian are chosen so that 
the kinetic-energy term of the fields in the Lagrangian have standard normalization.

The symmetry-breaking mass term in the QCD Lagrangian 
can be implemented in the EFT by using the trick of supposing that the mass matrix $m_q$ 
transforms under $SU(3)_R \times SU(3)_L$ as $m_q \rightarrow L \, m_q \, R^\dag$, so that
the mass term in the Lagrangian is invariant under $SU(3)_R \times SU(3)_L$. Then, to lowest order 
in $m_q$, the symmetry-breaking term in the EFT is 
\be
{\cal L}_{m_q} = \overline{q}(x) m_q q(x) = \overline{q}_R(x) \, m_q \,
q_L(x) + \overline{q}_L(x) \, m_q \, q_R(x) \rightarrow 
{\cal L}_{{\rm GB}-m_q} = B \, {\rm Tr} \left[ U^\dag(x) m_q + m^\dag_q U(x) \right]. 
\label{L-GB-m}
\ee

The lowest order Lagrangian of this EFT is the sum of the two contributions, Eqs.~(\ref{L-GB})
and (\ref{L-GB-m}). Higher order terms can be constructed in a similar fashion. There is a huge 
literature associated with the use of this EFT to study low-energy phenomena involving 
(pseudo)Goldstone bosons. The reader interested in knowing more on the techniques and applications
of this and other chiral EFTs can find good guidance in the book of Ref.~\cite{Scherer:2012xha}. 
Of interest for the present review are two applications of this chiral EFT. One is related to the 
restoration of chiral symmetry at finite temperature $T$ and baryon density $\rho_B$. The other is 
in connection with the derivation of the long-range part of the quarkonium-quarkonium interaction, 
that we discuss in subsection~\ref{sub:pNRQCD}. 

At very large temperatures, the relevant scale in QCD is the temperature and due to
asymptotic freedom the coupling strength $\alpha_s$ becomes small and so the quark condensate. 
The same can be expected for very large baryon densities, when the relative quark distances
become small. For zero baryon density, recent lattice QCD simulations~\cite{Bazavov:2011nk}, employing
a pion mass of $m_\pi = 161$~MeV, have shown that there is a drastic decrease of the quark condensate 
around a temperature of $T_{\rm pc} = 154 \pm 9$~MeV. For finite baryon densities, the combined 
$T$ and $\rho_B$ behavior of the condensate is presently unknown in lattice QCD. However, for 
sufficiently low temperatures, the lowest mass excitations are pions and the chiral EFT just 
described can be used~\cite{{Gasser:1986vb},{Gerber:1988tt}} to obtain the temperature dependence of the 
condensate; to leading order, $\langle \overline{q}q\rangle_{T}$ is given by (for two massless quark 
flavors):
\be
\dfrac{\langle \overline{q}q\rangle_T}{\langle \overline{q}q\rangle} 
= 1 - \dfrac{T^2}{8f^2_\pi}.
\label{cond-T}
\ee
For nonzero $\rho_B$, there are results from calculations using QCD sum rules that
generalize this result for small values of $\rho_B$ to~\cite{{Drukarev:1991fs},{Cohen:1991nk}}: 
\begin{equation}
\dfrac{\langle\overline{q}q\rangle_{T\rho_B}}{\langle \overline{q} q\rangle } 
= 1 -  \sum_{\rm h} \dfrac{\Sigma_h}{f^2_\pi m^2_\pi} \, \rho^{\rm h}_s
= 1 - \dfrac{T^2}{8f^2_\pi} - \dfrac{1}{3} \, \dfrac{\rho_B}{\rho_0} \, ,
\label{cond-Trho}
\end{equation}
where $\Sigma_h = m_q \, \partial m_h/\partial m_q$, $\rho^{\rm h}_s$ is the scalar 
density of a hadron ${\rm h}$ in matter and $\rho_0$ is the baryon saturation 
density of nuclear matter. These results point to a partial restoration o chiral
symmetry and a large body of work has been devoted to extract signals of changes of 
the intrinsic structure of hadrons in nuclei due to this. In another direction, through 
experiments of heavy ion collisions, new phases of QCD matter are expected to be
created which can teach us about the first stages of the evolution of the Universe. 
Ref.~\cite{Krein:2016fqh} is a recent review on the theoretical and phenomenological 
understanding of selected topics on hadrons in medium. 

The classical Lagrangian in Eq.~(\ref{L-RL}) has two additional global $U(1)$ symmetries: one 
is a vector $U(1)$ symmetry where right- and left-handed quarks transform by a common phase, 
and the other is an axial $U(1)$ symmetry where right- and left-handed quarks transform 
by the opposite phase. While the first remains a symmetry in the quantum version of
the theory and is associated with baryon number conservation, the second does not. More
specifically, the axial transformation is of the form
\be
q_R(x) \rightarrow  q'_R(x) = e^{-i\,\theta } \, q_R(x), \hspace{1.0cm}
q_L(x) \rightarrow  q'_L(x) = e^{+i\,\theta } \, q_L(x), 
\label{axial-RL}
\ee
or, equivalently
\be
q(x) \rightarrow  q'(x) = e^{-i\,\theta\,\gamma_5 } \, q(x).
\label{axial-q}
\ee
Noether's theorem implies conservation of the flavor singlet current
\be
\partial^{\mu} J^{(0)}_{5\mu}(x) = 0,
\ee
where
\be
J^{(0)}_\mu(x) = \overline{q}(x)\,\gamma_\mu\gamma_5\, q(x) = \overline{u}(x)\gamma_\mu\gamma_5 u(x) 
+ \overline{s}(x)\gamma_\mu\gamma_5 s(x) + \overline{s}(x)\gamma_\mu\gamma_5 s(x).
\ee 
There are various ways to demonstrate that this symmetry does not happen in the quantum
version of the theory~\cite{{Adler:1969gk},{Bell:1969ts}}. The simplest to state but more elaborate to 
demonstrate is the assertion that the transformations in Eq.~(\ref{axial-RL}) (or in Eq.~(\ref{axial-q})) 
change the measure in the path integral~\cite{Fujikawa:1979ay}. One important consequence of this axial 
anomaly (also known as chiral anomaly) is that divergence of the flavor-singlet current is given by 
(now including the explicit symmetry-breaking mass term)
\be
\partial^{\mu} J^{(0)}_{5\mu}(x) = \dfrac{3\alpha_s}{8\pi} \epsilon^{\mu\nu\alpha\beta} 
G^{a}_{\mu\nu} G^{a}_{\alpha\beta} + 2i \left[m_u \overline{u}(x) \gamma_5 u(x) + 
m_d \overline{d}(x) \gamma_5 d(x) + m_s \overline{s}(x) \gamma_5 s(x)\right].
\label{flav-curr-nonc}
\ee
We refer the reader to Ref.~\cite{Donoghue:1992dd} for a modern textbook demonstration. As in the
case of the trace anomaly, one sees that the nonconservation of the flavor-singlet current is
driven by gluon dynamics. 

In the absence of this axial anomaly, one would have one more pseudo Goldstone
boson with flavor content $\overline{u}u + \overline{d}d + \overline{s}s \equiv \eta_0$, whose mass 
would vanish in the chiral limit. But the lightest candidate for a ninth pseudoscalar particle 
is $\eta'(960)$, which is much heavier the other real eight pseudo Goldstone bosons~\cite{PDG}.  
Current understanding is that the physical pseudoscalar particles with an $\overline{s}s$ content are $\eta(549)$
and $\eta'(960)$, which are mixtures of the flavor singlet combination $\eta_0$ and flavor octet
$\eta_8 = \overline{u}u + \overline{d}d - 2 \overline{s}s$ combination, with $\eta'(960)$ being 
predominantly composed by the $\eta_0$ component. In section~\ref{etaod} we review studies of
the formation of $\eta$ and $\eta'$ nuclear bound states. These are motivated~\cite{{Bass:2013nya},
{Bass:2010kr},{Bass:2005hn}} by the fact that the in-medium mass of the $\eta$ and $\eta'$ are 
modified by the interaction with the nuclear mean fields, in particular by the attractive 
scalar-isoscalar component. The study of these nuclear bound states provides a unique
opportunity to learn about the interplay between dynamical chiral symmetry breaking and the 
axial anomaly.

We also review recent progress in studies of nuclear bound states of heavy-flavor hadrons.
Hadrons containing light quarks and only one heavy quark (or antiquark) are very 
interesting QCD bound states. They are interesting as their internal dynamics is still determined 
by the light quarks and the presence of a heavy quark gives rise to a symmetry which is not apparent 
in the QCD Lagrangian. In the $m_Q \rightarrow \infty$ limit, the heavy quark does not 
contribute with a (positive) kinetic energy to the bound state and the mass and typical size 
of the hadron are still governed by $\Lambda_{\rm QCD}$. Moreover, the average velocity 
$v_Q$ of the heavy quark in a heavy-light bound state is changed very little by the interactions, 
as $\Delta v_Q = \Delta p_Q/m_Q \sim \Lambda_{\rm QCD}/m_Q \ll 1$ and the interactions 
of a heavy quark, regardless of its flavor, within the hadron become independent 
of its spin. This gives rise to a spin-flavor heavy quark $U(2N_Q)$ symmetry{\textemdash}$\,2N_Q 
= 2\,({\rm spin})\times N_Q\, (\rm{heavy\;flavors})$. This means that the light quark dynamics occurs 
in the background of a strong color field of an essentially static spectator. An EFT, known as 
HQET, can be constructed from the QCD Lagrangian by a implementing a suitable $1/m_Q$ 
expansion~\cite{Manohar:2000dt}.   

On the other hand, the internal dynamics of hadrons containing more than one heavy quark is 
very different from that of the previous two cases. Let us take the case of a heavy quarkonium 
$Q\overline{Q}$, which is one of the heavy hadrons of  interest to the present review. For large
values of $m_Q$, the bound-state is controlled by short-distance dynamics, which is weakly coupled 
and nonrelativistic. 
The interaction between the $Q$ and $\overline{Q}$ is essentially an attractive Coulomb force and the 
heavy quark must have a kinetic energy to balance this force and form a bound state. The mass and size of 
the $Q\overline{Q}$ quarkonium are not at all controlled by $\Lambda_{\rm QCD}$. Nonrelativistic QCD 
(NRQCD)~\cite{Caswell:1985ui} and potential nonrelativistic QCD 
(pNRQCD)~\cite{{Pineda:1997bj},{Brambilla:1999xf}} are examples of EFTs that have been used to 
treat heavy quarkonia{\textemdash}for a comprehensive review,
see Ref.~\cite{Brambilla:2004jw}. These EFTs provide an adequate first-principles framework for studying 
the low energy interactions of quarkonia in nuclear matter as their Lagrangian (or Hamiltonian) can be
cast into the form usually used to treat the nuclear many-body problem. We discuss this in the next section.

To finalize this section, we mention that the top quark $t$ is of no interest in the present review.
The top quark is the heaviest quark, its lifetime is very short, $\tau_t = 5 \times 
10^{-25}~{\rm s}$, it decays with a probability close to 100\% to a $W$ boson and $b$ quark, 
and has a width much larger than $\Lambda_{\rm QCD}$, namely $\Gamma_t \simeq 1.4$~GeV~\cite{PDG}
and, because of this, it cannot form a QCD bound state like a toponium $t\overline{t}$ or a heavy-light
$q\overline{t}$ meson~\cite{Bigi:1986jk}.

\subsection{Nonrelativistic QCD (NRQCD)}
\label{sub:NRQCD}

In this and in the following sections we outline the construction of nonrelativistic approximations 
to QCD that lead to EFTs adequate for studying quarknonium in free space. It is not our aim to give a 
full account of the field of nonrelativistic QCD or of the results obtained within such approximations;
we direct the interested reader to excellent review articles available in the literature. We focus 
on concepts and techniques that we believe are useful for building EFTs for studying quarkonium in 
matter. A theoretical framework rooted in QCD and built on controllable approximations that can be 
systematically improved is not yet available for quarknonium in matter. It is our hope that the
present and the following chapters will motivate readers to contribute to this endeavor.

As pointed out previously, when $m_Q \gg \Lambda_{\rm QCD}$, the quark and antiquark in the 
rest frame of a heavy quarkonium are nonrelativistic, with relative velocities 
$v \ll 1$. This implies that relative momenta are given by $|\p| \sim  m_Q v \ll m_Q$. 
A nonrelativistic approximation involves an expansion in powers of $|\p|/m_Q$; since $|\p|/m_Q  
\sim v \ll 1$, the velocity $v$ can be used as a small expansion parameter. The binding energy $E_B$ 
is another important scale in the problem. From the virial theorem, $E_B \sim m_Q v^2 \ll |\p|$. 
Therefore, the heavy quarkonium bound state is charaterized by three well-separated scales, traditionally 
named {\em hard}, {\em soft} and {\em ultrasoft}:

\begin{itemize}

\item Hard: $m_Q$, the heavy quark mass, it is the largest scale;

\item Soft: $|\p| \sim m_Q v$, the relative momentum;

\item Ultrasoft: $E_B \sim m_Q v^2$, the binding energy.  

\end{itemize}

NRQCD is an EFT designed to accurately reproduce the low momentum behavior of QCD in the heavy quark 
sector, where $m_Q \gg \Lambda_{\rm QCD}$. NRQCD is obtained from QCD by {\em integrating out} 
the hard scale $m_Q$. The heavy-quark field operators 
are two-component nonrelativistic Pauli spinors, $\psi$ for the quark and $\chi$ for the antiquark, 
and gluons are kept fully relativistic. The field $\psi$ annihilates a heavy quark and $\chi$ creates
a heavy antiquark. For simplicity, we consider a single heavy flavor quark, so there is no need
for flavor indices in  $\psi$ and $\chi$ and $m_Q$. Here we will follow the simplest path to construct 
the lowest-order terms of the Lagrangian: one writes down the most general gauge-invariant, 
rotationally symmetric and Hermitian local Lagrangian that respects P (parity), T (time-reversal) 
and Galilean invariances{\textemdash}here we follow closely the presentation in Ref.~\cite{Paz:2015uga}. 
Moreover, the Lagrangian is built using a power counting scheme, whose meaning will become clear in 
the next discussions. We concentrate on the part of the effective Lagrangian that describes the couplings 
of the heavy quark and antiquark fields $(\psi, \chi)$ to gluon fields; they are written as an expansion 
in powers $1/m_Q$. The NRQCD Lagrangian density can be written generically as
\beq
{\cal L}_{\rm NRQCD} =  {\cal L}_{\rm light} + {\cal L}_{\psi\psi} 
+ {\cal L}_{\chi\chi} + {\cal L}_{\psi\psi\chi\chi} + \cdots,
\label{L-NRQCD}
\eeq 
where ${\cal L}_{\rm light}$ collects the contributions containing only soft and ultrasoft 
light-quark and gluon fields. Let us start discussing the quadratic terms in $(\psi, \chi)$; they are of the form
\bea
{\cal L}_{\psi\psi} &=& \sum_{n} C_n \, \psi^\dagger(x) \, \dfrac{O_n(x)}{m^n_Q} \, \psi(x) 
\nn \\[0.25true cm]
&=& {\cal L}^{(0)}_{\psi\psi}  + {\cal L}^{(1)}_{\psi\psi}  + {\cal L}^{(2)}_{\psi\psi} + \cdots ,
\label{L-psipsi}
\eea
with the $O_n$ built from gluon fields, their derivatives with respect to time and space coordinates 
and contractions with spin matrices $\vsigma$. The Lagrangian ${\cal L}_{\chi\chi}$ is of same form 
as Eq.~(\ref{L-psipsi}), but with $\psi \rightarrow \chi$ and a global sign change in some of the terms 
due to charge-conjugation symmetry. In addition, the $\cdots$ in Eq.~(\ref{L-NRQCD}) include pure-gluon terms 
and of course multiquark interactions, beyond the four-quark term. The mass dimension of fermion fields 
is $[\psi] = [\chi] = 3/2$. The factors of $m_Q$ are introduced in order the coefficients 
$C_n$ be dimensionless. Since $[{\cal L}] = 4$, therefore $[O_n] = n+1$. The coefficients $C_n$ are 
functions of $\alpha_s$, the quark mass $m_Q$ and a regularization cutoff like the scale $\mu$ of
dimensional regularization; they are determined by a {\em matching procedure}: this amounts to 
calculating a low-energy heavy-quark on-shell scattering amplitude in NRQCD at a given order in 
$\alpha_s$ and up to some order in the small parameter expansion $v$, and equating the result to the same
calculation in full QCD at some matching scale, e.g. at $\mu = m_Q$. That perturbation theory can be used 
follows from the fact that corrections to the coefficients from quantum fluctuations are from short-distances. 

To build gauge-invariant expressions, one can use the covariant derivatives, written as Hermitean operators 
as  $iD_t = i \partial/\partial t - g A_0$, $i\D = i\vec{\nabla} + g \vec{A}$. Whenever convenient, 
one can use the relation of the covariant derivatives to the color electric and magnetic fields: 
$E^i = G^{i0} = - (i/g) [D_t,D^j]$ and $B^i = \epsilon^{ijk} \, G_{jk} = (i/g) \epsilon^{ijk} \, [D_j,D_k]$. 
Their mass dimensions are $[D_t] = [\D] = 1$, and $[\El] = [\B] = 2$. One last ingredient 
in the construction of the effective Lagrangian is field redefinition, in that some terms in $O_n$ that 
are in principle allowed by the symmetries can be eliminated by redefining the fields~\cite{Manohar:1997qy}
{\textemdash}for a very recent discussion on field redefinitions in EFTs, see Ref.~\cite{Passarino:2016saj}. 
Table~\ref{tab:dimPT} summarizes the dimensions and transformation properties under $P$ and $T$ of all the 
building blocks.

\begin{table}[htb]
\caption{\label{tab:dimPT} Mass dimension and transformation properties under parity $P$ and time-reversal
$T$ of the basic quantities used to build the Lagrangian of NRQCD.}
\begin{center}
\begin{tabular}{cccccc}
\hline
{}        & {}        & {}         & {}        & {}        & {}  \\[-0.4true cm]
          &    $iD_t$ & $i\D$ & $\El$ & $\B$ & $\vsigma$\\
\hline
{}        &     {}    & {}         & {}        & {}        & {}  \\[-0.4true cm]
Mass dim. &       1   &     1      &     2     &     2     &     0    \\[0.2true cm]
P         &      $+$  &    $-$     &    $-$    &    $+$    &    $+$   \\[0.2true cm]
T         &      $+$  &    $-$     &    $+$    &    $-$    &    $-$   \\[0.10true cm] 
\hline
\end{tabular}
\end{center}
\end{table}

Starting with the lowest-dimension operator $O_0$, one has two operators that can be used: $m_Q$ and $iD_t$. 
The first can be eliminated by using the field redefinition $\psi \rightarrow e^{im_Qt}\psi$; the presence of
such a term would spoil the power-counting scheme in $1/m_Q$. What remains is then the second operator, 
leading to  $C_0 \, \psi^\dag(x) \, iD_t \, \psi(x)$. To obtain the canonical $i\partial/\partial t$ term in 
the equation of motion for $\psi$ (i.e. the Schr\"odinger equation), one needs to choose $C_0 = 1$. Therefore,
the zeroth-order Lagrangian is given by
\beq
{\cal L}^{(0)}_{\psi\psi} = \psi^\dag(x) \, iD_t \, \psi(x). 
\label{L-psipsi-0}
\eeq
The Lagrangian ${\cal L}^{(0)}_{\chi\chi}$ is of the same form as this with $\psi \rightarrow \chi$.

At the next order, there are in principle three terms of dimension~2 contributing to $O_1$: $(iD_t )^2$, 
and terms proportional to the product of $iD^i$ and $iD^j$. The first leads to an interaction of the form 
$ C_1 \psi^\dag (iD_t)^2\psi$. This term can be eliminated by using the field redefinition 
$\psi \rightarrow \psi - C_1 (iD_t/2m_Q)\psi$, as can be verified very easily. To obtain an Hermitean
operator, the $iD^i$ and $iD^j$ must be combined in the form of a commutator $i [iD^i,iD^j]$ or
anticommutador $\{iD^i,iD^j\}$; these must be contracted with $\sigma^i$, $\delta^{ij}$ and $\epsilon^{ijk}$ 
to obtain operators respecting rotational and $T$ symmetries. Since the commutator gives the magnetic field, 
one can write the first-order Lagrangian as 
\beq
{\cal L}^{(1)}_{\psi\psi} = \psi^\dag(x)  \left( \dfrac{\D^2}{2m_Q} 
+ c_F \, g \dfrac{\vsigma\cdot\B}{2m_Q}\right) \psi(x),
\label{L-psipsi-1}
\eeq
where we used the fact the coefficient of the first term in equal to unity~\cite{Manohar:1997qy} 
and the index $F$ in $c_F$ stands for Fermi, motivated by the equivalent term that gives the
Fermi hyperfine splitting in QED~\cite{Kinoshita:1995mt}. ${\cal L}^{(1)}_{\chi\chi}$ is of the
same form but with an overall minus sign multiplying both terms.

Construction of $O_2$ is also straightforward, though a bit more involved~\cite{Paz:2015uga}. 
We will not use them here, but they are given next for completeness~\cite{Bodwin:1994jh}:
\bea
{\cal L}^{(2)}_{\psi\psi} = \psi^\dag(x) \left[ c_D g \, \dfrac{\left(\D\cdot\El 
- \El\cdot\D\right)}{8m^2_Q}
+ c_S g \,\dfrac{i\vsigma\cdot\left(\D\times\El-\El\times\D\right)}{8m^2_Q}\right]
\psi(x),
\label{L-psipsi-2}
\eea
where here $D$ and $S$ in $c_D$ and $c_S$ stand for Darwin and spin-orbit. ${\cal L}^{(2)}_{\chi\chi}$ 
is the same as this with $\psi \rightarrow \chi$. The coefficients $c_s$ and $c_F$ are related to each other
by~\cite{Manohar:1997qy} $c_S = 2 c_F -1$.

From dimensional analysis, one has that the four-fermion contact interaction terms 
${\cal L}_{\psi\psi\chi\chi}$ demand a factor $1/m^2_Q$ to obtain a dimensionless coefficient; 
we write these constact interactions as~\cite{Pineda:1998kj}:
\bea
{\cal L}_{\psi\psi\chi\chi} &=& \dfrac{d_{ss}}{m^2_Q} \, \psi^\dag(x)\psi(x) \, \chi^\dag(x)\chi(x) 
+ \dfrac{d_{sv}}{m^2_Q} \, \psi^\dag(x)\vsigma\psi(x) \cdot \chi^\dag(x) \vsigma \chi(x) 
\nn \\ 
&& + \, \dfrac{d_{vs}}{m^2_Q} \, \psi^\dag(x)T^a\psi(x) \, \chi^\dag(x)T^a\chi(x) 
+ \dfrac{d_{vv}}{m^2_Q} \, \psi^\dag(x)T^a\vsigma\psi(x) \cdot \chi^\dag(x)T^a\vsigma\chi(x) .
\eea
We refer to Ref.~\cite{Manohar:1997qy} for a complete list of operators up to $1/m^3_Q$ in the
NRQCD Lagrangian.

As mentioned previously, the dimensionless parameters $c_F, c_D, d_{ss}, \cdots$ are determined by 
a matching procedure. The factors of 2 and 8 in Eqs.~(\ref{L-psipsi-2}) and (\ref{L-psipsi-2}) were
introduced to reproduce tree-level matching, so that a generic coefficient is given as
$C_n \sim 1 + {\cal O}(\alpha_s)$ when loop corrections are calculated. As examples, we quote 
the ${\cal O}(\alpha_s)$ expressions for the coefficients of the bilinear terms, determined in
Ref.~\cite{Manohar:1997qy} by using dimensional regularization
\beq
c_F = 1 + \dfrac{\alpha_s}{2\pi} \left(C_A + C_F\right), \hspace{1.0cm}
c_D = 1 + \dfrac{\alpha_s}{2\pi} C_A, 
\label{cF-cD}
\eeq
where $C_F = 4/3$ and $C_A = 3$ are the eigenvalues in the fundamental representation of the SU(3) 
Casimir operators. The four-fermion contact interactions were calculated at ${\cal O}(\alpha^2_s)$ 
in Ref.~\cite{Pineda:1998kj}, also in dimensional regularization:
\bea
d_{ss} &=& \dfrac{2}{3} C_F \left( \dfrac{C_A}{2} - C_F\right) \alpha^2_s, \hspace{1.0cm}
d_{sv} = C_F \left( \dfrac{C_A}{2} - C_F\right) \alpha^2_s, \nn \\
d_{vs} &=& \left(\dfrac{4}{3} C_F + \dfrac{11}{12} C_A\right) \alpha^2_s, \hspace{1.0cm}
d_{vv} = \left(2C_F - \dfrac{1}{4} C_A\right) \alpha^2_s.
\label{dss-et-al}
\eea
In the above, we have chosen the matching scale to be $\mu = m_Q$. 

It is important to note that we have not written down the contributions of the 
light quark degrees in the above. They, of course, cannot be neglected in any realistic
QCD calculation. As we discuss in the next section, in some applications their contributions
to quarkonium properties can be introduced by coupling the nonrelativistic field to effective
hadron degrees of freedom, like Goldstone boson fields. 

NRQCD is rigorously derived from QCD in a systematic manner and has been adapted to 
lattice QCD~\cite{Thacker:1990bm}, where it has shown to provide an efficient framework for 
simulating heavy quarks~\cite{Hashimoto:2004fv}. For analytical calculations, however, NRQCD 
still has too many active degrees of freedom, that is, it contains degrees that never appear 
as asymptotic states, only through virtual fluctuations. For example, while the binding energy, 
the ultrasoft scale, is a property of the asymptotic state, the relative momentum, the soft scale, 
is not; they are entangled in NRQCD. This makes difficulties with the power counting, as discussed
in-depth in the reviews in Refs.~\cite{{Brambilla:2004jw},{Pineda:2011dg}}. The solution within the
framework of an EFT is that if one is interested in physics at the scale of the binding 
energy, the degrees of freedom at scales higher than the binding energy should be integrated out. 
This idea was implemented in Ref.~\cite{Pineda:1997bj} and further elaborated in 
Ref.~\cite{Brambilla:1999xf}. The resulting EFT is named potential nonrelativistic QCD, pNRQCD,
and will be discussed next. 

\subsection{Potential nonrelativistic QCD (pNRQCD) and van der Waals forces}
\label{sub:pNRQCD}

pNRQC gives a natural connection of QCD with a nonrelativistic Hamiltonian 
and the Schr\"odinger equation and it has proven to provide an adequate framework for describing
heavy quarkonium dynamics. Its resemblance to nuclear many-body potential models makes it 
an adequate starting point for constructing EFTs for heavy quarkonia in a nuclear medium, 
particularly when treating nuclei within an independent-particle approximation. 

As discussed in the previous section, the largest scale in a quarkonium bound state is 
the heavy quark mass $m_Q$. Other two relevant scales are the relative momentum 
$|\p| \sim m_Q v$ and the relative kinetic energy $E \sim m_Q v^2$, where $v$ is the 
relative velocity $v$ between $Q$ and $\overline{Q}$. These scales must be compared
to $\Lambda_{\rm QCD}$, which sets the scale of strong coupling. If one {\em assumes} 
that they satisfy the following hierarchy
\beq
m_Q \gg m_Q v \gg m_Q v^2 \gg \Lambda_{\rm QCD},
\label{scales}
\eeq
then $v \sim \alpha_s$ and the integration of degrees of freedom at the scale $m_Q v$ out can be carried 
out using perturbation theory. NRQCD can be matched to an EFT whose fields describe the ultrasoft (US) 
degrees of freedom. This gives rise to pNRQCD in the weak coupling regime, the 
matching coefficients being potentials that depend on the relative coordinate and momentum between 
$Q$ and $\overline{Q}$. Moreover, the pNRQCD Lagrangian density can be written in terms of color singlet 
and color octet fields representing the $Q\overline{Q}$ pair, and fields representing the US gluons and 
light quarks. We refer the reader to the reviews in Refs.~\cite{{Brambilla:2004jw},
{Pineda:2011dg}} for the technical aspects involved in the  derivations. The $Q\overline{Q}$ 
fields depend on two coordinates, which are conveniently chosen to be the center-of-mass $\R$ 
and relative $\r$ coordinates. The US gauge and light-quark fields depend only on $\R$; 
in practice, this corresponds to a multipole expansion of the US fields. At leading order 
in $1/m_Q$ and at ${\cal O}(R)$ in the multipole expansion, one has that the action of pNRQCD 
can be written as~\cite{Pineda:1997bj,Brambilla:1999xf}
\beq
S_{\rm pNRQCD} = \int dt \, d^3R \; {\cal L}_{pNRQCD} ,
\eeq 
with ${\cal L}_{\rm pNRQCD}$ given by
\bea
\hspace{-0.5cm}
{\cal L}_{\rm pNRQCD} &=&  {\cal L}_{\rm light} + 
\int d^3r\, \Bigl\{ {\rm Tr}\left[{\rm S}^{\dagger}\left(i\partial_0 - h_{\rm s}\right) {\rm S} 
+ {\rm O}^{\dagger} \left(iD_0-h_{\rm o}\right) {\rm O}\right] \nn \\
&& + \, g V_A(r) \, {\rm Tr}\left[ {\rm O}^{\dagger} (\r\cdot\El) {\rm S} 
+ {\rm S}^{\dagger} (\r\cdot\El) {\rm O} \right] 
+ \dfrac{g}{2} V_B(r) \, {\rm Tr}\left[ {\rm O}^{\dagger} (\r\cdot\El) {\rm O}
+ {\rm O}^{\dagger} {\rm O} (\r\cdot\El) \right] 
\Bigr\} ,
\label{pnrqcdla}
\eea
where ${\rm S}={\rm S}(t,\R,\r)$ and  ${\rm O}={\rm O}(t,\R,\r)$ are respectively 
quark-antiquark matrix-valued singlet and octet fields, normalized with respect to color, 
${\rm S} = {\bfm 1} S/\sqrt{N_c}$ and ${\rm O} 
= T^a O^a/\sqrt{T_F}$, with $N_c = 3$ the number of colors and $T_F = 1/2$, and $h_{\rm s}$ and 
$h_{\rm o}$ are the singlet and octet Hamiltonians 
\be
h_{\rm s}=-\dfrac{{\nabla_r}^2}{m_Q}-\dfrac{{\nabla_R}^2}{4m_Q} + V_{\rm s}(r) , \hspace{1.0cm}
h_{\rm o}=-\dfrac{{\nabla_r}^2}{m_Q}-\dfrac{{D_R}^2}{4m_Q} + V_{\rm o}(r) ,
\ee
and 
\be
V_{\rm s}(r)= - C_F \dfrac{\alpha_s}{r}\,, \hspace{0.5cm} V_{\rm o}(r) = \dfrac{\alpha_s}{2N_c} \dfrac{1}{r} ,
\hspace{0.5cm}V_A(r) = V_B(r) = 1 .
\ee
We recall that the light-quark and gauge fields, including those gauge fields in ${\cal L}_{\rm pNRQCD}$, 
are functions of time and $\R$: $q_i = q_i(t,\R)$, $\bar q_i(t,\R)$, $G^{\mu \nu}
= G^{\mu \nu}(t,\R)$, $\El^i  \equiv G^{i0}(\R,\,t)$, $iD_0 O \equiv i\partial_0O 
- g\left[A_0(\R,\,t),O\right]$. At order $1/m^2_Q$ there appear retardation terms and spin-dependent 
potentials, all of them calculable analytically.

A heavy quarkonium injected in a nucleus at low energies interacts by exchanging gluons with the light 
quarks of the nucleons. When this energy is much smaller than the quarkonium binding energy $E_B \sim m_Q v^2 
\approx m_Q \alpha^2_s \gg \Lambda_{\rm QCD}$, the internal structure of the quarkonium is not 
resolved. Therefore, one can treat the quarkonium state in terms of an independent color-singlet field. 
This is equivalent to the statement that one can integrate out the 
ultra-soft scale $m_Q\alpha^2_s$ and match pNRQCD to another EFT whose degrees of freedom are color-singlet 
quarkonium states, described by a field generically denoted by $\varphi$ and  gluon fields at the scale 
$\Lambda_{\rm QCD}$~\cite{Vairo:2000ia}. Such an EFT is known as gluonic van der Waals EFT (gWEFT), the 
name being motivated by its original construction~\cite{Vairo:2000ia} aiming at deriving long range color 
van der Waals interactions in quarkonium-quarkonium interactions. Very recently, gWEFT was applied in 
Ref.~\cite{Brambilla:2015rqa} to the concrete case of the $\eta_b$-$\eta_b$ system. A similar van der Waals
EFT was recently developed in Ref.~\cite{Brambilla:2017ffe} to describe the low-energy dynamics of an atom 
pair. In an earlier publication~\cite{Holstein:2008fs}, effective field theory theories were used
to treat the electromagnetic scattering of two massive particles, wherein one particle (or both) is 
electrically neutral.  

Gluons at the scale $\Lambda_{\rm QCD}$ interact nonperturbatively with quarkonium states and the quarks
of light hadrons. In the case of quarkonium-quarkonium interactions, such ultrasoft (US) gluons can be hadronized  
in terms of (pseudo) Goldstone bosons (GB) and gWEFT can be matched to a chiral EFT whose degrees of freedom 
are decribed by the $\varphi$ fields and pseudoscalar GB fields~\cite{Brambilla:2015rqa}. The values of the 
couplings of the matching to the chiral EFT can be determined from the anomaly in the trace of the QCD 
energy-momentum tensor~\cite{Chanowitz:1972vd,Chanowitz:1972da,Crewther:1972kn, Freedman:1974gs,Collins:1976yq,
Voloshin:1980zf, Novikov:1980fa,Fujii:1999xn,Voloshin:2007dx}. Although the quarkonium-quarkonium 
interaction is not our focus in this review, we discuss in the following a systematic procedure for
obtaining its long-range van der Waals component by matching gWEFT Lagrangian to a chiral 
EFT~\cite{Brambilla:2015rqa}. Our aim in presenting such a discussion here is twofold. First, to motivate 
its possible extension to study quarkonium interactions with light hadrons that couple to pions and other 
hadronic degrees of freedom, with couplings fixed phenomenologically or by lattice QCD 
results~\cite{jaume}. Second, to emphasize the close connection of such an EFT approach with model 
calculations of quarkonium binding in medium using phenomenological Lagrangians; in particular with 
those based on evaluations meson-antimeson loop contributions to quarkonium in-medium 
self-energies that we discuss in sections ahead~\cite{{Ko:1992tp},{Ko:2000jx},{Krein:2010vp},{Tsushima:2011fg},
{Tsushima:2011kh},jpsi4,{Cobos-Martinez:2017vtr}}. 

The following discussion is based on the presentation in Ref.~\cite{Krein:2017njw}; further details can
be found in Ref.~\cite{Brambilla:2015rqa}. As mentioned above, gWEFT is an EFT at the scale 
$\Lambda_{\rm QCD}$ and it is not difficult to convince ourselves by using dimensional analysis that 
its lowest order Lagrangian can be written as (for simplicity of notation, we omit labels referring 
to spin and other eventually required quantum numbers):
\bea 
{\cal L}_{\rm gWEFT} =  {\cal L}_{\rm light} + 
\varphi^{\dagger}(t,\vec{R})\left(i\partial_0  - E_{\varphi} + \dfrac{{\nabla_{\vec{R}}^2}}{4m_Q}
+ \dfrac{1}{2}\alpha_\varphi g^2 \El^2 +\cdots\right)\varphi(t,\vec{R})\,,
\label{LgWEFT}
\eea
where $\cdots$ stands for relativistic kinetic corrections or other higher-order operators coupling $\varphi$ 
to gluons and $\beta$ is the matching coefficient, which is the chromopolarizability of the quarkonium, 
given by~\cite{Voloshin:1978hc,Peskin:1979va,Bhanot:1979vb}
\be
\alpha_\varphi = - \dfrac{2V^2_A T_F}{3N_c} \langle \varphi|{r}^i\dfrac{1}{E_{\varphi}-h_{\rm o}}{r}^i|\varphi\rangle , 
\label{epol}
\ee
with $|\varphi\rangle$ being the quarkonium bound-state wave function. If the hierarchy in Eq.~(\ref{scales})
is strictly valid, then, at order $1/m_Q$ and at ${\cal O}(R)$, $|\varphi\rangle$ is an $1S$ Coulombic state
and the polarizability $\beta$ can be evaluated in closed form~\cite{Voloshin:1978hc,Leutwyler:1980tn}. 
Note that in Ref.~\cite{Brambilla:2015rqa} the polarizability is denoted by $\beta$. In the evaluation of 
$\alpha_\varphi$, one inserts a complete set of intermediate states that are eigenstates of the 
octet Hamiltonian $h_{\rm o}(\vec{r})$. Since $h_{\rm o}(\vec{r})$ is a repulsive Coulomb potential, the octet 
intermediate states correspond to Coulombic continuum eigenstates. A commonly employed 
approximation~\cite{Ko:2000jx,Peskin:1979va,Bhanot:1979vb,{Kaidalov:1992hd}} in the evaluation of 
$\alpha_\varphi$ is to use plane waves in place of the Coulombic continuum eigenstates. 
This corresponds to the $N_c \rightarrow \infty$ limit. 

At energies of order $m_\pi \ll \Lambda_{QCD}$, the relevant degrees of freedom are the quarkonium 
states and the pseudo GB and it is natural to integrate out the gluons in favor of pions, 
and match gWEFT to a chiral EFT, denoted $\chi EFT$, in which the pseudo GB enter as explicit 
degrees of freedom. Specifically, limiting the discussion to the case that quarkonium field $\varphi$ 
is a scalar under chiral symmetry, the interaction operators with Goldstone bosons can be easily 
constructed making use as building blocks the unitary matrix $U(x)$, that parametrizes the Goldstone 
boson fields and that involving the light-quark masses, Eqs.~(\ref{L-GB}) and (\ref{L-GB-m}).
To leading order, the Lagrangian density of  $\chi$EFT can be written as (we limit the
discussion to the pion sector of the $U(x)$ matrix)~\cite{Brambilla:2015rqa}
\bea
\hspace{-0.5cm}
{\cal L}_{\rm \chi EFT}^{\varphi} &=& \varphi^{\dagger}\left(i\partial_0 
+ \dfrac{{\nabla^2}}{2m_{\varphi}}\right)\varphi + {\cal L}_{GB} + {\cal L}_{{\rm GB}-m_q} 
\nn \\
&& + \,  \dfrac{f^2}{4} \, \varphi^{\dagger}\varphi
\left[
c_{d0} {\rm Tr}\left( \partial_0 U \partial_0 U^\dagger \right) 
+ c_{di} {\rm Tr}\left( \partial_i U \partial^i U^\dagger \right)
+ c_{m}  {\rm Tr}\left( \chi^\dagger U+\chi U^\dagger\right) 
\right] + V_{\rm cont.}(\varphi^\dag,\varphi) ,
\label{chi-EFT}
\eea
where $V_{\rm cont.}(\varphi^\dag,\varphi)$ are $\varphi$-contact interactions{\textemdash{they do not play 
any role in the long-distance properties of the quarkonium-quarkonium interaction in lowest order,
although they are needed in the renormalization of ultraviolet divergences coming from 
chiral loops}}. The $c_{d0}$, $c_{di}$ and $c_{m}$ are matching coefficients that 
can be obtained from the matrix element~\cite{Voloshin:2007dx}  
\be
g^2 \langle \pi^+(p_1) \pi^-(p_2)|\El_a^2|0\rangle = \dfrac{8\pi^2}{b}\left(\kappa_1 \, E_1 E_2
- \kappa_2 \, \p_1 \cdot \p_2 + 3m^2_{\pi}\right),
\label{trace-anomaly}
\ee
where $p_1 = (E_1,\p_1)$ and $p_2= (E_2,\p_2)$ are the pion four-momenta, 
$\kappa_1=2-9\kappa/2$, $\kappa_2=2+3\kappa/2$, $b= b_0/4\pi$, with $b_0$ given in Eq.~(\ref{b_0}),
and $\kappa \simeq 0.2$, as extracted from pionic transitions of quarkonium 
states~\cite{Voloshin:2007dx}. More specifically, the matching coefficients can be obtained by 
equating the amplitudes for two-pion production calculated in gWEFT and $\chi$EFT~\cite{Brambilla:2015rqa}:
\be
\dfrac{4\pi^2\alpha_\varphi}{b}\left(\kappa_1 E_1 E_2 - \kappa_2 \, \p_1\cdot\p_2 + 
3m^2_{\pi}\right)
= - c_{d0} \, E_1 E_2 + c_{di} \, \p_1 \cdot \p_2 - c_m \, m^2_{\pi}\,, 
\ee
which yields
\be
c_{d0} = - \dfrac{4\pi^2\alpha_\varphi}{b}\kappa_1\,, \qquad
c_{di} = - \dfrac{4\pi^2\alpha_\varphi}{b}\kappa_2\,, \qquad
c_{m}  = - \dfrac{12\pi^2\alpha_\varphi}{b}\,. 
\label{mcm}
\ee
 
Once the matching coefficients appearing in Eq.~(\ref{chi-EFT}) are determined, one can
obtain an effective potential describing $\varphi\varphi$ interactions at low momentum 
transfers. For momentum transfers of the order of the pion mass, $\k_{\varphi\varphi} \sim m_\pi$, 
the relative $\varphi\varphi$ kinetic energy is much smaller than the pion mass: 
$\k^2_{\varphi\varphi}/m_\varphi \sim m^2_{\pi}/m_\varphi \ll m_\pi$. Under such circumstances,
the pions can be integrated out and the $\varphi\varphi$ potential appears as a matching coefficient 
in an effective $\varphi\varphi$ Lagrangian, the leading order of which can be written as
\bea
L_{\varphi\varphi} &=& \int d^3R \, \varphi^{\dagger}(t,\R) \left(i\partial_0 
+ \dfrac{{\nabla^2}}{2m_{\varphi}}\right)\varphi(t,\R) \nn \\
&& - \, \dfrac{1}{2} \int d^3{R} d^3{R'}\, \varphi^{\dagger}(t,\R) \varphi^{\dagger}(t,\R')
\,W_{\varphi\varphi}(\R, \R')\,\varphi(t,\R') \varphi(t,\R),
\label{L-WEFT}
\eea
where $W_{\varphi\varphi}(\R, \R')$ contains short- and long-distance contributions. 
Details on the derivations and explicit formulae for the pion contribution to $W(\R, \R')$ 
can be found in Ref.~\cite{Brambilla:2015rqa}; here we simply quote the final result for its 
long-distance part ($r = |\R_1-\R_2|$):
\be
W^{\rm vdW}_{\varphi\varphi}(r) = \displaystyle{\lim_{r \gg 1/2m_\pi}} 
W_{\varphi\varphi}(r) 
= -\dfrac{3(3+\kappa_2)^2\pi^{3/2}\alpha_\varphi^2}{4b^2}\dfrac{m^{9/2}_{\pi}}{r^{5/2}} \, e^{-2m_{\pi} r},
\label{longrangevdW}
\ee
which is identified as a color van der Waals contribution to the $\varphi\varphi$ interaction. 
An earlier result derived in Ref.~\cite{Fujii:1999xn} using a similar method to ours is contained 
in Eq.~(\ref{longrangevdW}) if one takes $\kappa_2=2$ in that equation and also neglects 
contributions proportional to $m_\pi^2$ in the expression coming from the trace anomaly, 
Eq.~(\ref{trace-anomaly}). Neglecting such terms does not change the functional dependence 
on $r$ and $m_\pi$ of $W^{\rm vdW}_{\varphi\varphi}(r)$, but does make it weaker by a factor 
of $16/25$. Explicit numerical results for the case of the $\eta_b$-$\eta_b$ system
are presented in Ref.~\cite{Brambilla:2015rqa}. In particular, for future reference, we mention
that the numerical value for the polarizability of $\eta_b$, for a bottom mass of 
$m_Q = 5$~GeV, was obtained to be 
\be
\alpha_{\eta_b} = 0.50^{+0.42}_{-0.38}~{\rm GeV}^{-3}\,.
\label{betaresult}
\ee
where the central value refers to $\alpha_{\rm s}($1.5~GeV$) = 0.35$, the largest to  
$\alpha_{\rm s}($2~GeV$) = 0.3$ and the smallest to $\alpha_{\rm s}($1~GeV$) = 0.5$.
We mention that the effect of using continuum Coulomb wave functions instead of plane 
waves~\cite{Ko:2000jx,Peskin:1979va,Bhanot:1979vb,{Kaidalov:1992hd}} in the evaluation 
of the integral in Eq.~(\ref{epol}) is to increase $\alpha_{\eta_b}$ by at 
most~5\%~\cite{{Brambilla:2015rqa},{Krein:2017njw}}.

The results show very clearly that as the attraction from the Coulomb potential weakens, 
the polarizability increases. This is because the size of the color charge distribution
increases. One consequence of the larger value of the polarizability is that the van
der Waals interaction becomes stronger. Excited quarkonia states are expected to have
larger polarizabilities and therefore interact with stronger van der Waals forces. 
On the other hand, a calculation of $\alpha_\varphi$ for such states with Eq.~(\ref{epol}) might
not be reliable, as excited states of charmonia or bottomonia are not Coulomb bound 
states and, most likely, nonperturbative physics plays an important role in the
determination of the polarizability of such states. 

It is remarkable that one is able to derive from QCD an analytic expression for a
hadron-hadron interaction. One should, however, recall that such a derivation was made
possible by assuming the validity of the hierarchy of scales in Eq.~(\ref{scales}). Amongst
the most important implications of the hierarchy is the weakly coupled nature of the
quarkonium bound states, which allows the use of perturbation theory to integrate out
the high-energy scales. Now, when the relative position of the scales $m_Q v$ and 
$\Lambda_{\rm QCD}$ are inverted in Eq.~(\ref{scales}), QCD becomes strongly coupled 
and an entire new strategy is required to derive EFT for treating low-energy heavy 
quarkonium interactions. To describe the physics below $\Lambda_{\rm QCD}$, the use of 
explicit hadronic degrees of freedom is one possibility and hence the appropriate degrees 
of freedom in this regime are described by quarkonium singlet fields, (pseudo) Goldstone 
boson fields and other hadron fields representing heavy-light mesons and light baryons. 
An advantage of using explicit hadron degrees of freedom is that the correct chiral 
non-analytic behavior of observables as a function of the pion mass is obtained more
easily~\cite{{Thomas:1999mu},{Thomas:2000fa},{Bicudo:2001cg},{Perry:2017ypk}}.
The construction of such an EFT would need input from phenomenology and lattice QCD simulations 
to fix parameters~\cite{jaume}. In particular, to describe e.g. the quarkonium states $\eta_c$ 
and $J/\Psi$, one would need to include in addition to the color Coulomb potential discussed above, 
nonperturbative interactions associated with confinement. While great progress has been achieved 
recently in constructing such an EFT to describe heavy quarkonium hybrids by using as degrees of 
freedom heavy quarks and excited glue degrees of freedom~\cite{Berwein:2015vca}, the treatment of light
quarks remains challenging. The use of models is an essential part of contemporary research in
the field. Recent studies of interactions of heavy quarkonia wherein nonperturbative forces are 
involved have been conducted in Refs.~\cite{{Giordano:2015oza},{Karliner:2016zzc},{Richard:2017vry},
{Bai:2016int}}. 

In summary, the study of heavy quarkonium in a nucleus requires the use of different pieces
of theoretical tools. The use of EFTs for treating heavy quarkonium states in free space allows 
us to constrain many of their properties, but to describe their interactions with light
hadrons requires the use of models. In particular, for making reliable predictions for experimental
searches of possible bound states of heavy quarkonium with atomic nuclei, no matter how well
a heavy quarkonium is understood in free space, one still needs a well-constrained nuclear 
many-body model. Most of the predictions presented in this review rely on one of such models, 
the quark-meson coupling (QMC) model, a self-consistent quark-based model that has proven to 
successfully describe a great variety of nuclear phenomena. The model is particularly suitable,
as we discuss in sections ahead, for studying nuclear binding of the light quarkonium $\phi$ and
also the $\eta$ and $\omega$ mesons, as well as of heavy-flavored hadrons, the $D$ and $\overline{D}$ mesons, 
and $\Lambda_c$ and $\Lambda_b$ baryons. The foundations of QMC model and its predictions for hadron
properties in medium are the subject of the next section. 
%
%
%
%\input{sec_qmc.tex}
%
%
%
%%%%%%%%%%%%%%%%%%%%%%%%%%%%%%%%%%%%%%%%%%%%%%%%%%%
%%%%%%%%%% Section 2 
%%%%%%%%%%%%%%%%%%%%%%%%%%%%%%%%%%%%%%%%%%%%%%%%%%%
\section{Quark-meson coupling (QMC) model}
\label{qmc}

In this section we briefly review the quark-meson coupling (QMC) model, 
the quark-based model for nuclear matter and finite nuclei  
invented by Guichon~\cite{QMCGuichon} and further developed and 
applied to finite nuclei in Ref.~\cite{Guichonfinite}.

Many nuclear phenomena now seem to indicate that 
the traditional approach, omitting any consideration of the underlying 
quark and gluon degrees of freedom, may have its limitations, 
and suggest a need for subnucleonic and subhadronic degrees of freedom.  
There is no doubt that hadrons consist of quarks, antiquarks 
and gluons and that they can respond to the 
environment and change their character in matter.  
The basic working hypothesis and assumption of the QMC model  
is that quarks play an important role in nuclei and nuclear matter.

Based on the QMC model, various nuclear phenomena have been successfully 
studied~\cite{QMCreview} starting at the quark level, using a self-consistent model 
for nuclear physics. Although there are many kinds of relativistic mean-field 
theory for nuclear physics, very few are built from the quark level. 
We emphasise especially the wide success of the QMC model 
which has been applied systematically to many nuclear phenomena. Specially 
relevant for the present review, are nuclear bound states of heavy and light 
quarkonium and light hadrons, and heavy-baryon hypernuclei. 
The model incorporates explicit quark degrees of freedom into nuclear many-body 
systems.  It is shown that at the {\em hadronic} level, 
it is certainly possible to cast the QMC model into 
a form similar to that of a quantum hadrodynamics (QHD)~\cite{QHD1,QHD2},  
or a similar type of mean-field model by re-defining the scalar field.  
However, at the same time, the QMC model can describe 
how the internal structure of hadrons changes in 
a nuclear medium. That is the greatest advantage of the QMC model and
it has opened a tremendous number of new lines of investigation.

Since the discovery of QCD as the fundamental theory of the strong interaction, 
numerous attempts have been made to derive the nuclear force within quark models. 
The QMC model stands between the traditional 
meson-exchange picture and the hard core 
quark models, namely, it is a mean-field model in the sense of 
QHD but with the couplings of $\sigma$ and $\omega$ mesons 
to confined quarks, rather than to the point-like 
nucleon.  After a considerable amount of work, 
one finds that the effect of the internal, quark structure 
of the nucleon is absorbed into the scalar polarizability 
in the effective nucleon mass in matter. 
It is the dependence of the scalar polarizability on the 
scalar field in matter (or it is numerically equivalent 
to the dependence on nuclear density) that is the heart of the 
QMC model and leads to the novel saturation 
mechanism of the binding energy of nuclear matter as a function of density. 

Because the scalar polarizability plays such an important 
role in the QMC model, it is of great interest 
to study whether the dependence of the scalar 
polarizability on the scalar field can be extracted from the 
fundamental theory, i.e., QCD.  
In Ref.~\cite{TONY}, it is shown that 
the remarkable progress in resolving the problem of 
chiral extrapolation of lattice QCD 
data gives one confidence that the pion loop contributions are under control. 
In the case of the nucleon, one can then use this 
control to estimate the effect of applying 
a chiral invariant scalar field to the nucleon, i.e., to estimate the 
scalar polarizability of the nucleon. The resulting value is in excellent 
agreement with the range found in the QMC model, which is vital to 
describe many phenomena in nuclear physics. 
Thus, in a very real sense, the results presented 
in Ref.~\cite{TONY} provide a direct connection 
between the growing power to compute hadron 
properties from QCD itself and fundamental 
properties of atomic nuclei. Further work   
in this direction is certainly necessary.

%%%%%%%%%%%%%%%%%%%%%%%%%%%%%%%%%%%%%%%%%%%%%%%%%%%%%%%%%%%%%%%%%%%%%%%%%%%%
\subsection{Nuclear matter and finite (hyper)nucleus}
\label{qmc_finite}

Using the Born-Oppenheimer approximation, a relativistic Lagrangian 
density which gives the same mean-field equations
of motion for a nucleus or a hypernucleus, in which the quasi-particles moving
in single-particle orbits are three-quark clusters with the quantum numbers
of a strange, charm or bottom hyperon or a nucleon, 
when expanded to the same order in velocity, 
is given by QMC~\cite{Guichonfinite,QMCreview,Saitofinite,QMChyp,QMCbc,QMChypbc}:  
%%%
\begin{eqnarray}
{\cal L}^{Y}_{QMC} &=& {\cal L}^N_{QMC} + {\cal L}^Y_{QMC},
\label{eq:LagYQMC} \\
{\cal L}^N_{QMC} &\equiv&  \overline{\psi}_N(\vec{r})
\left[ i \gamma \cdot \partial
- m_N^*(\sigma) - (\, g_\omega \omega(\vec{r})
+ g_\rho \dfrac{\tau^N_3}{2} b(\vec{r})
+ \dfrac{e}{2} (1+\tau^N_3) A(\vec{r}) \,) \gamma_0
\right] \psi_N(\vec{r}) \quad \nn \\
  & & - \dfrac{1}{2}[ (\nabla \sigma(\vec{r}))^2 +
m_{\sigma}^2 \sigma(\vec{r})^2 ]
+ \dfrac{1}{2}[ (\nabla \omega(\vec{r}))^2 + m_{\omega}^2
\omega(\vec{r})^2 ] \nn \\
 & & + \dfrac{1}{2}[ (\nabla b(\vec{r}))^2 + m_{\rho}^2 b(\vec{r})^2 ]
+ \dfrac{1}{2} (\nabla A(\vec{r}))^2, \label{eq:LagN} \\
{\cal L}^Y_{QMC} &\equiv&
\overline{\psi}_Y(\vec{r})
\left[ i \gamma \cdot \partial
- m_Y^*(\sigma)
- (\, g^Y_\omega \omega(\vec{r})
+ g^Y_\rho I^Y_3 b(\vec{r})
+ e Q_Y A(\vec{r}) \,) \gamma_0
\right] \psi_Y(\vec{r}), 
\nn\\
& & (Y = \Lambda,\Sigma^{0,\pm},\Xi^{0,+},
\Lambda^+_c,\Sigma_c^{0,+,++},\Xi_c^{0,+},\Lambda_b),
\label{eq:LagY}
\end{eqnarray}
%%%
where, for a normal nucleus, ${\cal L}^Y_{QMC}$ in Eq.~(\ref{eq:LagYQMC}), 
namely Eq.~(\ref{eq:LagY}) is not needed.
In the above $\psi_N(\vec{r})$ and $\psi_Y(\vec{r})$
are respectively the nucleon and hyperon (strange, charm or bottom baryon) fields. 
The mean meson fields represented by, $\sigma, \omega$ and $b$ which 
are directly coupled to the quarks self-consistently, are  
the scalar-isoscalar, vector-isoscalar and third component of  
vector-isovector fields, respectively, while $A$ stands for the Coulomb field.
When we consider the situation where a hadron $h$ is embedded in a nucleus or 
in nuclear matter, we may add the corresponding Lagrangian ${\cal L}^h$ 
instead of ${\cal L}^Y_{QMC}$. 

In an approximation where the $\sigma$, $\omega$ and $\rho$ mean fields couple
only to the $u$ and $d$ light quarks, the coupling constants for the hyperon  
are obtained as $g^Y_\omega = (n_q/3) g_\omega$, and
$g^Y_\rho \equiv g_\rho = g_\rho^q$, with $n_q$ being the total number of
valence light quarks in the hyperon $Y$, where $g_\omega$ and $g_\rho$ are 
the $\omega$-$N$ and $\rho$-$N$ coupling constants. $I^Y_3$ and $Q_Y$
are the third component of the hyperon isospin operator and its electric
charge in units of the proton charge, $e$, respectively.
The field dependent $\sigma$-$N$ and $\sigma$-$Y$
coupling strengths respectively for the nucleon $N$ and hyperon $Y$,  
$g_\sigma(\sigma) \equiv g^N_\sigma(\sigma)$ and  $g^Y_\sigma(\sigma)$,
appearing in Eqs.~(\ref{eq:LagN}) and~(\ref{eq:LagY}), are defined by
\bg
m_N^*(\sigma) &\equiv& m_N - g_\sigma(\sigma)
\sigma(\vec{r}),  
\label{effnmass}
\\
m_Y^*(\sigma) &\equiv& m_Y - g^Y_\sigma(\sigma)
\sigma(\vec{r}), 
\label{effymass}
\en
where $m_N$ ($m_Y$) is the free nucleon (hyperon) mass. 
Note that the dependence of these coupling strengths on the applied
scalar field ($\sigma$) must be calculated self-consistently within the quark
model~\cite{QMCGuichon,Guichonfinite,QMCreview,Saitofinite,QMChyp,QMCbc,QMChypbc,Whittenbury:2016sma}.
Hence, unlike quantum hadrodynamics (QHD)~\cite{QHD1,QHD2}, even though
$g^Y_\sigma(\sigma) / g_\sigma(\sigma)$ may be
2/3 or 1/3 depending on the number of light quarks $n_q$ in the hyperon 
in free space, $\sigma = 0$ (even this is true only when their bag 
radii in free space are exactly the same), this will not necessarily 
be the case in a nuclear medium.

The Lagrangian density Eq.~(\ref{eq:LagYQMC}) [or (\ref{eq:LagN}) and (\ref{eq:LagY})] 
leads to a set of equations of motion for the finite (hyper)nuclear system:
\begin{eqnarray}
& &[i\gamma \cdot \partial -m^*_N(\sigma)-
(\, g_\omega \omega(\vec{r}) + g_\rho \dfrac{\tau^N_3}{2} b(\vec{r})
 + \dfrac{e}{2} (1+\tau^N_3) A(\vec{r}) \,)
\gamma_0 ] \psi_N(\vec{r}) = 0, \label{eqdiracn}\\
& &[i\gamma \cdot \partial - m^*_Y(\sigma)-
(\, g^Y_\omega \omega(\vec{r}) + g_\rho I^Y_3 b(\vec{r})
+ e Q_Y A(\vec{r}) \,)
\gamma_0 ] \psi_Y(\vec{r}) = 0, \label{eqdiracy}\\
& &(-\nabla^2_r+m^2_\sigma)\sigma(\vec{r}) =
- \left[\dfrac{\partial m_N^*(\sigma)}{\partial \sigma}\right]\rho_s(\vec{r})
- \left[\dfrac{\partial m_Y^*(\sigma)}{\partial \sigma}\right]\rho^Y_s(\vec{r}),
\nn \\
& & \hspace{7.5em} \equiv g_\sigma C_N(\sigma) \rho_s(\vec{r})
    + g^Y_\sigma C_Y(\sigma) \rho^Y_s(\vec{r}) , \label{eqsigma}\\
& &(-\nabla^2_r+m^2_\omega) \omega(\vec{r}) =
g_\omega \rho_B(\vec{r}) + g^Y_\omega
\rho^Y_B(\vec{r}) ,\label{eqomega}\\
& &(-\nabla^2_r+m^2_\rho) b(\vec{r}) =
\dfrac{g_\rho}{2}\rho_3(\vec{r}) + g^Y_\rho I^Y_3 \rho^Y_B(\vec{r}),
 \label{eqrho}\\
& &(-\nabla^2_r) A(\vec{r}) =
e \rho_p(\vec{r})
+ e Q_Y \rho^Y_B(\vec{r}) ,\label{eqcoulomb}
\end{eqnarray}
where, $\rho_s(\vec{r})$ ($\rho^Y_s(\vec{r})$), $\rho_B(\vec{r})=\rho_p(\vec{r})+\rho_n(\vec{r})$
($\rho^Y_B(\vec{r})$), $\rho_3(\vec{r})=\rho_p(\vec{r})-\rho_n(\vec{r})$, 
$\rho_p(\vec{r})$ and $\rho_n(\vec{r})$ are the nucleon (hyperon) scalar, 
nucleon (hyperon) baryon, third component of isovector,
proton and neutron densities at the position $\vec{r}$ in
the (hyper)nucleus. On the right hand side of Eq.~(\ref{eqsigma}),
$- [{\partial m_N^*(\sigma)}/{\partial \sigma}] \equiv
g_\sigma C_N(\sigma)$ and
$- [{\partial m_Y^*(\sigma)}/{\partial \sigma}] \equiv
g^Y_\sigma C_Y(\sigma)$, where $g_\sigma \equiv g_\sigma (\sigma=0)$ and
$g^Y_\sigma \equiv g^Y_\sigma (\sigma=0)$.
At the hadronic level, the entire information
on the quark dynamics is condensed into the effective couplings
$C_{N,Y}(\sigma)$ of Eq.~(\ref{eqsigma}), which are characteristic new 
features of QMC. 
Furthermore, when $C_{N,Y}(\sigma) = 1$, which corresponds to
a structureless nucleon or hyperon, the equations of motion
given by Eqs.~(\ref{eqdiracn})-(\ref{eqcoulomb})
can be identified with those derived from QHD~\cite{QHD1,QHD2}.

We note here that, for the Dirac equation Eq.~(\ref{eqdiracy})
for $Y = \Lambda_{c,b}$ baryons to be discussed in section~\ref{bchyp},  
the effects of Pauli blocking at the quark level 
is introduced by adding a repulsive potential. 
This is the same as that used for the strange $\Lambda$-hyperon case. 
This was extracted by the fit to the $\Lambda$- and $\Sigma$-hypernuclei taking 
into account the $\Sigma N - \Lambda N$ channel coupling~\cite{QMChyp}. 
The modified Dirac equation for the $Y = \Lambda_{c,b}$ is, 
\begin{equation}
 [i\gamma \cdot \partial - M_Y(\sigma)-
(\, \lambda_Y \rho_B(\vec{r}) + g^Y_\omega \omega(\vec{r}) 
+ g_\rho I^Y_3 b(\vec{r}) 
 + e Q_Y A(\vec{r}) \,) \gamma_0 ] \psi_Y(\vec{r}) = 0,
\label{Pauli}
\end{equation}
where $\rho_B(\vec{r})$ is the baryon density at the position $\vec{r}$ in 
the $\Lambda_{c,b}$-hypernucleus. The value of $\lambda_Y = \lambda_{c,b}$ is 
60.25 MeV (fm)$^3$. The details about the effective Pauli blocking at 
the quark level can be found in Refs.~\cite{QMCreview,QMChyp}.

The effective mass of the nucleon $N$ and hyperon $Y$ are calculated by Eq.~(\ref{hmass}) 
to be shown later, by replacing $h \to N$, and $h \to Y$, respectively.
The explicit expressions for $C_{N,Y}(\sigma) \equiv S_{N,Y}(\sigma) / S_{N,Y}(0)$, are related by,
%%%%%%
\begin{equation}
\dfrac{\partial m_{N,Y}^*(\sigma)}{\partial \sigma}
= - n_q g_{\sigma}^q \int_{bag} d\vec{y} 
\ {\overline \psi}_q(\vec{y}) \psi_q(\vec{y})
\equiv - n_q g_{\sigma}^q S_{N,Y}(\sigma) = - \dfrac{\partial}{\partial \sigma}
\left[ g^{N,Y}_\sigma(\sigma) \sigma \right],
\label{Ssigma}
\end{equation}
%%%%%%
where $g^q_\sigma$ is the light-quark-$\sigma$ coupling constant.
By the above relation, we may define the $\sigma$-$N$ and $\sigma$-$Y$ coupling 
constants:
%%%%%%
\begin{equation}
g^{N,Y}_\sigma = n_q g^q_\sigma S_{N,Y} (0).
\label{sigmacc}
\end{equation}
Note that, the same as that for $C_{N,Y}(\sigma)$, $S_N(0)$ and $S_Y(0)$ 
values are different, because the ground state light quark wave functions 
in the nucleon $N$ and hyperon $Y$ are different in vacuum as well as in medium.
(That is, the bag radii of the $N$ and $Y$ are different in vacuum as well as in medium.)

In the calculations summarised here the parameters which are used for  
the study of infinite nuclear matter and finite nuclei~\cite{Saitofinite},
are $m_\omega = 783$ MeV, $m_\rho = 770$ MeV, $m_\sigma = 418$ MeV and 
$e^2/4\pi = 1/137.036$. The corresponding meson-nucleon coupling constants 
are given in next subsection.

%%%%%%%%%%%%%%%%%%%%%%%%%%%%%%%%%%%%%%%%%%%%%%%%%%%%%%%%%
\subsection{Hadron masses in nuclear matter}
\label{hadronmass}
%%%%%%%%%%%%%%%%%%%%%%%%%%%%%%%%%%%%%%%%%%%%%%%%%%%%%%%%

Below we consider the rest frame of infinitely large 
symmetric nuclear matter, a spin and isospin saturated system 
with only strong interactions.
To do so, one first keeps only ${\cal L}^N_{QMC}$ in Eq.~(\ref{eq:LagYQMC}), 
or correspondingly drops all the quantities with the super- and under-script $Y$, 
and set the Coulomb field $A(\vec{r})=0$ in Eqs.~(\ref{eqdiracn})-(\ref{eqcoulomb}). 
Next one sets all the terms with any derivatives of the density to be zero.  
Then, within Hartree mean-field approximation, 
the nuclear (baryon), $\rho_B$, and scalar, $\rho_s$, densities 
are respectively given by,
\begin{eqnarray}
\rho_B &=& \dfrac{4}{(2\pi)^3}\int d\vec{k}\ \theta (k_F - |\vec{k}|)
= \dfrac{2 k_F^3}{3\pi^2},
\label{rhoB}
\\
\rho_s &=& \dfrac{4}{(2\pi)^3}\int d\vec{k} \ \theta (k_F - |\vec{k}|)
\dfrac{m_N^*(\sigma)}{\sqrt{m_N^{* 2}(\sigma)+\vec{k}^2}} \, .
\label{rhos}
\end{eqnarray}
Here, $m^*_N(\sigma)$ is the value (constant) of the effective nucleon mass at the given density 
(see also Eq.~(\ref{effnmass})) and $k_F$ the Fermi momentum.
In the standard QMC model~\cite{QMCGuichon,QMCreview}, the MIT bag model is used 
for describing nucleons and hyperons (hadrons). The use of this quark model is 
an essential ingredient for the QMC model, as similar results can be obtained 
using the Nambu--Jona-Lasinio model~\cite{Whittenbury:2016sma} or a constituent quark 
model~\cite{Bracco:1998qw}. 

Then, the Dirac equations for the quarks and antiquarks 
in nuclear matter, in bags of hadrons, $h$, ($q = u$ or $d$, and $Q = s,c$ or $b$, hereafter) 
neglecting the Coulomb force in nuclear matter, are given by  
($|\mbox{\boldmath $x$}|\le$ bag radius)~\cite{QMCbc,Tsushimak,etap,qmcapp,Tsushimad}:
\begin{eqnarray}
\left[ i \gamma \cdot \partial_x -
(m_q - V^q_\sigma)
\mp \gamma^0
\left( V^q_\omega +
\dfrac{1}{2} V^q_\rho
\right) \right] 
\left( \begin{array}{c} \psi_u(x)  \\
\psi_{\bar{u}}(x) \\ \end{array} \right) &=& 0,
\label{Diracu}\\
\left[ i \gamma \cdot \partial_x -
(m_q - V^q_\sigma)
\mp \gamma^0
\left( V^q_\omega -
\dfrac{1}{2} V^q_\rho
\right) \right]
\left( \begin{array}{c} \psi_d(x)  \\
\psi_{\bar{d}}(x) \\ \end{array} \right) &=& 0,
\label{Diracd}\\
\left[ i \gamma \cdot \partial_x - m_{Q} \right]
\psi_{Q} (x)\,\, ({\rm or}\,\, \psi_{\Qbar}(x)) &=& 0. 
\label{DiracQ}
\end{eqnarray}
The (constant) mean-field potentials for a bag in nuclear matter
are defined by $V^q_\sigma \equiv g^q_\sigma \sigma$, 
$V^q_\omega \equiv g^q_\omega \omega$ and
$V^q_\rho \equiv g^q_\rho b$,
with $g^q_\sigma$, $g^q_\omega$ and
$g^q_\rho$ the corresponding quark-meson coupling
constants. 
We assume $SU$(2) symmetry, $m_{u,\bar{u}}=m_{d,\bar{d}} \equiv m_{q,\bar{q}}$. 
The corresponding effective quark masses are defined  
by, $m^*_{u,\bar{u}}=m^*_{d,\bar{d}}=m^*_{q,\bar{q}} \equiv m_{q,\bar{q}}-V^q_{\sigma}$.
In symmetric nuclear matter within Hartree approximation, 
the $\rho$-meson mean field is zero, $V^q_{\rho}=0$~in Eqs.~(\ref{Diracu}) and~(\ref{Diracd})     
and we ignore it. (This is not true in a finite nucleus, even with equal numbers of protons and neutrons, 
since the Coulomb interactions among the protons induces an asymmetry  
between the proton and neutron densities to give $\rho_3 = \rho_p - \rho_n \ne 0$.)

The same meson mean fields $\sigma$ and $\omega$ for the quarks  
in Eqs.~(\ref{Diracu}) and~(\ref{Diracd}), satisfy self-consistently 
the following equations at the nucleon level (together with the effective 
nucleon mass $m_N^*(\sigma)$ of Eq.~(\ref{effnmass}) 
calculated by Eq.~(\ref{hmass}) below):
\begin{eqnarray}
{\omega}&=&\dfrac{g_\omega \rho_B}{m_\omega^2},
\label{omgf}\\
{\sigma}&=&\dfrac{g_\sigma }{m_\sigma^2}C_N({\sigma})
\dfrac{4}{(2\pi)^3}\int d\vec{k} \ \theta (k_F - |\vec{k}|)
\dfrac{m_N^*(\sigma)}{\sqrt{m_N^{* 2}(\sigma)+\vec{k}^2}}
=\dfrac{g_\sigma }{m_\sigma^2}C_N({\sigma}) \rho_s,
\label{sigf}
%C_N(\sigma) &=& \dfrac{-1}{g_\sigma(\sigma=0)}
%\left[ \dfrac{\partial m^*_N(\sigma)}{\partial\sigma} \right],
%\label{CN}
\end{eqnarray}
where 
\be
C_N(\sigma) \equiv \dfrac{-1}{g_\sigma(\sigma=0)}
\left[ {\partial m^*_N(\sigma)}/{\partial\sigma} \right].
\ee
Because of the underlying quark structure of the nucleon used to calculate
$m^*_N(\sigma)$ in nuclear medium, $C_N(\sigma)$ decreases significantly as $\sigma$ increases,
whereas in the usual point-like nucleon-based models it is constant, $C_N(\sigma) = 1$.
It is this variation of $C_N(\sigma)$ (or equivalently dependence of the scalar coupling on density,
 $g_\sigma (\sigma)$) that yields a novel saturation mechanism for nuclear matter 
in the QMC model, and contains the important dynamics which originates in the quark structure
of the nucleon and hyperon. It is this variation through the scalar polarisability which yields 
three-body or density dependent effective forces, as has been demonstrated by constructing 
an equivalent energy density functional~\cite{Guichon:2006er}.   As a 
consequence of the {\em derived}, nonlinear
couplings of the meson fields in the Lagrangian density at the nucleon (hyperon) and meson level,
the standard QMC model yields the nuclear incompressibility of $K \simeq 280$~MeV with 
$m_q=5$ MeV. This is in contrast to a naive version of QHD~\cite{QHD1,QHD2}
(the point-like nucleon model of nuclear matter),
results in the much larger value, $K \simeq 500$~MeV;
the empirically extracted value falls in the range $K = 200 - 300$ MeV.
(See Ref.~\cite{Stone} for a recent, extensive analysis of this issue.)

%%%%%%%%%%%%%%%%%%%%%%%%%%%%%%%%%%%%%%%%%%%%%%%%%%%%%%%%%%%%%%%%%%%%%%%
\begin{table}[htb]
\begin{center}
%\begin{minipage}[t]{16.5cm}
\caption{Current quark masses (input), coupling constants 
and the bag constant obtained with the nucleon bag 
radius $R_N = 0.8$ fm in vacuum.
\vspace{1ex}
}
\label{qmasscc}
%\end{minipage}
\begin{tabular}{r|r||l|l}
\hline
$m_{u,d}$ &5    MeV &$g^q_\sigma$ &5.69\\
$m_s$     &250  MeV &$g^q_\omega$ &2.72\\
$m_c$     &1300 MeV &$g^q_\rho$   &9.33\\
$m_b$     &4200 MeV &$B^{1/4}$    &170 MeV\\
\hline
\end{tabular}
\end{center}
\end{table}
%%%%%%%%%%%%%%%%%%%%%%%%%%%%%%%%%%%%%%%%%%%%%%%%%%%%%%%%%%%%%%%%%%%%%%%

Once the self-consistency equation for the ${\sigma}$ field, 
Eq.~(\ref{sigf}), has been solved, one can evaluate the total energy per nucleon:
\begin{equation}
E^\mathrm{tot}/A=\dfrac{4}{(2\pi)^3 \rho}\int d\vec{k} \
\theta (k_F - |\vec{k}|) \sqrt{m_N^{* 2}(\sigma)+
\vec{k}^2}+\dfrac{m_\sigma^2 {\sigma}^2}{2 \rho}+
\dfrac{g_\omega^2\rho}{2m_\omega^2} .
\label{toten}
\end{equation}
We then determine the coupling constants, $g_{\sigma}$ and $g_{\omega}$, so as
to fit the binding energy of 15.7~MeV at the saturation density $\rho_0$ = 0.15 fm$^{-3}$
($k_F^0$ = 1.305 fm$^{-1}$) for symmetric nuclear matter.
The determined quark-meson coupling constants, and the current quark mass values 
used are listed in Tab.~\ref{qmasscc}.
The coupling constants at the nucleon level are  
$g^2_\sigma/4\pi = 3.12$, $g^2_\omega/4\pi = 5.31$ and $g^2_\rho/4\pi = 6.93$.
(See Eq.~(\ref{sigmacc}).)

\begin{figure}[htb]
\centering
\epsfig{figure=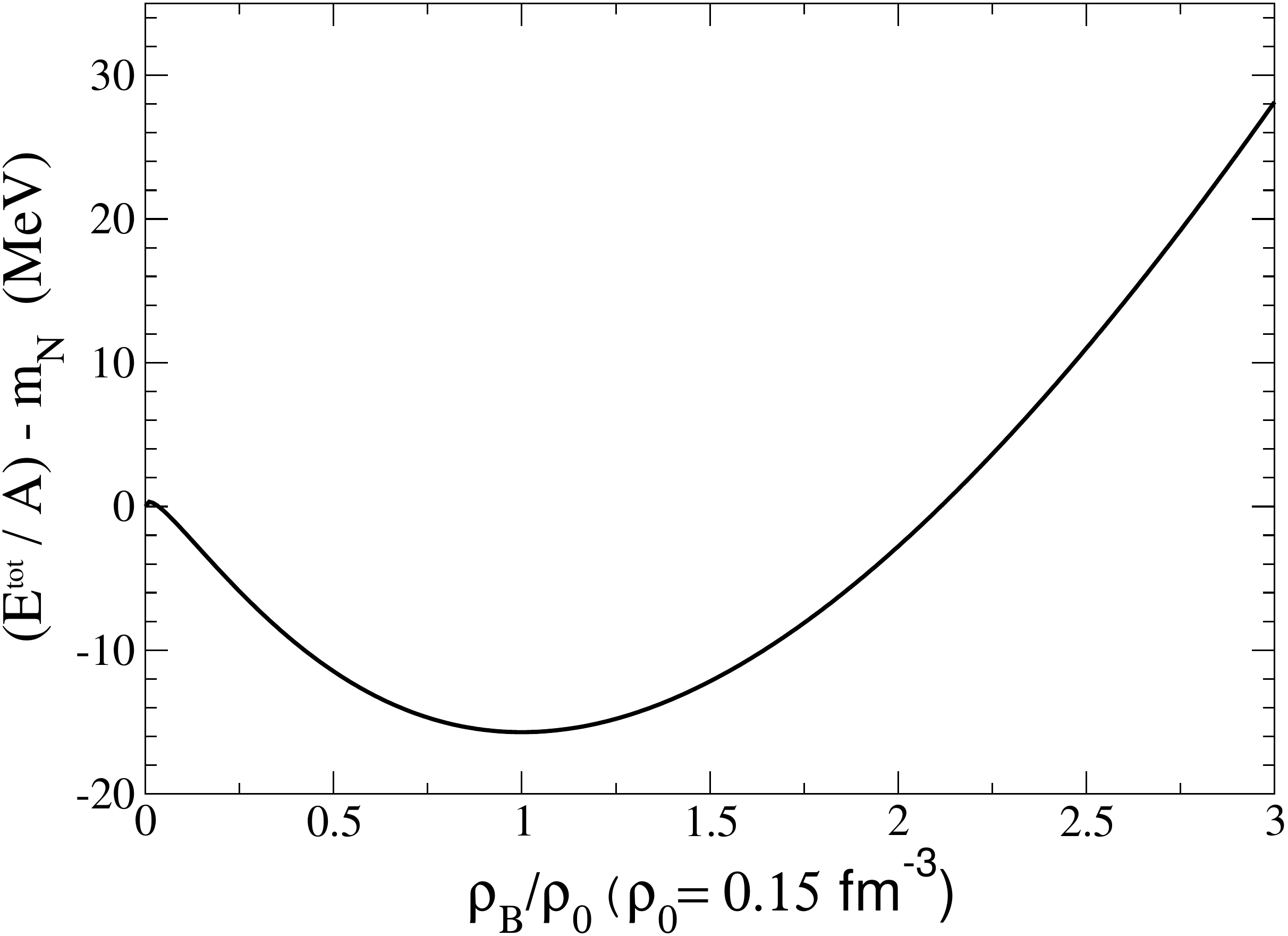,width=7cm}
\hspace{2ex}
\epsfig{figure=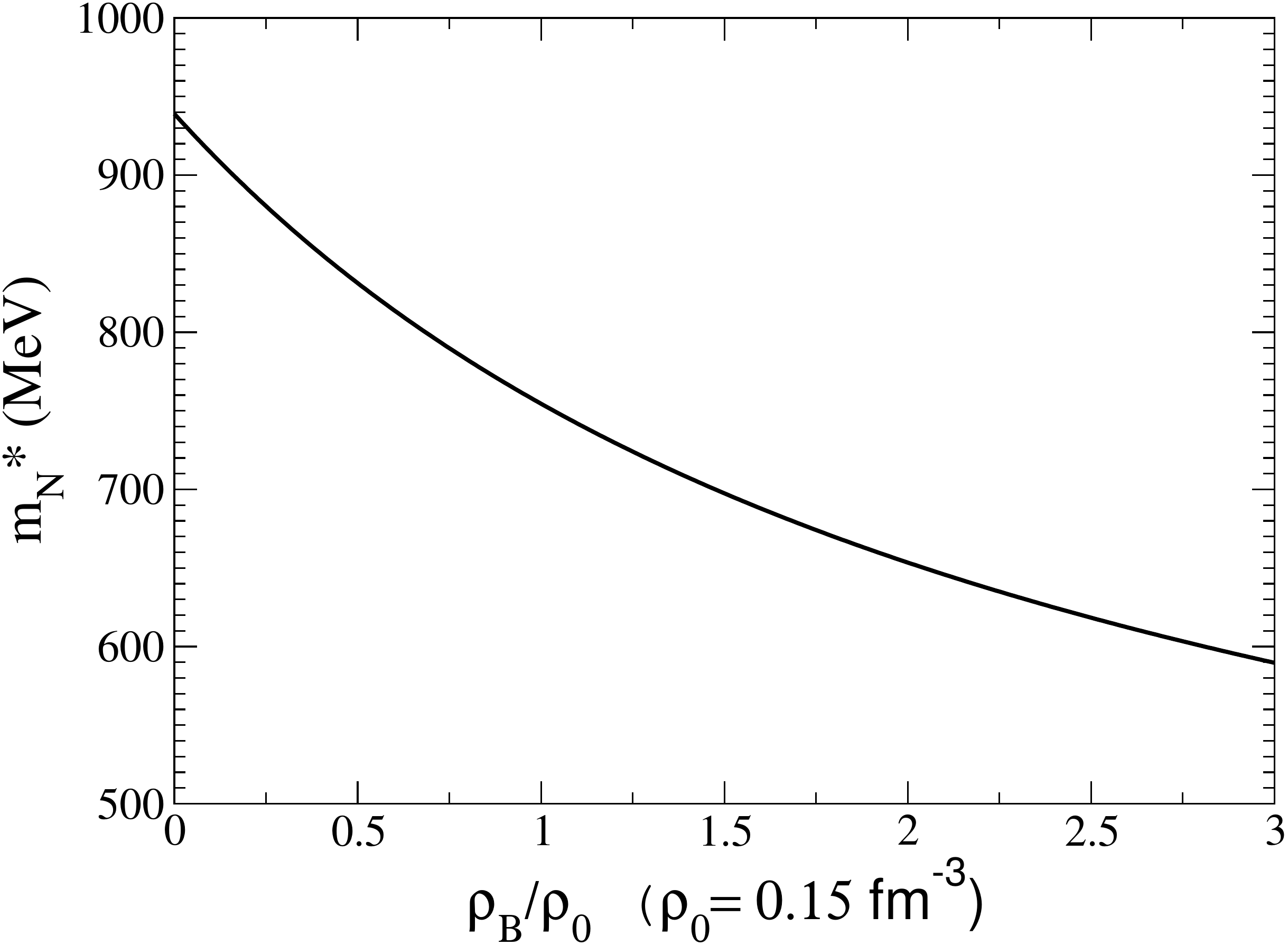,width=7cm}
\caption{Negative of the binding energy per nucleon for symmetric nuclear matter,  
$(E^{\rm tot}/A)-m_N$, v.s. $\rho_B/\rho_0$ ($\rho_0=0.15$ fm$^{-3}$) 
with the vacuum light quark mass $m_q=m_{\bar{q}}=5$~MeV ($q=u,d$), calculated using the 
QMC model~(left panel), as well as the effective nucleon mass~$m^*_N$~(right panel).  
The incompressibility $K$ obtained is $K=279.3$ MeV. 
\label{E/AmNstar}
} 
\end{figure}

\begin{figure}[tb]
\centering 
%\hspace*{-1.4cm}
\epsfig{figure=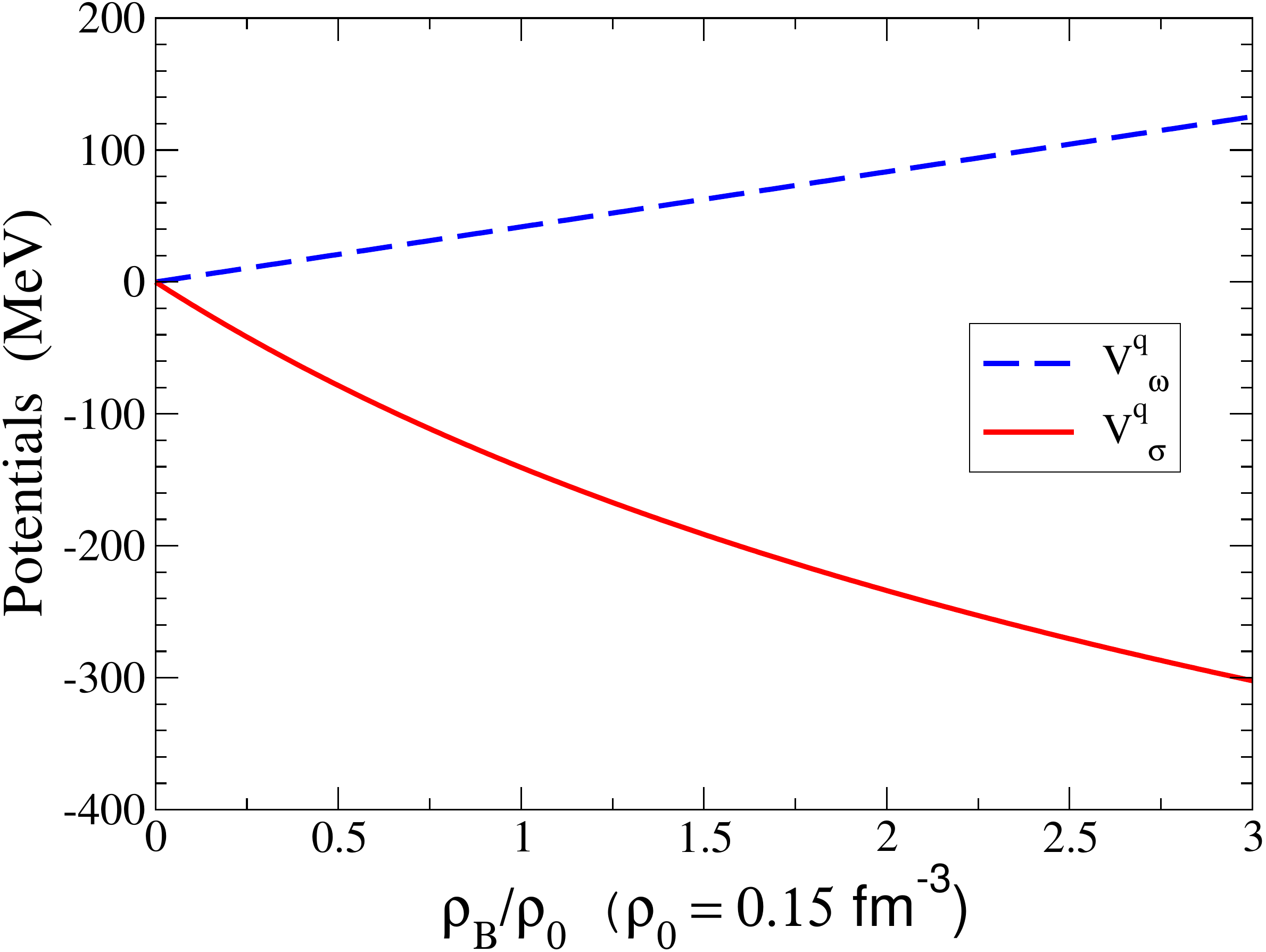,width=7cm} %% ,angle=90}
\caption{
Light quark potentials.
}
\label{qpot}
\end{figure}

We show in Fig.~\ref{E/AmNstar} the modulus of the binding energy per nucleon  
$E^{\rm tot}/A - m_N$ (left panel) and the effective nucleon mass, $m^*_N$ (right panel) 
obtained using the determined quark-meson coupling constants. 
The corresponding mean field potentials felt by the light quarks, 
$V^q_\omega$ and $V^q_\sigma$, are shown in Fig.~\ref{qpot}.

%%%%%%%%%%%%%%%%%%%%%%%%%%%%%%%%%%%%%%%%%%%%%
%%%%%%%%%%%%%%%%%%%%%%%%%%%%%%%%%%%%%%%%%%%%%%

%%%%%%%%%%%%%%%%%%%%%%%%%%%%%%%%%%%%%%%%%%%%%%
%%%%%%%%%%%%%%%%%%%%%%%%%%%%%%%%%%%%%%%%%%%%%%

In the following, let us consider the situation that a hadron $h$ is immersed 
in nuclear matter.
The normalized, static solution for the ground state quarks or antiquarks
with flavor $f$ in the hadron $h$ may be written,  
$\psi_f (x) = N_f e^{- i \epsilon_f t / R_h^*}
\psi_f (\r)$,
where $N_f$ and $\psi_f(\r)$
are the normalization factor and
corresponding spin and spatial part of the wave function. 
The bag radius in medium for hadron $h$, denoted $R_h^*$, 
is determined through the
stability condition for the mass of the hadron against the
variation of the bag 
radius~\cite{QMCGuichon,QMCreview,Guichonfinite,Saitofinite}
(see also Eq.~(\ref{hmass})). 
The eigenenergies in units of $1/R_h^*$ are given by, 
%%%%%%%%%
\bge
\left( \begin{array}{c}
\epsilon_u \\
\epsilon_{\bar{u}}
\end{array} \right)
= \Omega_q^* \pm R_h^* \left(
V^q_\omega
+ \dfrac{1}{2} V^q_\rho \right),\,\,
\left( \begin{array}{c} \epsilon_d \\
\epsilon_{\bar{d}}
\end{array} \right)
= \Omega_q^* \pm R_h^* \left(
V^q_\omega
- \dfrac{1}{2} V^q_\rho \right),\,\,
\epsilon_{Q}
= \epsilon_{\Qbar} =
\Omega_{Q}.
\label{energy}
\ene
%%%%%%%%%
The hadron masses
in a nuclear medium $m^*_h$ (free mass will be denoted by $m_h$),
are calculated by
\begin{eqnarray}
m_h^* &=& \sum_{j=q,\bar{q},Q,\Qbar} 
\dfrac{ n_j\Omega_j^* - z_h}{R_h^*}
+ {4\over 3}\pi R_h^{* 3} B,\quad
\left. \dfrac{\partial m_h^*}
{\partial R_h}\right|_{R_h = R_h^*} = 0,
\label{hmass}
\end{eqnarray}
%%%%%%
where $\Omega_q^*=\Omega_{\bar{q}}^*
=[x_q^2 + (R_h^* m_q^*)^2]^{1/2}\,(q=u,d)$, with
$m_q^*=m_q{-}g^q_\sigma \sigma$,
$\Omega_Q^*=\Omega_{\Qbar}^*=[x_Q^2 + (R_h^* m_Q)^2]^{1/2}\,(Q=s,c,b)$,
and $x_{q,Q}$ are the bag eigenfrequencies.
$B$ is the bag constant, $n_q (n_{\qbar})$ and $n_Q (n_{\Qbar})$ 
are the lowest mode quark (antiquark) 
numbers for the quark flavors $q$ and $Q$ 
in the hadron $h$, respectively, 
while $z_h$ parametrizes the sum of the
center-of-mass and gluon fluctuation effects, 
which (following Ref.~\cite{Guichonfinite}) are assumed to be
independent of density. Concerning the sign of $m_q^*$ in nuclear 
medium, it reflects nothing but the strength 
of the attractive scalar potential 
as in Eqs.~(\ref{Diracu}) and~(\ref{Diracd}), 
and thus naive interpretation of the mass for a (physical) particle, 
which is positive, should not be applied. 
The parameters are determined to
reproduce the corresponding masses in free space.
The quark-meson coupling constants, $g^q_\sigma$, $g^q_\omega$
and $g^q_\rho$, have been already determined. 
Exactly the same coupling constants, $g^q_\sigma$, $g^q_\omega$ and
$g^q_\rho$, are used for the light quarks in the mesons and baryons as in 
the nucleon. (See Tab.~\ref{qmasscc}.)

However, in studies of the kaon system, we found that it was
phenomenologically necessary to increase the strength of the vector
coupling to the non-strange quark in the $K^+$ (by a factor of
$1.4^2$, i.e., $g_{K\omega}^q \equiv 1.4^2 g^q_\omega$) 
in order to reproduce the empirically extracted repulsive $K^+$-nucleus
interaction~\cite{Tsushimak}. This may be related to the fact that 
kaon is a pseudo-Goldstone boson, where treatment of the Goldstone 
bosons in a naive quark model should not be expected to be satisfactory. 
We also discuss this possibility, $g^q_\omega \to 1.4^2 g^q_\omega$, 
for the $D$- and $\Dbar$-meson nuclear bound states~\cite{Tsushimad} 
in subsection~\ref{DDbar_result}. 
The scalar ($V^{h}_s$) and vector ($V^{h}_v$) potentials 
felt by the hadrons $h$,  
in nuclear matter are given by,
\bg
V^h_s &=& m^*_h - m_h,\,\,\label{spot}
\\
V^h_v &=&
  (n_q - n_{\bar{q}}) {V}^q_\omega + I^h_3 V^q_\rho,  
\qquad (V^q_\omega \to 1.4^2 {V}^q_\omega\,\,
{\rm for}\, K,\Kbar), 
\label{vpot}
\en
where $I^h_3$ is the third component of isospin projection  
of the hadron $h$. Thus, the vector potential felt by a heavy baryon 
with a charm or bottom quark, is equal to that of the hyperon with 
the same light quark configuration in QMC.

In Tab.~\ref{bagparambc} we present the input (vacuum masses $m_h$) 
bag parameters $z_h$, and the bag radius obtained in vacuum (at $\rho_0 = 0.15$ fm$^{-3}$) 
$R_h$ ($R^*_h$) for various hadrons. Note that the bag radius in vacuum for the nucleon, 
$R_N = 0.8$ fm is the input. For a more recent study of the $\eta$-$\eta^\prime$ system, which takes 
into account the role of the $U_A$(1) axial anomaly, we refer to 
Refs.~\cite{Bass:2013nya,Bass:2010kr,Bass:2005hn}.

%%%%%%%%%%%%%%%%%%%%%%%%%%%%%%%%%%%%%%%%%%%%%%%%%%%%%%%%%%%%%%%%%%%%%%%
\begin{table}[htb]
\begin{center}
%\begin{minipage}[t]{16.5cm}
\caption{The bag parameters,
various hadron masses and the bag radii in free space
[at normal nuclear matter density, $\rho_0=0.15$ fm$^{-3}$]
$z_h, R_h$ and $m_h$ [$m_h^*$ and $R_h^*$].
$m_h$ and $R_N = 0.8$ fm in free space are inputs.
Note that the quantities for the physical 
$\omega$, $\phi$, $\eta$ and $\eta'$ 
are calculated including the octet-singlet mixing effect (see subsection \ref{massetaod}), 
and that $\omega$ and $\rho$ 
below stand for the physical particles, and are different from those 
appearing in the Lagrangian density of QMC. 
\vspace{1ex}
}
\label{bagparambc}
%\end{minipage}
\begin{tabular}{c|ccc|cc}
\hline
%\hline
$h$ &$z_h$ &$m_h$ (MeV) &$R_h$ (fm) &$m_h^*$ (MeV) &$R_h^*$ (fm)\\
\hline
%%%& & & & &\\[-0.30true cm]
$N$           &3.295 &939.0  &0.800 &754.5  &0.786\\
$\Lambda$     &3.131 &1115.7 &0.806 &992.7  &0.803\\
$\Sigma$      &2.810 &1193.1 &0.827 &1070.4 &0.824\\
$\Xi$         &2.860 &1318.1 &0.820 &1256.7 &0.818\\
$\Lambda_c$   &1.766 &2284.9 &0.846 &2162.5 &0.843\\
$\Sigma_c$    &1.033 &2452.0 &0.885 &2330.2 &0.882\\
$\Xi_c$       &1.564 &2469.1 &0.853 &2408.0 &0.851\\
$\Lambda_b$   &-0.643&5624.0 &0.930 &5502.9 &0.928\\
\hline
$\omega$      &1.866 &781.9  &0.753 &658.7  &0.749\\
$\rho$        &1.907 &770.0  &0.749 &646.2  &0.746\\
$K$           &3.295 &493.7  &0.574 &430.4  &0.572\\
$K^*$         &1.949 &893.9  &0.740 &831.9  &0.738\\
$\eta$        &3.131 &547.5  &0.603 &483.9  &0.600\\
$\eta'$       &1.711 &957.8  &0.760 &896.5  &0.758\\
$\phi$        &1.979 &1019.4 &0.732 &1018.9 &0.732\\
$D$           &1.389 &1866.9 &0.731 &1804.9 &0.730\\
$D^*$         &0.849 &2000.8 &0.774 &1946.7 &0.772\\
$B$           &-1.136&5279.2 &0.854 &5218.1 &0.852\\
%\hline
\hline
\end{tabular}
\end{center}
\end{table}
%
%%%%%%%%%%%%%%%%%%%%%%%%%%%%%%%%%%%%%%%%%%%%%%%%%%%%%%%%%%%%%%%%%%%%%%%

%%%%%%%%%%%%%%%%%%%%%%%%%%%%%%%%%%%%%%%%%%%%%%%%%%%%
\begin{figure}[hbt]
\centering
\epsfig{file=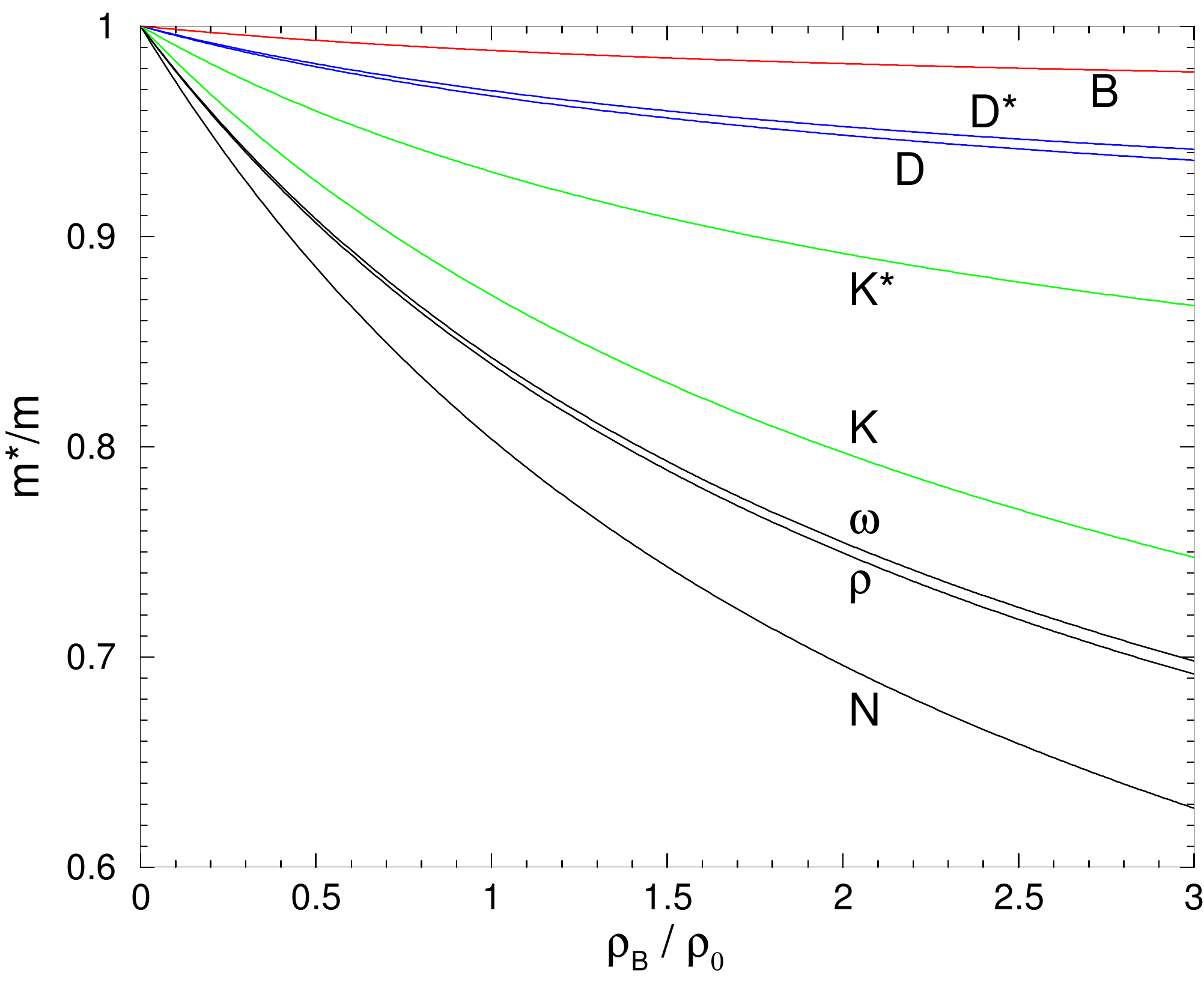,height=8cm,angle=-90}
\hspace{1cm}
\epsfig{file=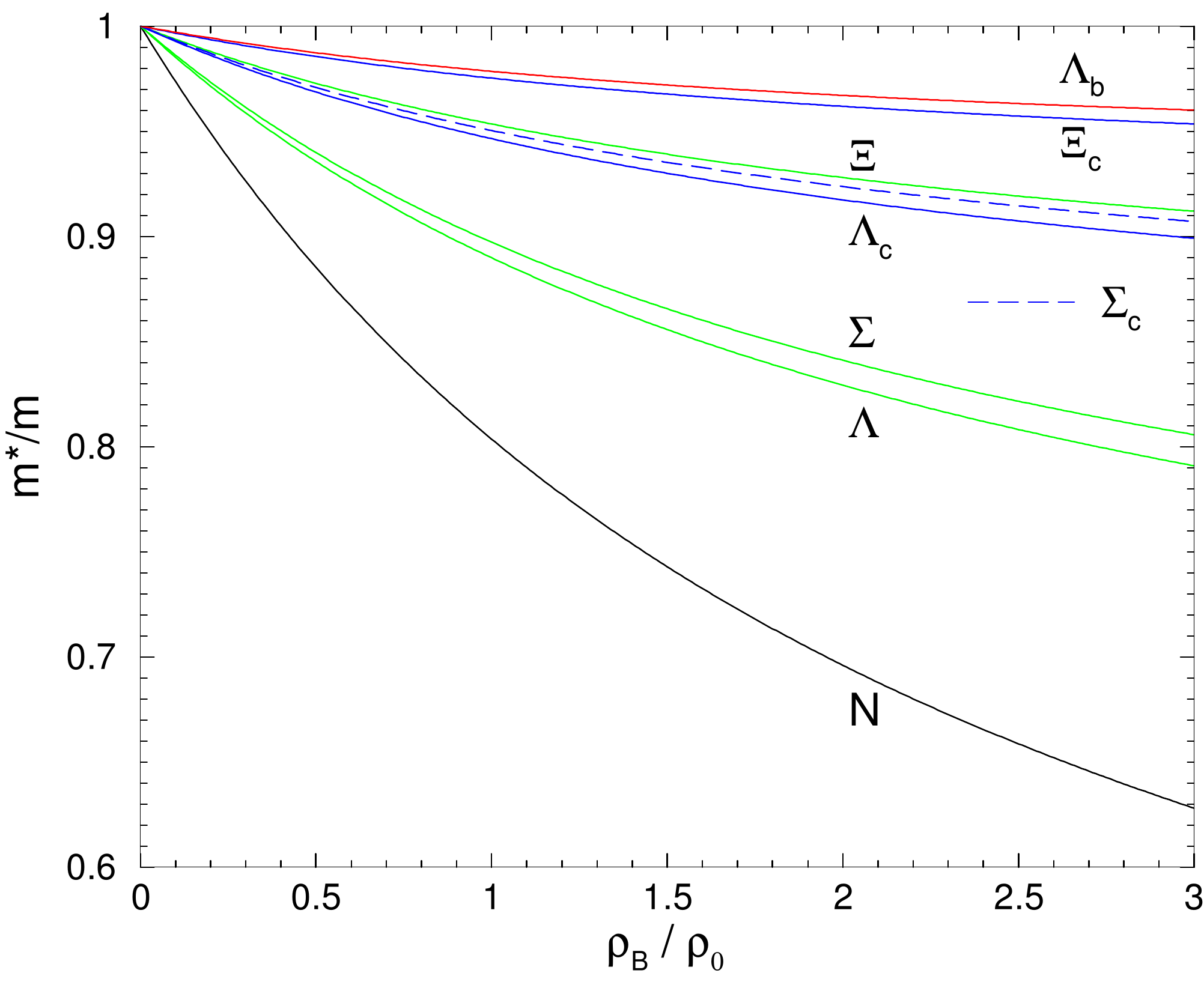,height=8cm,angle=-90}
\caption{Effective mass ratios for mesons (left panel) and baryons (right panel) 
in symmetric nuclear matter, where, $\rho_0 = 0.15$ fm$^{-3}$.
$\omega$ and $\rho$ stand for physical mesons which are treated in 
the quark model, and should not be confused with the fields appearing 
in the QMC model.
\label{massratio}
}
\end{figure}
%%%%%%%%%%%%%%%%%%%%%%%%%%%%%%%%%%%%%%%%%%%%%%%%%%%%

%%%%%%%%%%%%%%%%%%%%%%%%%%%%%%%%%%%%%%%%%%%%%%%%%%
\begin{figure}[hbt]
\centering
\epsfig{file=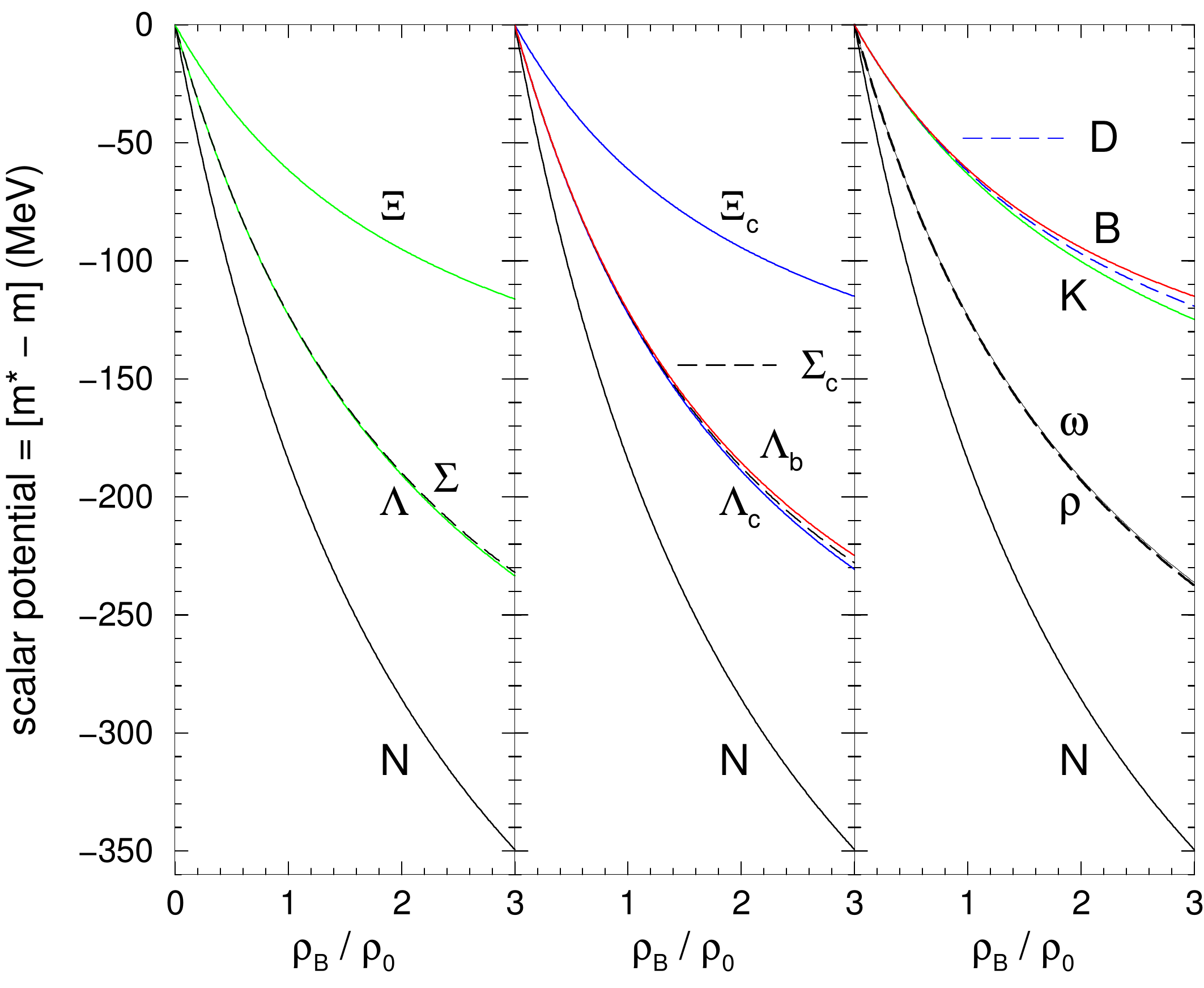,height=8cm,angle=-90}
\caption{Scalar potentials for various hadrons in nuclear matter, where, 
$\rho_0 = 0.15$ fm$^{-3}$.
(See also caption of Fig.~\ref{massratio}.)
\label{spotential}
}
\end{figure}
%%%%%%%%%%%%%%%%%%%%%%%%%%%%%%%%%%%%%%%%%%%%%%%%%%%%

In Fig.~\ref{massratio} we show ratios of 
effective masses (free masses + scalar potentials) 
versus those of the free particles, for 
mesons (left panel) and baryons (right panel), respectively.
With increasing density the ratios decrease as usually expected, 
but the magnitude of this reduction is reduced as we move from  
hadrons with only light quarks to those with one strange quark, with one charm quark, and with one 
bottom quark. This is because their masses in free space 
are in the order from light to heavy. Thus, the net ratios for the 
decrease in masses (developing of scalar masses) compared to that of
the free masses becomes smaller. 
In Fig.~\ref{massratio} one may notice the 
somewhat anomalous behavior of the ratio for the kaon ($K$).
This is related to what we meant by the pseudo-Goldstone boson nature, 
i.e., its mass in free space is relatively light, $m_K \simeq 495$ MeV, 
and the relative ratio for the reduction in mass in a nuclear medium is large.

Probably it is simpler and clearer to compare the   
scalar potentials felt by each hadron in nuclear matter.
Calculated results are shown in Fig.~\ref{spotential}.
From the results one can confirm that 
the scalar potential felt by the hadron $h$, $V_s^h$, 
follows a simple light quark number scaling rule:
\bge
V_s^h \simeq \dfrac{n_q + n_{\qbar}}{3} V_s^N, 
\ene
where $n_q$ ($n_{\qbar}$) is the number of light quarks (antiquarks) in 
the hadron $h$, and $V_s^N$ is the scalar potential felt by the nucleon. 
It is interesting to notice that baryons with a charm and a bottom quark
($\Xi_c$ is a quark configuration, $qsc$), show very similar 
features to those of hyperons with one or two strange quarks.

It has been found that the function $C_B({\sigma})
(B = N,\Lambda,\Sigma,\Xi,\Lambda_c,\Sigma_c,\Xi_c,\Lambda_b)$
(see Eq.~(\ref{Ssigma}) and two lines above), 
can be parameterized as a linear
form in the $\sigma$ field, $g_{\sigma}{\sigma}$, for practical
calculations~\cite{Guichonfinite,Saitofinite,QMChyp,QMCbc,QMChypbc}:
\begin{equation}
C_B ({\sigma}) = 1 - a_B
\times (g_{\sigma} {\sigma}),\hspace{1em}
(B = N,\Lambda,\Sigma,\Xi,\Lambda_c,\Sigma_c,\Xi_c,\Lambda_b).
\label{cynsigma}
\end{equation}
The values obtained for $a_B$ are listed in Tab.~\ref{slope}.
%
%%%%%%%%%%%%%%%%%%%%%%%%%%%%%%%%%%%%%%%%%%%%%%%%%%%%%%%%%%%%%%%%%%%%%%%
\begin{table}
\begin{center}
\caption{The slope parameters, $a_B\,\,
(B = N,\Lambda,\Sigma,\Xi,\Lambda_c,\Sigma_c,\Xi_c,\Lambda_b)$.
\vspace{1ex}
}
\label{slope}
\begin{tabular}[t]{c|c||c|c}
\hline
%\hline
$a_B$ &$\times 10^{-4}$ MeV$^{-1}$ &$a_B$ &$\times 10^{-4}$ MeV$^{-1}$ \\
\hline
$a_N$           &8.8  &$a_{\Lambda_b}$   &10.9 \\
$a_\Lambda$     &9.3  &$a_{\Lambda_c}$   &9.8 \\
$a_{\Sigma}$    &9.5  &$a_{\Sigma_c}$    &10.3 \\
$a_{\Xi}$       &9.4  &$a_{\Xi_c}$       &9.9 \\
%\hline
\hline
\end{tabular}
\end{center}
\end{table}
%%%%%%%%%%%%%%%%%%%%%%%%%%%%%%%%%%%%%%%%%%%%%%%%%%%%%%%%%%%%%%%%%%%%%%%
%
This parameterization works very well up to
about three times of normal nuclear matter density $3 \rho_0$.
Then, the effective masses for the baryons, $B$, in nuclear matter
are well approximated by:
\bge
m^*_B \simeq m_B - \dfrac{n_q}{3} g_\sigma 
\left[1-\dfrac{a_B}{2}(g_\sigma {\sigma})\right]\sigma,
\hspace{2ex}(B = N,\Lambda,\Sigma,\Xi,\Lambda_c,\Sigma_c,\Xi_c,\Lambda_b),
\label{Mstar}
\ene
with $n_q$ the number of light quarks in the baryon $B$.
%
%
%
%\input{sec_HQ-medium.tex}
%
%
%
% % % % % % % % % % % % % % % % % % % % % 
\section{Nuclear-bound $\eta_c$ and $J/\Psi$}
\label{sec:HQ}

As remarked in the Introduction, the mean fields generated by light meson exchange, 
which provide a natural explanation of the binding of atomic nuclei, cannot 
bind heavy $c\overline{c}$ (or $b\overline{b}$) quarkonia. Their interactions with 
the medium necessarily involve other mechanisms, including those based
on mutigluon-gluon exchange~\cite{Peskin:1979va,Brodsky:1989jd,Brodsky:1997gh} and excitation of
charmed hadronic intermediate states with light quarks created from the
vacuum~\cite{{Ko:2000jx},{Krein:2010vp},{Tsushima:2011fg},
{Tsushima:2011kh},{jpsi4}}. Reference~\cite{Voloshin:2007dx} presents 
a recent review of the properties of charmonium states and compiles a
fairly complete list of references on theoretical studies concerning a
great variety of physics issues related to these states. On the experimental
side, one of the major challenges is to find appropriate kinematical conditions
to produce these hadrons essentially at rest, or with small momentum
relative to the nucleus, as effects of the nuclear medium are driven
by low energy interactions. The studies of in-medium charmonia present advantages 
compared to  those of the $\phi$-meson, to be discussed in section~\ref{phi-meson}.
First, they have a very small decay width in vacuum, and they are expected 
to have a small width in a nucleus too. Second, since they are heavier than the
$\phi$-meson, they are expected to move slower than the $\phi$-meson once 
produced near threshold in a nucleus. Therefore, one can hope that they
can form nuclear bound states more easily than the $\phi$-meson.  

Since the earlier suggestion~\cite{Brodsky:1989jd,Wasson:1991fb} that heavy charmonia 
states may form bound states with nuclei, a large literature on this subject has 
accumulated along the years. Many different methods have been used to investigate the 
possible existence of such exotic states. These include QCD-based approaches, the most 
prominent examples being calculations based on the color polarizability of 
quarkonium~\cite{{Voloshin:2007dx},{Ko:2000jx},{Kaidalov:1992hd},{Brodsky:1997gh},
{Luke:1992tm},{deTeramond:1997ny},{Sibirtsev:2005ex}}, QCD sum rules~\cite{{Klingl:1998sr},
{Hayashigaki:1998ey},{Kim:2000kj}}, meson loops~\cite{{Ko:2000jx},{Krein:2010vp},{Tsushima:2011fg},
{Tsushima:2011kh},{jpsi4}}, phenomenological approaches~\cite{{Belyaev:2006vn},{Yokota:2013sfa}}, 
and, more recently, lattice QCD simulations~\cite{{Beane:2014sda},{Alberti:2016dru}}.

Knowledge of the low-energy quarkonium-nucleon interaction in free space is an important 
input for the study of quarkonium binding to nuclei. As for now, there is no direct experimental 
information on the interaction of heavy quarkonium with the nucleon; practically all of 
our knowledge on this interaction comes from lattice QCD simulations. All available lattice 
results on quarkonium-nucleon interaction~\cite{{Alberti:2016dru},{Yokokawa:2006td},
{Liu:2008rza},{Kawanai:2010ev},{Kawanai:2010ru}} indicate that it is 
attractive and not very strong. The crucial question is whether the strength 
is strong enough to form a bound state with the nucleon; if not, the important 
issue then becomes whether it is strong enough to bind a quarkonium to a sufficiently 
large nucleus. 

In this section, we present predictions for $\eta_c$ and $J/\Psi$ bound states 
with nuclei of different sizes. We concentrate on two main approaches to the problem
of quarkonium binding to nuclei. One approach is 
based on nonrelativistic effective quarkonium-nucleon potentials. Such potentials are either obtained 
by making use, in one way or another, of the ideas underlying pNRQCD through the multipole expansion 
of color fields and the concept of chromopolarizability of quarkonium, or by fits from lattice QCD 
simulations. The other approach considers the self-energy of a charmonium due to $D\overline{D}$ 
loops in nuclear matter. The self-energy depends on the nuclear matter density which, through a 
local density approach, provides an effective potential for the charmonium in a finite nucleus.  

Both approaches make use in an essential way of the independent-particle nature of the nucleus. 
The average baryon number density in the center of a large nucleus is close to the saturation 
density of nuclear matter, $\rho_0 = 0.16~$fm$^{-3}$ (in QMC $\rho_0 = 0.15~$fm$^{-3}$ 
has commonly been adopted); thereby the average separation distance 
of two two nearest neighbour nucleons in such a nucleus is $d_{\rm av} \sim \rho^{-1/3} \sim 1.8$~fm. 
Taking for the typical size of a nucleon the r.m.s. charge radius of the proton, $r_p \equiv \langle r^2_p
\rangle^{1/2} \simeq 0.88$~fm $\sim \Lambda^{-1}_{\rm QCD}$, as extracted from measurements 
of electric form factors in electron-proton scattering experiments~\cite{Bernauer:2013tpr}, 
one obtains $2 r_p \sim d_{\rm av}$. On the other hand, the nucleon-nucleon interaction has 
a strong short-range repulsion (hard-core), which prevents substantial superposition of the 
quark cores of different nucleons in the nucleus. The interplay between the hard-core repulsion 
and the Pauli Exclusion Principle is at the basis of the nuclear shell model~\cite{weisskopf}, 
in that nucleons move almost independently from each other with a well-defined angular momentum
in an average mean field. The fact that $2 r_p \sim d_{\rm av}$ does not spoil this picture of 
the nucleus~\cite{{Betz:1985fz},Krein:1987sg} as being a close-packed 
system of nucleons moving almost independently from each other.

\subsection{Quarknonium-nucleon potentials}
\label{sub:HQvdw}

Let us imagine a heavy quarkonium $Q\bar{Q}$ injected with low momentum 
into a close-packed nucleus. When the quark mass $m_Q$ is large, i.e $m_Q \gg \Lambda_{\rm QCD}$, 
the heavy quarkonium is, as discussed in section~\ref{sec:QCD}, essentially a Coulomb bound state, 
with a Bohr radius $r_0 = (m_Q \alpha_s)^{-1}$, with $\alpha_s = \alpha_s(m_Q) \ll 1$.  
The quarkonium interacts by exchanging gluons with the light quarks of the nucleons, the 
typical wavelengths of the gluons being $\lambda_g \sim r_p$. For $J/\Psi$, for example, 
which is not a Coulomb bound state, potential models inspired by the Cornell 
model~\cite{{Eichten:1974af},{Eichten:1978tg}} predict a r.m.s. radius of the order of 
$r_{J/\Psi} \sim 0.2~{\rm fm} \ll r_p$\rm. Therefore, since $\lambda_g \gg r_{J/\Psi}$, 
the $J/\Psi$ in the nucleus behaves like a small color dipole interacting with a uniform gluon 
field. 

Having such a picture in mind for a quarkonium in a nucleus, one can describe the quarkonium-nucleus 
system using standard mean-field techniques of nonrelativistic many-body physics~\cite{Negele:1988vy}.  
Given a nonrelativistic two-body quarkonium-nucleon potential, $W_{\varphi N}$, motivated either 
by a pNRQCD calculation or by phenomenological fits to lattice results, one can embed such a 
potential in a nonrelativistic many-body Hamiltonian of the form 
\be
H  = H_N + H_{\varphi N},
\label{phi-nucleus}
\ee
where $H_N$ contains the kinetic energy of the nucleons and nucleon-nucleon (and multi-nucleon) 
interactions, and  
\bea
H_{\varphi N} &=& \int d^3r \, 
\varphi^\dag(t,\r) \left( - \dfrac{1}{2m_\varphi} \nabla^2  \right)\varphi(t,\r) \nn \\
&& + \, \int d^3r d^3r' \, N^\dag(t,\r) \varphi^\dag(t,{\r}\,') \, W_{\varphi N}(\r - {\r}\,') \,
\varphi(t,{\r}\,') N(t,\r), 
\label{H-phiN}
\eea
where $N(t,\r)$ and $\varphi(t,\r)$ are the nucleon and quarkonium nonrelativistic quantum 
field operators{\textemdash}for simplicity of presentation, we omit flavor and spin indices.  Let 
$\{\varphi_\alpha (\r)\}$ be a complete set of single-particle states of a quarknonium $\varphi$, and 
$\{N_n(\r)\}$ a corresponding set for the nucleons, where $\alpha$ and $n$ collect all quantum numbers 
necessary to specify the corresponding single-particle states. Taking the expectation value of 
$H$ in a state with $A$ independent nucleons and one quarkonium and varying the expectation value 
with respect to a $\varphi^*_\alpha (\r)$, one obtains~\cite{Negele:1988vy}
\beq
- \dfrac{1}{2m_\varphi}  \nabla^2  \varphi_\alpha (\r) + W_{\varphi A}(\r) 
\varphi_\alpha (\r)  = \epsilon_\alpha \varphi_\alpha (\r) ,
\label{Schr-phi}
\ee
where $\epsilon_\alpha$ are the single-particle energy states of a quarkonium in the system, 
$W_{\varphi A}(\r)$ is the $\varphi$-nucleus mean-field potential, given by
\beq
W_{\varphi A}(\r) = \int d^3r' \, W_{\varphi N}(\r - {\r}\,')  \,  \rho_A ({\r}\,'),
\label{W-phi-nuc}
\ee
with $\rho_A(\r)$ being the nuclear density
\beq
\rho_A(\r) = \langle A|N^{\dag}(\r) N(\r)|A\rangle = \sum^A_{n=1} N^*_n(\r) N_n(\r).
\eeq
If one makes the assumption that the quarkonium in the nucleus does not change the density 
of nucleons, i.e. that the single-nucleon states are not modified by the interactions of the 
quarkonium with the nucleons, we have that, given a nuclear density profile $\rho_A(\r)$, from a model
or from experiment, Eq.~(\ref{Schr-phi}) is an ordinary Schr\"odinger equation
for a particle $\varphi$ in a potential $W_{\varphi A}(\r)$.  

For nuclei composed of two or three nucleons only, the mean field treatment discussed above
is not appropriate. For those nuclei, few-body methods are required. A few of these include
the Green's function Monte Carlo, Faddeev-Yakubovsky equations, hyperspherical coordinates, 
Gaussian-basis variational, and the stochastic variational 
methods{\textemdash}Ref.~\cite{Kamada:2001tv} 
presents a comparison of results obtained with those methods for four-body nuclei using a particular 
$NN$ potential. In this review, the smallest nucleus considered is $^4{\rm He}$; although it is a 
very dense nucleus, the mean field approximation might give only a first estimate for what one 
can expect with an accurate few-body calculation. We refer the reader to the recent review
in Ref.~\cite{Hosaka:2016ypm} for a presentation on the use of Gaussian expansion 
method~\cite{{Kamimura:1988zz},{Hiyama:2003cu}} to charmonium binding to two- and four-nucleon 
nuclei~\cite{Yokota:2013sfa}. We also remark that in a mean field treatment of a many-body system,
the center of mass motion is not removed, but for a large nucleus this is not a severe problem.
For $^4{\rm He}$ we follow the prescription used in the QMC model, where we use the 
$\varphi$-$^{4}{\rm He}$
reduced mass in Eq.~(\ref{Schr-phi}) instead of $m_\varphi$.

A quarkonium-nucleon potential $W_{\varphi N}(\r)$ can be constructed from the forward scattering 
amplitude of a quarkonium off a nucleon. The scattering amplitude can be written as a product of two 
terms: (1) a matrix element of gluon fields in the nucleon state, which can be obtained from 
the anomaly in the trace of the QCD energy-momentum 
tensor~\cite{{Chanowitz:1972vd},{Crewther:1972kn}}, and (2) a quarkonium-gluon interaction, which can be 
evaluated using the multipole 
expansion~\cite{{Voloshin:2007dx},{Kaidalov:1992hd},{deTeramond:1997ny},{Sibirtsev:2005ex}}, 
or the operator product expansion of a current-current correlation function 
in an one-nucleon state~\cite{Kharzeev:1995ij}, or an effective Lagrangian for quarkonium-gluon 
fields~\cite{Luke:1992tm}. Specifically, the forward elastic $\varphi$N scattering matrix can 
be evaluated from the gWEFT Lagrangian in Eq.~(\ref{LgWEFT}) as the expectation value of the 
$\varphi$-gluon term in an one-nucleon 
state 
\be
{\cal A}_{\varphi N} =  \dfrac{1}{2} \alpha_\varphi \, \langle N|\left(g\El\right)^2|N\rangle ,
\label{A-phiN}
\ee
where we use a nonrelativistic normalization for the states $|N\rangle$, and the expression for 
the quarkonium polarizability $\beta$ is given in Eq.~(\ref{epol}). From Eq.~(\ref{def-AdS}), one 
then has that this corresponds to a scattering length $a_{\varphi N}$ given 
by% ** beginning of footnote
\footnote{Our conventions for the relationships involving the (on-shell) $S-$matrix, scattering 
amplitude ${\cal A}(k)$, phase shift $\delta(k)$, scattering length $a$, and effective range $r_e$,
for $s$-waves are 
\be
S = 1 + i  \dfrac{2 \mu k}{2\pi} {\cal A}(k) = e^{2i\delta(k)},      \hspace{0.5cm}
{\cal A}(k)  = \dfrac{2\pi/\mu}{ k \; {\rm cotan} \, \delta (k) - ik}, \hspace{0.5cm}
k \;{\rm cotan}\,\delta(k) = - \dfrac{1}{a} + \dfrac{1}{2} r_e k^2 + \cdots , 
\label{def-AdS}
\ee
where $\mu$ is the reduced mass of the two-body system. }
% ** end of footnote
\be
a_{\varphi N} = - \left(\dfrac{\mu_{\varphi N}}{2\pi} \right) \, {\cal A}_{\varphi N} 
=   - \left(\dfrac{\mu_{\varphi N}}{4\pi} \right) \, \alpha_\varphi \, 
\langle N|\left(g\El\right)^2|N\rangle .
\label{a-phiN}
\ee
where $\mu_{\varphi N}$ is the reduced mass 
\be
\mu_{\varphi N} = \dfrac{m_\varphi m_N}{m_\varphi + m_N}.
\label{red-mass}
\ee
One can obtain an estimate for the the matrix element $\langle N|\left(g\El\right)^2|N\rangle$ 
by relating it to the nucleon mass via the anomaly in the trace of the energy momentum 
tensor (in the chiral limit). First, note that~\cite{Voloshin:2007dx}: 
\bea
\langle N|\left[ \left(g\El\right)^2 - \left(g{\B}\right)^2 \right] |N\rangle 
= - \dfrac{1}{2} \langle N|g^2G^a_{\mu\nu} G^{a \mu \nu}|N\rangle
= \dfrac{16\pi^2}{9} m_N \leq \langle N|\left(g\El\right)^2 |N\rangle,
\eea
where we used Eq.~(\ref{mass-anomaly}) and the last inequality follows from the fact that 
$\langle N|\left(g{\B}\right)^2|N\rangle \ge 0$. Therefore, one obtains the following
estimate for the scattering length
\be
a_{\varphi N} \leq - \left(\dfrac{\mu_{\varphi N}}{4\pi} \right) \dfrac{16 \pi^2}{9} 
m_N \alpha_\varphi  
= - \dfrac{4\pi m_N}{9} \mu_{\varphi N}  \alpha_\varphi.
\label{a-fin}
\ee
This shows that the color polarization force is attractive. 

To investigate the binding of $\varphi$ in a nucleus due to such a polarization force 
one would need a potential to be used in Eq.~(\ref{W-phi-nuc}). Knowledge of the 
momentum-independent forward scattering amplitude ${\cal A}_{\varphi N}$ is not enough 
for constructing such a potential. One can, however, construct a $\varphi N$ contact 
interaction, which we denote by $W^{\rm pol}_{\varphi N}(\r)$, that reproduces that 
forward scattering amplitude; namely:
\beq
W^{\rm pol}_{\varphi N}(\r) =  \dfrac{4\pi}{2\mu_{\varphi N}} \, a_{\varphi N} \, 
\delta(\r) = - \dfrac{8\pi^2}{9} m_N \alpha_\varphi \, \delta(\r),
\label{phiN-contact}
\eeq 
where, for definiteness, we have taken the equality in Eq.~(\ref{a-fin}). Using this into 
Eq.~(\ref{W-phi-nuc}), one obtains for the $\varphi$-nucleus potential due to the color 
polarizability force the following expression:
\be
W^{\rm pol}_{\varphi A}(\r) = \dfrac{4\pi}{2\mu_{\varphi N}} \, a_{\varphi N} \, 
\rho_A(\r) = - \dfrac{8\pi^2 }{9}  m_N \alpha_\varphi \, \rho_A(\r).
\label{phiA-pol}
\ee

Although the full $\varphi N$ potential is of finite range and includes a long-range
tail of the form of a van der Waals force, this interaction already 
provides interesting insight into the problem of quarkonium binding to nuclei. It is
clear that the strength of the polarization force in Eq.~(\ref{phiA-pol}) is controlled 
by the value of the polarizability $\alpha_\varphi$. The estimates in 
Refs.~\cite{Voloshin:2007dx,{Sibirtsev:2005ex}} find that the binding energy of $J/\Psi$ 
in nuclear matter (taking $\rho_0 = 0.17~{\rm fm}^{-3}$), is $21$~MeV for $\alpha_{J/\Psi} 
= 2~{\rm GeV}^{-3}$, which corresponds to a scattering length of $a_{J/\Psi N} = - 0.37$~fm. 
This is a rather 
strong binding energy on the nuclear scale; recall that the binding energy of a nucleon 
in nuclear matter is $16$~MeV. On the other hand, a very recent extraction~\cite{Gryniuk:2016mpk} 
based on a global fit to both differential and total cross sections from available data on 
$J/\Psi \, p$ scattering leads for the spin-averaged $J/\Psi \, p$ $s$-wave scattering length 
the value $a_{J/\Psi\,  p} = - 0.046 \pm 0.005$~fm, which is consistent with the value 
$a_{\eta_c N} = - 0.05$~fm from Ref.~\cite{Kaidalov:1992hd} but smaller by almost an order 
of magnitude than the value estimated in Refs.~\cite{Voloshin:2007dx,{Sibirtsev:2005ex}}.
A scattering length of $a_{J/\Psi\,  p} = - 0.046$~fm leads to a $J/\Psi$ 
binding energy in nuclear matter (taking $\rho_0 = 0.17~{\rm fm}^{-3}$) of $2.7 \pm 0.3$~MeV, 
which is of the order of the deuteron binding energy. Estimates based on lattice hybrid 
potentials~\cite{Lakhina:2003pj} lead to even smaller values for the strength of the $\varphi N$ 
interaction. Reference~\cite{Xiao:2015fia} investigated the existence of
bound states in elastic and inelastic channels of charmonium-nucleon and bottomonium-nucleon systems 
using an extended local hidden gauge formalism and incorporating constraints from heavy quark 
spin-flavor symmetry, but no scattering lengths and effective range parameters were provided.

A first estimate on the required attraction to bind a quarkonium to a nucleus can be
obtained from the condition for the existence of a nonrelativistic $s$-wave bound state 
of a particle of mass $m$ trapped in an attractive spherical well of radius $R$ and 
depth $V_0$~\cite{Schiff}:
\begin{equation}
V_0 > \dfrac{\pi^2 \hbar^2}{8 m R^2}.
\label{bound}
\end{equation}
Taking $m = m_{J/\Psi}$ and $R = 5$~fm (radius of a relatively large nucleus),
one obtains $V_0 > 1$~MeV. Based on such an estimate, one may say that the prospects for a 
quarkonium to form bound states with a nucleus seem to be very good. Real nuclei, 
however, have a surface and, as we show shortly, the above estimate can be quite 
misleading. A full calculation, using realistic density profiles of nuclei is required for 
a more reliable estimate. Single-particle bound-state energies will be presented
for a few nuclei in subsection~\ref{jpsiresult}   

Next, we consider the charmonium-nucleon system from the perspective of the lattice results 
of Refs.~\cite{{Kawanai:2010ev},{Kawanai:2010ru}}. Those references present a fit of the lattice 
data in terms of a finite-range nucleon-charmonium effective nonrelativistic potential 
$W_{\varphi N}(\r)$. In addition, results are presented for $\eta_c N$ and $J/\Psi N$ scattering
lengths and effective range parameters extracted from evaluations of scattering phase shifts 
through L\"uscher's method~\cite{Luscher:1990ux} adapted to twisted boundary 
conditions~\cite{Bedaque:2004kc}. Most of the results are from quenched lattice data,
but Ref.~\cite{Kawanai:2010ru} also presents selected results for the 
$\eta_c N$ system using unquenched lattice data. The quenched results are obtained with a 
lattice size of $L^3\times T=32^3\times 48$ and lattice spacing $a \approx 0.094$~{\rm fm}. 
The simulations are for three pion masses, $m_\pi={640,\ 720,\ 870}$~MeV, which correspond to 
nucleon masses $m_N = {1430,\ 1520,\ 1700}$~MeV, quarkonium masses $m_{\eta_c}=2920$~MeV and 
$m_{J/\Psi}=3000$ MeV. For the $J/\Psi N$ system, there are two spin states, spin-1/2 and 
spin-3/2. The $s$-wave scattering lengths found are~\cite{Kawanai:2010ru}: 
$(a_{J/\Psi N})_{\rm SAV} \sim 0.35~{\rm fm} > a_{\eta_c N} \sim 0.25~{\rm fm}$, 
where ${\rm SAV}$ means spin average, with very little dependence on the pion mass 
within the range of masses used. In addition, the $s$-wave effective range $r_e$ is 
of similar value for both $\eta_c N$ and $J/\Psi N$ systems, $r_e \sim 1.0$~fm, also 
with very little pion mass dependence but with errors of the order of 50\%.

The extraction of an effective potential in Refs.~\cite{{Kawanai:2010ev},{Kawanai:2010ru}} 
is based on the method introduced in Ref.~\cite{Ishii:2006ec} for the $NN$ interaction. 
In general terms, an equal-time Bethe-Salpeter amplitude is calculated on the lattice 
through nucleon and charmonium interpolators. An appropriate projection with respect 
to the discrete elements of cubic rotation group is made to obtain a Bethe-Salpeter 
wave function that corresponds in the continuum to an $s$-wave. The effective 
charmonium-nucleon potential is then defined as the equivalent appearing in a 
Schr\"odinger equation with stationary energy $E$, namely:
\beq
V_{\varphi N}(r) = E_{\varphi N} + {1 \over 2 \mu_{\varphi N}} {\nabla^2_{\rm latt} 
\phi_{\varphi N}(r) 
\over \phi_{\varphi N}(r) },
\label{lattVNpsi}
\eeq
where $\nabla^2_{\rm latt}$ is a lattice Laplacian and $\mu_{\varphi N}$ is the reduced 
mass of the $\varphi N$ two-body system. The fit provided for the $\eta_c N$ potential from 
simulations using $m_\pi = 640$~MeV gives a very good fit of the data in the range $r \sim  
0.3$~fm to $r = 2.5$~fm. It is given in the form of a Yukawa potential with a strength 
$\gamma$ and range~$\alpha$:
\beq
V^{\rm fit}_{\eta_c N}(r) = - \gamma \, {e^{-\alpha r} \over r},
\label{VetacN-fit}
\eeq
with $\gamma = 0.1$ and~$\alpha = 0.6~{\rm GeV} \sim 3~{\rm fm}^{-1}$. 
Interestingly, the lattice results for large $r$ cannot be fitted with 
a $1/r^n$ falloff, with $n > 1$, as one would expect from a typical 
van der Waals force. The authors of Ref.~\cite{Kawanai:2010ru} argue
that this might be indication of a screening effect of nonperturbative
nature. 

As mentioned, the lattice values for the scattering lengths reported in 
Ref.~\cite{Kawanai:2010ru} are extracted from L\"uscher's formula for the
phase shifts and, therefore, they contain information on the full $\varphi N$ interaction 
and not only from the fitting region $r \sim  0.3$~fm to $r = 2.5$~fm. One can assess
the importance of the interaction coming from the region $r <  0.3$~fm by calculating the 
scattering length with $V^{\rm fit}_{\eta_c N}(r)$ and comparing the result with the 
value from the lattice simulation.  One can extract the scattering length and effective range 
by calculating the scattering phase shift $\delta(k)$ and fitting $k \, {\rm cotan}\, \delta(k)$ 
for small values of $k$ to the formula given in Eq.~(\ref{def-AdS}). To calculate the phase shifts
we have used the variable phase approach~\cite{VPA}.
We have calculated the scattering length with the fitted potential given in Eq.~(\ref{VetacN-fit}),
using the lattice values $m_N = 1430$~MeV and $m_{\eta_c}=2920$~MeV. We obtained for the 
$s$-wave scattering length $a_{\eta_c N} = - 0.13$~fm, which is smaller by a factor of two 
in comparison with the value obtained with L\"uscher's formula. For orientation, we mention that 
in Born approximation, the scattering length from the Yukawa fit in Eq.~(\ref{VetacN-fit}) 
is $a^{\rm Born} = - \gamma \, 
(2\mu_{\eta_c N}/\alpha^2) = - 0.1$~fm. This indicates that the full $\eta_c N$ potential 
receives substantial contributions from relative distances shorter than $r < 0.3$~fm,
i.e. when the $\eta_c$ is immersed in the nucleon, a situation resembling very much the 
hadrocharmonium picture suggested in Ref.~\cite{Dubynskiy:2008mq}. In this picture, a
heavy quarkonium is bound as a compact object within the volume of a light hadron; in the 
case the light hadron is a nucleon, the hadrocharmonium would be a charmed pentaquark. A recent
lattice QCD simulation tested this picture with a variety of light quark hadrons and found 
binding energies ranging from $-1$ to $-2.5$~MeV. The simulations were performed with pion 
and kaon masses given by $m_\pi = 223~{\rm MeV}$ and $m_K = 476~{\rm MeV}$, and the charm quark 
mass was taken $m_c = 1269$~MeV.

Because the data of Refs.~\cite{{Kawanai:2010ev},{Kawanai:2010ru}} show a very mild pion mass 
dependence within the range of masses used in the lattice simulations, the extrapolation of the 
potential extracted in that references to physical masses and short distances with the use of a 
chiral EFT is virtually impossible. In view of this, here we simply 
take a phenomenological approach, in that we cut off the lattice potential of Eq.~(\ref{VetacN-fit}) 
at some prescribed distance, that we suggestively denote by $r_{\rm vdW}$, and add to it a 
constant piece, $W_0$, which extends from $r = 0$ to $r_{\rm vdW}$. The constant potential
mimics the feature of the interaction being of a small color dipole interacting with a uniform 
gluon field. The value of $W_0$ is fixed by fitting the lattice value for the scattering 
length{\textemdash}we make no effort to fit the effective range $r_e$ due to the large errors 
associated with its extraction from the lattice data. For the $J/\Psi N$ system we refit $W_0$ 
to reproduce the spin-averaged lattice value for $a_{J/\Psi N}$. Since we are unable to perform 
an extrapolation of the potentials to physical masses due to the large pion masses used
in the lattice simulations, we simply use the physical masses of the nucleon and quarkonia in 
the fitting procedure. Specifically, we write for the full $\varphi N$ potential 
\be
W^{\rm latt}_{\varphi N}(r) = - W_0 \, [1-f(r,r_{\rm vdW})] + V^{\rm fit}_{\eta_c N}(r) \, 
f(r,r_{\rm vdW}),
\label{phiN-latt}
\ee
with a $f(r,r_{\rm vdW})$ that resembles a step function: 
\be
f(r,r_{\rm vdW}) = \dfrac{1}{1 + \left(r_{\rm vdW}/r\right)^{10}}.
\label{step}
\ee
To avoid proliferation of uncontrolled parameters, in fitting $W_0$ for the 
$J/\Psi N$ system we use the $\eta_c N$ values for $\gamma$~and~$\alpha$ given above. 
The $\varphi-$nucleus potential is then obtained by inserting Eq.~(\ref{phiN-latt}) in 
Eq.~(\ref{W-phi-nuc}); we denote this potential by $W^{\rm latt}_{\varphi A}(\r)$ 
\be
W^{\rm latt}_{\varphi A}(\r) = \int d^3r'\,W^{\rm latt}_{\varphi N}(\r - \r\,') \rho_A(\r\,').
\label{phiA-latt}  
\ee

%%%%%%%%%%%%%%% Below causes bad table orders !!!!!
\begin{table}[b]
\begin{center}
\caption{\label{tab:W0}The value of $W_0$ in Eq.~(\ref{phiN-latt}) for two values of the cutoff 
parameter $r_{\rm vdW}$ for the $\eta_c N$ and $J/\Psi$ systems. Also shown are the effective 
range parameters that are obtained with the potential $W^{\rm latt}_{\varphi N}(r)$ when the 
scattering lengths are fitted to the lattice values. Physical masses of the nucleon and charmonia
are used. Units of $r_{\rm vdW}$ and $r_e$ are fm and of $W_0$ is MeV. \vspace{2ex}}
\begin{tabular}{cccccccc}
\hline \\[-0.4true cm]

 & \multicolumn{3}{c}{$\eta_c N$}  & & \multicolumn{2}{c}{$J/\Psi N$} & \\[0.1true cm]
\cline{2-3}\cline{6-7} \\[-0.4truecm]
$r_{\rm vdW}$ & $W_0$ & $r_e$ & & & $W_0$ & $r_e$ & \\[0.05true cm]
\hline \\[-0.4true cm]
0.3           &  252  & 1.4 & & &  288  &  1.2  & \\
\hline
0.5           &  74   & 1.7 & & &   95  &  1.4  & \\
\hline
\end{tabular}
\end{center}
\end{table}
%%%%%%%%%%%%%%%%%%%%%%%%%%%%%%%%%%%%%%%%%%

We have chosen two typical values for $r_{\rm vdW}$: one is 
$r_{\rm vdW} = 0.3$~fm, which corresponds roughly to the lowest value 
of $r$ used in the fitting of the lattice data with the Yukawa form 
$V^{\rm fit}_{\eta_c N}(r)$, and the other is $r_{\rm vdW} = 0.5$~fm, 
which is a little larger than one half of the radius of the proton. The resulting 
values for $W_0$ are shown in Tab.~\ref{tab:W0}; they reveal that as 
$r_{\rm vdW}$ decreases, $W_0$ 
must increase to fit the scattering length. This is expected, as the strength 
of the Yukawa, $\gamma = 0.1$, is not enough to describe the lattice values of the 
scattering lengths when $r_{\rm vdW} = 0$. The values of effective range 
parameter that are obtained with $W^{\rm latt}_{\varphi N}(r)$ for both 
$\eta_c N$ and $J/\Psi$ systems are compatible, within errors, with 
the lattice results. Of course, by adjusting simultaneously $W_0$ and $r_{\rm vdW}$,
both the scattering length and effective range could be fitted, but the erros are 
large for $r_e$ and there is not much to be gained with a fit to a quantity 
not well constrained by data. Given the potential, the important issue is to know 
how large must a nucleus be to bind a charmonium.  This will be discussed in the 
first part of subsection~\ref{jpsiresult}, where we present numerical results for 
the $\varphi-$nucleus potentials and single-particle energy levels.

%%%%%%%%%%%%%%%%%%%%%%%%%%%%%%%%%%%%%%%%%%%%%%%%%%%%%%%%%%%
\subsection{$J/\Psi$ self-energy and $D\overline{D}$ loop}
\label{DDloop}
%%%%%%%%%%%%%%%%%%%%%%%%%%%%%%%%%%%%%%%%%%%%%%%%%%%%%%%%%%

Based on the QMC model explained in section~\ref{qmc}, $J/\Psi$ mass shift in 
nuclear matter and $J/\Psi$-nucleus bound states were studied 
in Refs.~\cite{Krein:2010vp,{Tsushima:2011fg},{Tsushima:2011kh}},  
using the effective Lagrangian approach with the in-medium  
$D$ and $\overline{D}$ meson masses calculated by the QMC model~\cite{QMCreview}. 
In these studies the $J/\Psi$ self-energy in the intermediate 
states involved the $D, \overline{D}, D^*$ and $\overline{D^*}$ mesons.
It turned out that the $J/\Psi$ self-energy has larger contributions 
from the loops involving the $D^*$ and $\overline{D^*}$ mesons. 
This is unexpected, since the mass of the $D^*$ ($\overline{D^*}$) 
is heavier than that of $D$ ($\overline{D}$) by about 140 MeV in vacuum, 
and the decrease of the masses in nuclear matter are nearly the same for 
the $D$ and $D^*$ mesons~\cite{Krein:2010vp,Krein:2010vp,{Tsushima:2011kh}} 
(also to be shown later). Thus, the relative enhancement of these meson loop 
contributions to the $J/\Psi$ self-energy in medium should be similar to that 
in vacuum, or the loop involving only the $D^*$ and $\overline{D^*}$ mesons  
in the $J/\Psi$ self-energy is not expected to give a larger 
contribution than that involving only the $D$ and $\overline{D}$ mesons.
This is because, between the $D\overline{D}$- and $D^* \overline{D^*}$-loops, 
the energy necessary to excite the latter loop needs, at least twice 
the mass difference of the $D$ and $D^*$.

At least two issues may be considered for this unexpected result: 
(i) coupling constants used for the $J/\Psi$-$DD^*$ and $J/\Psi$-$D^*D^*$ 
were assumed to be the same as that for $J/\Psi$-$DD$, and (ii) ambiguity 
in the effective Lagrangian adopted. 
Considering these, we have chosen to update~\cite{jpsi4} 
our previous works of the $J/\Psi$-nucleus bound 
states~\cite{Krein:2010vp,{Tsushima:2011fg},{Tsushima:2011kh}} in this subsection.
Thus, similarly to the $\phi$-meson case to be discussed in section~\ref{phi-meson}, 
we update the $J/\Psi$ mass shift and $J/\Psi$-nuclear bound states
including only the $D\overline{D}$-loop for the $J/\Psi$ self-energy.

%%%%%%%%%%%%%%%%%%%%%%%%%%%%%%%%%%%%%%%%%%%%%%%%%%%%%%%%%%
%\subsection{$J/\Psi$ meson in nuclear matter and nuclei}
%\label{jpsi_potential}
%%%%%%%%%%%%%%%%%%%%%%%%%%%%%%%%%%%%%%%%%%%%%%%%%%%%%%%%%%

In Refs.~\cite{Krein:2010vp,{Tsushima:2011fg},{Tsushima:2011kh}} $J/\Psi$ mass shift
in nuclear matter was studied based on an effective Lagrangian
approach, including the $D, \overline{D}, D^*$ and $\overline{D^*}$ 
mesons in the intermediate states for the $J/\Psi$ self-energy.
We update the results by including only the $D\overline{D}$-loop contribution for 
the $J/\Psi$ self-energy as mentioned above. We report the updated results for the 
bound state single-particle energies, by solving the Schr\"{o}dinger equation for the 
$J/\Psi$ meson produced nearly at rest. 
The results for the $\phi$-nuclear bound states to be 
discussed in section~\ref{phi-meson}, both solving the Klein-Gordon equation 
and the Schr\"{o}dinger equation, have proven to be essentially the same. 
Thus, we solve the Schr\"{o}dinger equation, which makes the numerical 
procedure simpler and can compare with the nonrelativistic QCD approach. 
However the differences are expected to be even smaller    
for the results obtained by solving the Klein-Gordon and Schr\"{o}dinger equations 
for the $J/\Psi$ meson, since $J/\Psi$ is heavier than $\phi$.
Note that, the structure of heavy nuclei, as well as the
medium modification of the $D$ mass ($\overline{D}$ mass), 
are explicitly calculated by using the QMC model~\cite{QMCGuichon,QMCreview}.

\begin{figure}[t]
\centering
  \includegraphics[scale=0.6]{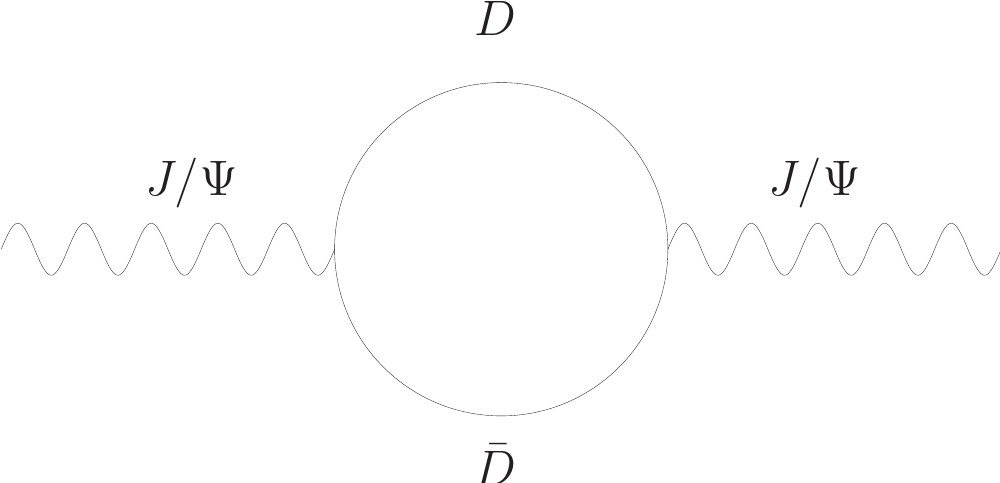}
  \caption{$D\overline{D}$-loop contribution to the $J/\Psi$ self-energy.}
  \label{fig:loop}
\end{figure}

A first estimate for the mass shifts for the
$J/\Psi$ meson (and $\Psi(3686)$ and $\Psi(3770)$) in nuclear medium
including the effects of $D\overline{D}$ loop in the $J/\Psi$
self-energy -- see Fig.~\ref{fig:loop} -- was made in Ref.~\cite{Ko:2000jx}.
Employing a gauged effective Lagrangian for the coupling of
$D$ mesons to the charmonia, the mass shifts
were found to be positive for $J/\Psi$ and $\psi(3770)$, and negative for
$\psi(3660)$ at normal nuclear matter density $\rho_0$. These results were
obtained for density-dependent $D$ and $\bar D$ masses that decrease linearly
with density, such that at $\rho_0$ they are shifted by $50$~MeV.
The loop integral in the self-energy (Fig.~\ref{fig:loop}) is divergent and
was regularized using form-factors derived from the $^3P_0$ decay
model with quark-model wave functions for $\Psi$ and $D$. The positive
mass shift is at first sight puzzling, since even with a $50$~MeV
reduction of the $D$ masses, the intermediate state is still
above threshold for the decay of $J/\Psi$ into a $D\overline{D}$ pair and
so a second-order contribution should be negative.
However, as we have shown in Ref.~\cite{Krein:2010vp},
this is a result of the interplay of the form factor used and the gauged nature
of the interaction used in Ref.~\cite{Ko:2000jx}.
Focusing on these points, we update the study with the ``the gauged term''  
more in detail. 
In the same way as done in Ref.~\cite{Krein:2010vp}, 
the mass shift of the $J/\Psi$ is estimated, but including 
only the $D\overline{D}$-loop in the $J/\Psi$ self-energy,  
using both non-gauged and gauged effective Lagrangians. 
It turns out that the sign of the $J/\Psi$ mass shift 
in nuclear matter due to the $D\overline{D}$-loop,    
depends on the value of the cutoff in the form factor 
when the gauged term is employed. Reasonably large cutoff values 
still give the negative $J/\Psi$ mass shift when 
including the gauged term.

%%%%%%%%%%%%%%%%%%%%%%%%%%%%%%%%
\vspace{0.5cm}
%%%%%%%%%%%%%%%%%%%%%%%%%%%%%%%%

We briefly discuss below how the $J/\Psi$ scalar potential (mass shift) 
is calculated in nuclear matter~\cite{Krein:2010vp} 
with the $D\overline{D}$-loop for the $J/\Psi$ self-energy.
We use the following phenomenological effective Lagrangian densities at the hadronic level, 
which are similar to those to be used for the $\phi$-meson 
in subsection~\ref{phi_vacuum}, correspond to  
replacing $\psi \to \phi$, $D \to K$ and $\overline{D} \to \overline{K}$ 
(in the following we denote by $\psi$ the field representing $J/\Psi$):
%
%\begin{widetext}
\bea
{\mathcal L}_{int} &=& {\mathcal L}_{\psi D D} + {\mathcal L}_{\psi\psi D D}, 
\\
{\mathcal L}_{\psi D D} &=& i g_{\psi D D} \, \psi^\mu
\left[\bar D \left(\partial_\mu D\right) -
\left(\partial_\mu \bar D\right) D \right] ,
\label{LpsiDDbar}\\
{\mathcal L}_{\psi\psi D D} &=& g^2_{\psi D D} \psi_\mu \psi^\mu \overline{D} D .
\label{Lpsigauge}
\eea
%\end{widetext}
%
Here our convention is
\bea
{D = \left(\begin{array}{c}
  D^0 \\
  D^+
\end{array}\right)}, 
\hspace{1.0cm}
\bar D = (\overline{D^0} \;\;\; D^{-} ).
\label{doublets}
\eea
We note that the Lagrangians are an $SU$(4) extension of light-flavor
chiral-symmetric Lagrangians of pseudoscalar and vector mesons. In the light
flavor sector, they have been motivated by a local gauge symmetry,
treating vector mesons either as massive gauge bosons or as dynamically
generated gauge bosons. Local gauge symmetry implies in the contact interaction
in Eq.~(\ref{Lpsigauge}) involving two pseudoscalar and two vector mesons.
In view of the fact that $SU$(4) flavor symmetry is strongly broken in nature, 
and in order to stay as close as possible to phenomenology, we use experimental values 
for the charmed meson masses and use the empirically known meson coupling constants. 
For these reasons we choose not to use gauged Lagrangians -- a similar attitude was 
followed in Ref.~\cite{Lin:1999ve} in a study of hadronic scattering of charmed mesons. 
However, in order to compare results with Ref.~\cite{Ko:2000jx} and assess the impact 
of a contact term of the form Eq.~(\ref{Lpsigauge}), we also present results for the 
$J/\Psi$ mass shift including such a term. More detailed, similar discussions will be 
made for the $\phi$-meson case, below Eq.~(\ref{KKbar}) in subsection~\ref{phi_vacuum}. 
We mention that recent investigations of $SU$(4) flavor symmetry breaking in hadron 
couplings of charmed hadrons to ligth hadrons are not conclusive; while two studies 
based on the Dyson-Schwinger equations of QCD find large deviations from $SU$(4) 
symmetry~\cite{{ElBennich:2011py},{El-Bennich:2016bno}}, studies using QCD sum 
rules~\cite{{Navarra:1998vi},{Khodjamirian:2011jp}}, a constituent quark 
model~\cite{Fontoura:2017ujf} and a holographic QCD model~\cite{Ballon-Bayona:2017bwk} 
find moderate deviations.

We are interested in the difference of the in-medium, $m^*_\psi$, and vacuum,
$m_\psi$,
\beq
\Delta m_\psi = m^*_\psi - m_\psi,
\label{Delta-m}
\eeq
with the masses obtained from
\beq
m^2_\psi = (m^0_\psi)^2 + \Sigma_{D\overline{D}}(k^2 = m^2_\psi)\, .
\label{m-psi}
\eeq
Here $m^0_\psi$ is the bare mass and $\Sigma_{D\overline{D}}(k^2)$ is the total $J/\Psi$
self-energy obtained from the $D\overline{D}$-loop contribution. 
The in-medium mass, $m^*_\psi$, is obtained likewise, with the
self-energy calculated with medium-modified $D$ meson mass 
calculated by the QMC model.

We take the averaged, equal masses for the neutral and charged $D$ mesons, i.e.
$m_{D^0} = m_{D^{\pm}}$. Averaging over the
three polarizations of $J/\Psi$, one can calculate the $D\overline{D}$-loop contribution
to the $J/\Psi$ self-energy $\Sigma_{D\overline{D}}$ as
\beq
\Sigma_{D\overline{D}}(m^2_\psi) = - \dfrac{ g^2_{\psi \, DD}}{3\pi^2}
\int^\infty_0 dq \, \vq^2 \, F_{D\overline{D}}(\vq^2) \, K_{D\overline{D}}(\vq^2),
\label{Sigma-l}
\eeq
where $F_{D\overline{D}}(\vq^2)$ is the product of vertex form-factors (to be discussed later) and the
$K_{D\overline{D}}(\vq^2)$ for the $D\overline{D}$-loop contribution is given by
\bea
K_{D\overline{D}}(\vq^2) &=& \dfrac{\vq^2}{\omega_D} \left( \dfrac{\vq^2}{\omega^2_D
- m^2_\psi/4} - \xi \right) , \label{KDD} 
\eea
where $\omega_D = (\vq^2+m^2_D)^{1/2}$, 
$\xi = 0$ for the non-gauged Lagrangian of Eq.~(\ref{LpsiDDbar}) 
and $\xi = 1$ with Eq.~(\ref{Lpsigauge}), 
for the gauged Lagrangian of Ref.~\cite{Ko:2000jx}.

Because of baryon number conservation, no vector potential should contribute to
the $D\overline{D}$-loop integral. 
Recall that, for the $K^+$ meson case, $g^q_\omega$ associated with the vector potential
had to be scaled $1.4^2$ times to reproduce an empirically extracted repulsive potential
of about 25 MeV at normal nuclear matter density~\cite{Tsushimak}.
The reason is that $K$-mesons may be regarded as pseudo-Goldstone bosons,
and they are therefore difficult to describe by naive
quark models as is also the case for pions.
For this reason, in earlier work we explored the possibility of also scaling the
$g^q_\omega$ strength by a factor $1.4^2$ for the
$D$ and $\overline{D}$ mesons~\cite{Tsushimad,Sibirtsev:1999js}.
In the present case, this possibility is irrelevant, 
since the vector potential does not contribute to
the final results. Thus, we may focus on the effective masses of the
$D$-mesons. 

\begin{figure}[htb]
\begin{center}
%\hspce*{-2cm}
\includegraphics[height=70mm]{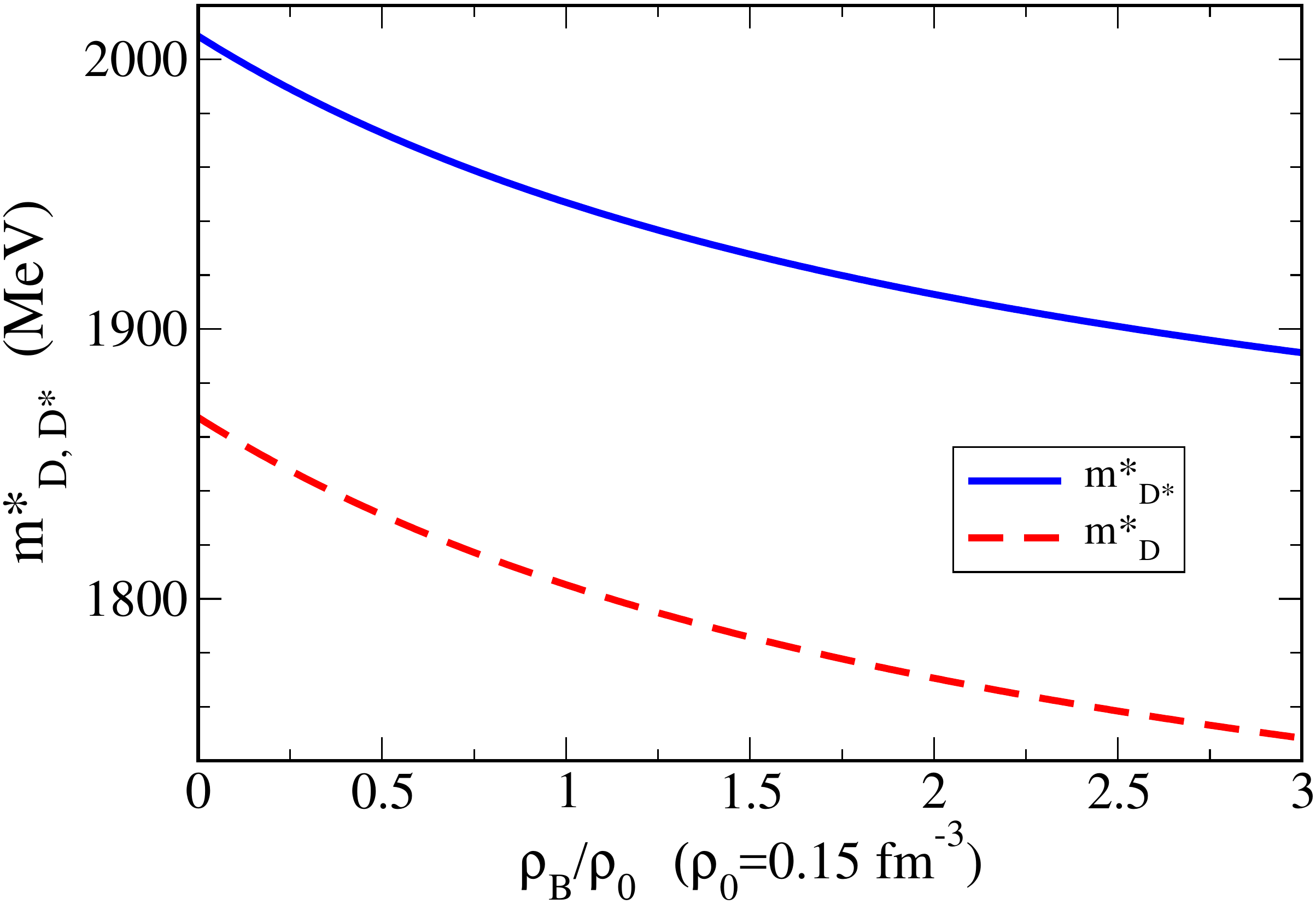}
\caption{
$D$ (lower dashed line) and $D^*$ (upper solid line) meson 
effective masses in symmetric nuclear matter.
}
\label{fig:DDsmass}
\end{center}
\end{figure}

For completeness we show the effective masses of
the $D$ and $D^*$ mesons calculated by the QMC model in Fig.~\ref{fig:DDsmass}.
The net reductions of the $D$ and $D^*$ masses  
are nearly equal for a given density,
as dictated by {\it the light quark number counting rule}~\cite{QMCbc}.
(See also subsection~\ref{hadronmass}.)
This fact supports the assumption applied here that 
the $J/\Psi$ self-energy involving any $D^*$ ($\overline{D^*}$) mesons 
intermediate states should be less enhanced than that involving 
only the $D$ ($\overline{D}$) mesons as in vacuum---the reason 
why we update and ignore any intermediate 
states involving the $D^*$ ($\overline{D^*}$).

%%%%%%%%%%%%%%%%%%%%%%%%%%%%%%%%%%%%%%%%%%%%%%
%%%%%%%%%%%%%%%%%%%%%%%%%%%%%%%%%%%%%%%%%%%%%

Amongst, one of the important ingredients in the present calculation 
is the phenomenological form factor, needed to regularize the self-energy 
loop integral in Eq.~(\ref{Sigma-l}). 
Following previous experience with a similar calculation
for the $\rho$ self-energy~\cite{Leinweber:2001ac} and   
the $\phi$-meson case, we use a dipole form for the vertex form factor,  
\begin{equation}
u_{D}(\vq^2) =  \left(\dfrac{\Lambda_{D}^2 + m^2_\psi}
{\Lambda_{D}^2 + 4\omega^2_{D}(\vq)}\right)^2,
\label{uff}
\end{equation}
so that the $F_{D\overline{D}}(\vq^2)$ in Eq.~(\ref{Sigma-l}) are given by
\bea
&& F_{D\overline{D}}(\vq^2)  = u^2_{D}(\vq^2), 
%\\
%&& F_{DD^*}(\vq^2) = u_D(\vq^2) \, u_{D^*}(\vq^2), \\
%&& F_{D^*D^*}(\vq^2) = u^2_{D^*}(\vq^2),
\label{ffs}
\eea
where $\Lambda_{D}$ is a cutoff mass. Obviously the
main uncertainty here is the value of this cutoff mass. In a simple-minded
picture of the vertex the cutoff mass is related to the extension of
the overlap region of $J/\Psi$ and $D$ meson at the vertex, and therefore
should depend upon the size of the wave function of this meson. One can
have a rough estimate of $\Lambda_{D}$ by using a quark
model calculation of the form factors. Using a $^3P_0$ model for quark-pair
creation~\cite{LeYaouanc} and Gaussian wave functions for the mesons,
the vertex form factor can be written as~\cite{Ko:2000jx}
\begin{equation}
u_{QM}(\vq^2)=e^{-\vq^2/4(\beta^2_D+2\beta^2_\psi)},
\label{ff}
\end{equation}
where $\beta_D$ and $\beta_\psi$ are respectively the Gaussian size parameters of the
$D$ and $J/\Psi$ wave functions. Demanding that the $u_D(\vq^2)$ of
Eq.~(\ref{uff}) and $u_{QM}(\vq^2)$ have the r.m.s. radii
$\langle r^2 \rangle^{1/2}$, with
\beq
\langle r^2 \rangle = - 6 \, \dfrac{d \ln u(q^2)}{dq^2}\Bigg|_{\vq^2=0},
\eeq
one obtains
\beq
\Lambda_D^2 = 32(\beta^2_D +2 \beta^2_\psi) - 4m^2_D.
\label{Lambda}
\eeq
Using $m_D = 1867.2$~MeV and for the $\beta_{D, \psi}$ the values used in Ref.~\cite{Ko:2000jx},
$\beta_D = 310$~MeV and $\beta_\psi = 520$~MeV, one obtains $\Lambda_D = 2537$~MeV.
Admittedly this is a somewhat rough estimate and it is made solely
to obtain an order of magnitude estimate,
since we do not expect that Gaussian form factors should be very accurate at high
$\vq^2$. In view of this and to gauge uncertainties
of our results, we allow the value
of $\Lambda_D$ vary in the range $2000~{\rm MeV} \leq \Lambda_D \leq 6000~{\rm MeV}$. 
We have studied much larger range of the values for $\Lambda_D$~\cite{jpsi4} 
than before~\cite{Krein:2010vp,{Tsushima:2011fg},{Tsushima:2011kh}}.

There remain to be fixed the bare $J/\Psi$ mass $m^0_\psi$ and the coupling constants.
The bare mass is fixed by fitting the physical mass $m_{J/\Psi} = 3096.9$~MeV using
Eq.~(\ref{m-psi}). For the coupling constants we use $g_{\psi DD} = 7.64$,  
which is obtained by the use of isospin symmetry~\cite{Lin:1999ad}.

\begin{figure}[htb]
\centering
\includegraphics[height=60mm]{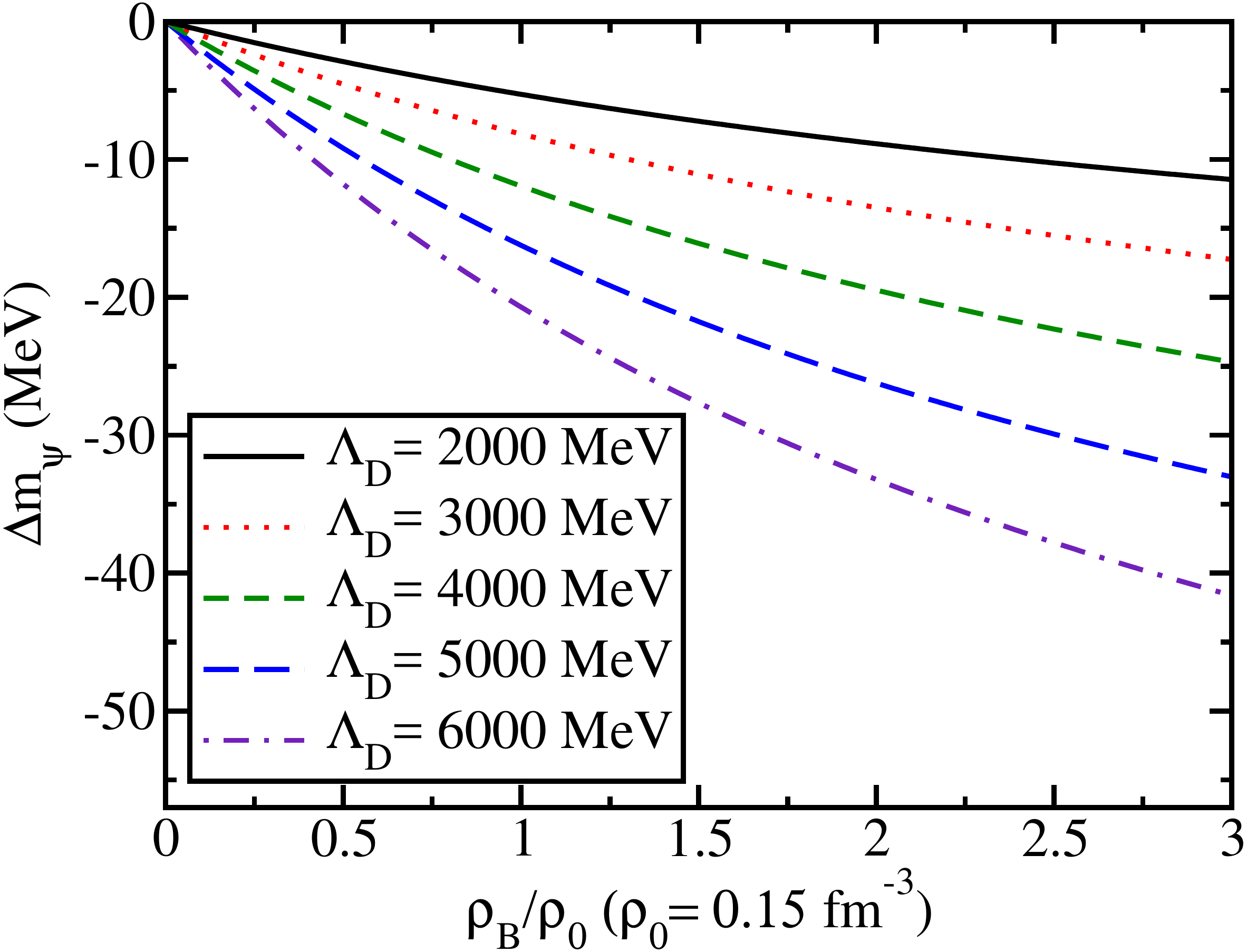}
\hspace{2ex}
\includegraphics[height=60mm]{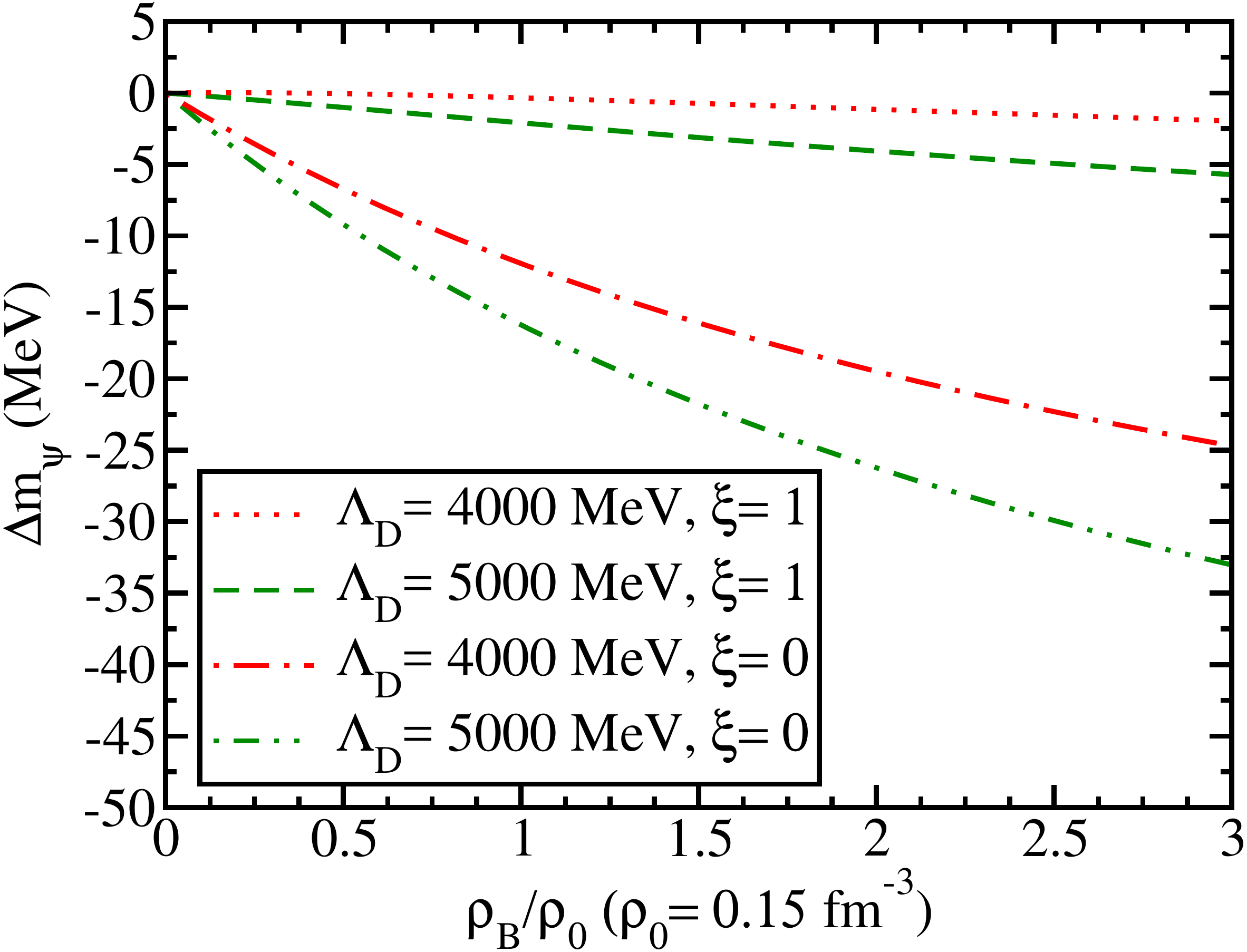}
\caption{Contribution from the $DD$-loop to the 
$J/\Psi$ mass shift $\Delta m_\psi = m^*_\psi - m_\psi$ in symmetric 
nuclear matter without the gauge term ($\xi = 0$) for five different 
values of the cutoff $\Lambda_D$ (left panel), 
and the comparison with including the gauge term ($\xi = 1$) 
for two values of $\Lambda_D$ (right panel).}
\label{fig:DD}
\end{figure}

We calculate the in-medium self-energy using the in-medium $D$ meson mass calculated 
by the QMC model presented in Fig.~\ref{fig:DDsmass}. 
We present results for both $\xi = 0$ (no gauge coupling)
and for $\xi = 1$ (with gauge coupling).
In Fig.~\ref{fig:DD} we show the contribution of
the $D\overline{D}$-loop to the $J/\Psi$ mass shift for $\xi = 0$ (left panel), 
and the comparison with $\xi = 0$ and $\xi = 1$ (right panel).
As the cutoff mass value increases in the form factor, obviously 
the $D\overline{D}$-loop contribution becomes larger.

%%%%%%%%%%%%%%%%%%%%%%%%%%%%%%%%%%%%%%%%%%%%%%%%%%%%%%%%%%%%%%%%%%%%%%%
%%%%%%%%%%%%%%%%%%%%%%%%%%%%%%%%%%%%%%%%%%%%%%%%%%%%%%%%%%%%%%%%%%%%%

First, from the result shown in the left panel in Fig.~\ref{fig:DD} 
without the gauge term ($\xi = 0$), one can see that 
the $J/\Psi$ gets the attractive potential for all the values of 
the cuttoff $\Lambda_D$, $2000 - 6000$ MeV.  
In contrast, one can see from the right panel in Fig.~\ref{fig:DD},  
that the effect of the gauge term tends to oppose the effect (repulsion) of the 
contribution of the $D\overline{D}$-loop as noticed in Ref.~\cite{Krein:2010vp}. 
When the value of $\Lambda_D$ is smaller, the mass shift becomes positive.
(Note that in Tab.~1 of Ref.~\cite{Krein:2010vp}, since the $J/\Psi$ bare mass 
$m^0_\psi$ was calculated including all the 
$D\overline{D}$-, $D\overline{D^*}$-, $\overline{D}D^*$- and 
$D^*\overline{D^*}$-loops, the dependence of the $J/\Psi$ mass shift on the applied 
$\Lambda_D$ values was slightly different from the present case with the inclusion of only 
the $D\overline{D}$-loop.)

The results shown in Fig.~\ref{fig:DD} reveal a negative mass shift (attractive potential) 
for the $J/\Psi$ in symmetric nuclear matter in all cases.
A negative self-energy for $\Delta m_\psi$ means that the
nuclear mean field provides attraction to $J/\Psi$. Using for 
$m = m^*_{\psi}$ and $R = 5$~fm in Eq.~(\ref{bound}), one sees
that the values of $\Delta m_\psi$ shown in Fig.~\ref{fig:DD}
are much larger than the lower bound minimum $V_0 > 1$~MeV. Of course,
as mentioned previously, the nuclear surface plays an important role 
and a detailed investigation is required.

%%%%%%%%%%%%%%%%%%%%%%%%%%%%%%%%%%%%%%%%%%%%%%%%%%%%%%%%%%%%%%%%%%%%
\subsection{Predictions for $\eta_c$- and $J/\Psi$-nucleus 
binding energies}
\label{jpsiresult}
%%%%%%%%%%%%%%%%%%%%%%%%%%%%%%%%%%%%%%%%%%%%%%%%%%%%%%%%%%%%%%%%%%%

We now present predictions for the binding energies of $\eta_c$ and $J/\Psi$ 
in selected nuclei, ranging from small- to large-sized: 
$^{4}{\rm He}$, $^{12}{\rm C}$, $^{40}{\rm Ca}$, $^{48}{\rm Ca}$, 
$^{90}{\rm Zr}$ and $^{208}{\rm Pb}$. We present first the predictions 
obtained with the quarkonium-nucleon potentials discussed in 
section~\ref{sub:HQvdw} and then those based on the potentials due 
to the $D\overline{D}$-loop for $J/\Psi$ self-energy discussed in 
section~\ref{DDloop}. The nucleon density distributions used for
$^{12}$C, $^{16}$O, $^{40}$Ca, $^{90}$Zr and $^{208}$Pb
are calculated by the QMC model~\cite{Saitofinite}. For $^4$He, 
we use the parametrization for the density distribution obtained in 
Ref.~\cite{Saito:1997ae}. The $J/\Psi$ potentials from the 
$D\overline{D}$-loop are calculated using a local density approximation.
Recall that also the in-medium mass of $D$ meson, which is necessary 
in this case, is consistently calculated in the QMC model without
changing any parameters. We stress that we limit the discussion to the 
situation that the quarkonium is produced nearly at rest in the 
interior of the nucleus (recoilless kinematics in experiments). 

Since in free space the width of $J/\Psi$ meson is $\sim 93$ keV~\cite{PDG},
we can ignore this tiny natural width in the following.
When the $J/\Psi$ meson is produced nearly at rest,
its dissociation process via $J/\Psi + N \to \Lambda_c^+ + \bar{D}$
is forbidden in the nucleus, as the threshold energy
in free space is about $115$ MeV above. This is because the same number of
light quarks three, participates in the initial and final states and hence,
the effects of the partial restoration of chiral symmetry
which reduces mostly the amount of the light quark condensates,
would affect a similar total mass reduction for
the initial and final states~\cite{QMCbc,Tsushimad}. 
(See subsection~\ref{hadronmass}, and Fig.~\ref{spotential}.)
Thus, the relative energy
($\sim$ 115 MeV) to the threshold would not be modified significantly.

We also note that once the $J/\Psi$ meson is bound in the nucleus,
the total energy of the system is below threshold for
nucleon knock-out and the whole system is stable.
The exception to this is the process $J/\Psi + N \to \eta_c + N$,
which is exothermic.
Reference~\cite{Sibirtsev:2000aw} has provided an estimate of this 
cross section from which we deduce a width for the bound $J/\Psi$ of order
$0.8$ MeV (nuclear matter density at $\rho_B = \rho_0/2$).
We therefore expect that the bound $J/\Psi$, while not being
completely stable under the strong interaction, should be
narrow enough to be clearly observed. It would be worthwhile
to investigate this further.
Then, provided that experiments can be performed
to produce the $J/\Psi$ meson in recoilless kinematics,
the $J/\Psi$ meson is expected to be captured by
the nucleus into one of the bound states, which has
no strong-interaction originated width. $\eta_c$ has a total decay 
width which is larger than that of $J/\Psi$, but still relatgively
small, $\sim 31.8$~MeV. Therefore, the situation of heavy quarkonia is 
completely different and advantageous compared to that of lighter
quarkonium $\phi$ and also $\eta$ and $\omega$, to be discussed in 
sections~\ref{phi-meson} and \ref{etaoetapresult}, respectively.
This justifies neglecting their widths in calculating the single-particle 
energy levels in a nucleus.

%%%%%%%%%%%%%%%%%%%%%%%%%%%%%%%%%%%%%%%%%%%%%%
\begin{figure}[htb]
\begin{center}
\includegraphics[height=80mm]{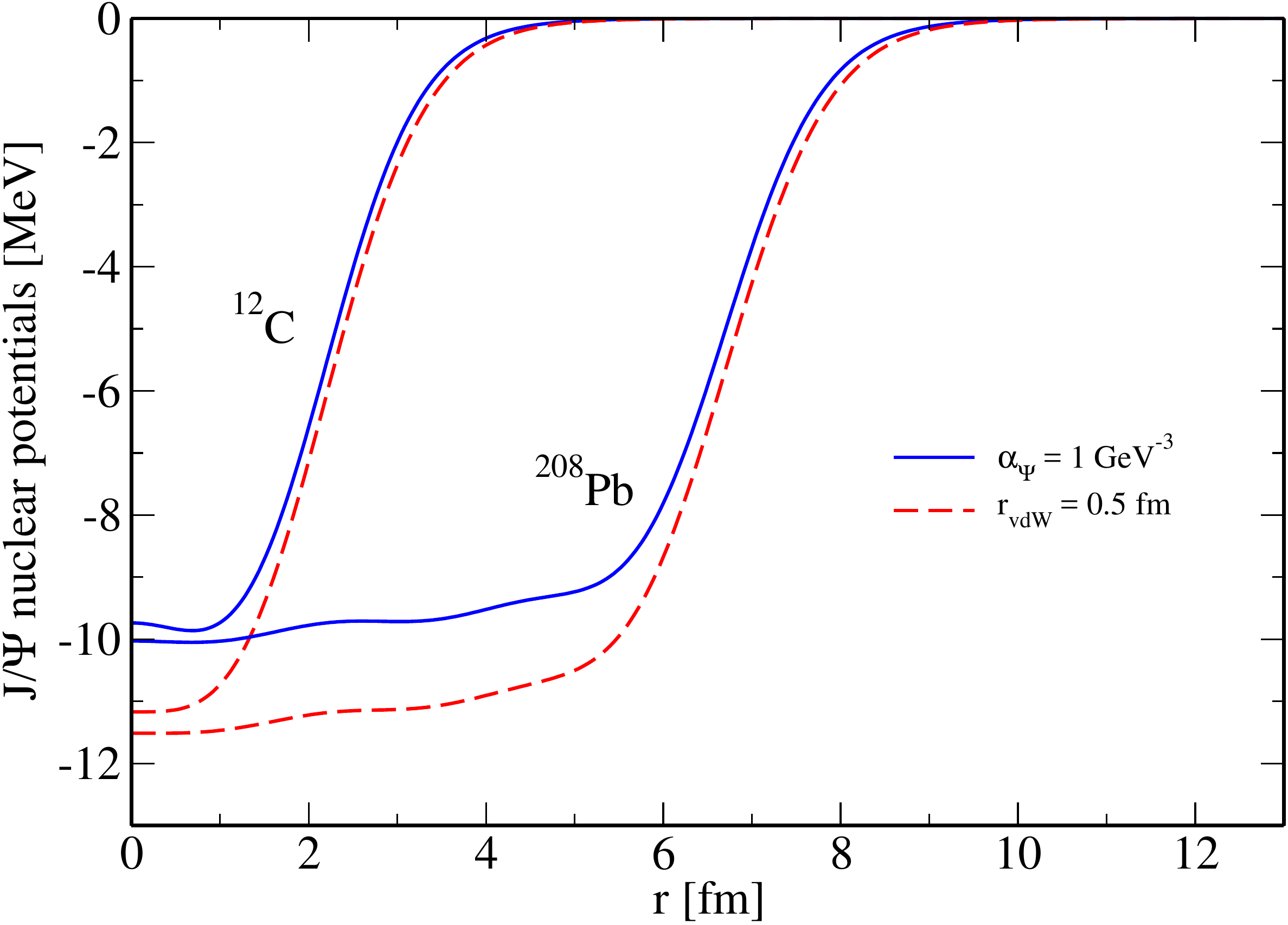}
\caption{$J/\Psi$ nuclear potentials $W^{\rm pol}_{J/\Psi A}(\r)$ (solid line) for 
a polarizability $\alpha_{J/\Psi} = 1~{\rm GeV}^{-3}$ and  
$W^{\rm latt}_{J/\Psi A}(\r)$ (dashed line) from a fit to the lattice data 
with a cutoff $r_{\rm vdW} = 0.5~{\rm fm}$. }
\label{fig:pot-pol-latt}
\end{center}
\end{figure}
%%%%%%%%%%%%%%%%%%%%%%%%%%%%%%%%%%%%%%%%%%%%%

Initially, we compare in Fig.~\ref{fig:pot-pol-latt} the potentials $W^{\rm pol}_{\varphi A}(\r)$ 
and $W^{\rm latt}_{\varphi A}(\r)$, defined respectively by Eqs.~(\ref{phiA-pol}) and 
(\ref{phiA-latt}). We make the comparison for $J/\Psi$ in the $^{12}{\rm C}$ and 
$^{208}{\rm Pb}$. We recall that while $W^{\rm pol}_{\varphi A}(\r)$ is proportional
to the nuclear density $\rho_A(\r)$, $W^{\rm latt}_{\varphi A}(\r)$ is a convolution
of the density with the finite-range potential that gives the lattice 
spin-averaged value for the $J/\Psi N$ interaction in free space. Not surprisingly,
the shapes of both nuclear potentials are similar, essentially following
the shape of the density.

\begin{table}[htb]
\begin{center}
\caption{\label{tab:pol} Predictions for $J/\Psi$ single-particle energies in several nuclei
obtained with the polarization potential $W^{\rm pol}_{J/\Psi A}(\r)$, defined in 
Eq.~(\ref{phiN-contact}). 
  \vspace{1ex}
  }
\begin{tabular}{crrrrrrr}
\hline \\[-0.3true cm]
             & $^{4}_{J/\Psi}{\rm He}$ & $^{12}_{J/\Psi}{\rm C}$ & $^{16}_{J/\Psi}{\rm O}$ &  $^{40}_{J/\Psi}{\rm Ca}$  
             & $^{48}_{J/\Psi}{\rm Ca}$ & $^{90}_{J/\Psi}{\rm Zr}$ & $^{208}_{J/\Psi}{\rm Pb}$ 
             %& 
             \\[0.15true cm]
\hline \\[-0.3true cm]
 & \multicolumn{7}{c}{$\alpha_{J/\Psi} = 1~{\rm GeV}^{-3}$}     \\[0.0true cm]
\cline{4-6} \\[-0.3truecm]
1s           &     n & -3.36 & -4.41 &  -6.77 &  -6.84 & -7.91 & -8.38     \\
1p           &     n &     n & -0.39 &  -3.47 &  -3.95 & -5.71 & -7.05     \\
2s           &     n &     n &     n &  -0.26 &  -0.59 & -2.70 & -5.01     \\
2p           &     n &     n &     n &      n &      n & -0.21 & -2.94    \\%[0.2true cm]
3s           &     n &     n &     n &      n &      n &     n & -0.70    \\[0.2true cm]
%
% & \multicolumn{10}{c}{$r_{\rm vdW} = 0.5~{\rm fm}$}     \\[0.0true cm]
%\cline{5-8} \\[-0.3truecm]
%
 & \multicolumn{7}{c}{$\alpha_{J/\Psi} = 2~{\rm GeV}^{-3}$}     \\[0.0true cm]
\cline{4-6} \\[-0.3truecm]
1s           & -4.49 & -10.76 & -12.62 & -16.41 & -16.16 & -17.70 & -17.27     \\
1p           &     n &  -3.98 &  -6.54 & -11.95 & -12.44 & -14.95 & -16.30     \\
2s           &     n &    n   &  -0.54 &  -6.74 &  -7.50 & -11.07 & -13.95     \\
2p           &     n &    n   &     n  &  -1.62 &  -2.52 &  -7.33 & -11.41    \\%[0.2true cm]
3s           &     n &    n   &     n  &     n  &      n &  -2.71 &  -8.28    \\[0.2true cm]
\hline
\end{tabular}
\end{center}
\end{table}
%%%%%%%%%%%%%%%%%%%%%%%%%%%%%%%%%%%%%%%%%%

Table~\ref{tab:pol} contains the predictions for $J/\Psi$ single-particle energies for 
several nuclei obtained by solving the Schr\"odinger equation, Eq.~(\ref{Schr-phi}), with the 
polarization potential given in Eq.~(\ref{phiA-pol}). The states are denoted by $n l$, 
where $n=1,2,3,\cdots$ is the principal quantum number, and $l = 0 ({\rm s}), 1 ({\rm p}),
2 ({\rm d}), \cdots$.  When the single-particle energy is less than $10^{-2}$ MeV,
we consider there is no bound state and denote this with ``n'' in the tables. 
We present results for two typical
values of the polarizability: $\alpha_{J/\Psi} = 2~{\rm GeV}^{-3}$, which is an estimate
made in Refs.~\cite{Voloshin:2007dx,{Sibirtsev:2005ex}}, and $\alpha_{J/\Psi} = 1~{\rm GeV}^{-3}$,
which is the upper value for $\alpha_{\eta_b}$ obtained in Ref.~\cite{Brambilla:2015rqa}
in the context of pNRQCD; they correspond respectively to scattering lengths 
$a_{J/\Psi N} = - 0.37$~fm and $a_{J/\Psi N} = - 0.19$~fm. The table reveals that while 
relatively deep $J/\Psi$ bound states, with more than $10$~MeV, can happen in all nuclei 
when $\alpha_{J/\Psi} = 2~{\rm GeV}^{-3}$, when $\alpha_{J/\Psi} = 1~{\rm GeV}^{-3}$ 
they can happen only for the largest nucleus, $^{208}{\rm Pb}$. We also mention that with
a scattering length $a_{J/\Psi N} = - 0.05$~fm, the value extracted from $J/\Psi \, p$ 
scattering~\cite{Gryniuk:2016mpk}, the binding of $J/\Psi$ to nuclei is about $1~{\rm MeV}$ 
for the largest nuclei, being close to $2~{\rm MeV}$ in $^{208}{\rm Pb}$.

Predictions for $\eta_c$ and $J/\Psi$ single-particle energies for several nuclei from the 
potential using lattice QCD input, given in Eq.~(\ref{phiA-latt}), are presented
in Tab.~\ref{tab:latt}. We recall that the $\eta_c N$ and $J/\Psi N$ potentials fit the lattice 
scattering lengths and incorporate the Yuakwa tail from the fit from lattice data. Results 
for $^{4}{\rm He}$ are not displayed because there are no bound states
and $^{12}{\rm C}$ are similar to those for $^{12}{\rm C}$ like in the previous case. 
One first conclusion can be drawn from the results displayed is that the nuclear potentials 
are rather weak. The finite range of the quarkonium-nucleon interaction clearly influences
the nuclear potentials; although they give a larger quarkonium-nucleon scattering length that
the contact-interaction with $\alpha_{J/\Psi} = 1~{\rm GeV}^{-3}$ 
(i.e. $a_{J/\Psi N} = - 0.19~{\rm fm}$), the nuclear surface cuts off the tail of the
finite-range potentials.   

\begin{table}[htp]
\begin{center}
\caption{\label{tab:latt} Single-particle energies of $\eta_c $ and $J/\Psi$ in
selected nuclei. The $\eta_c N$ and $J/\Psi N$ potentials fit the  lattice scattering 
lengths and incorporate the Yuakwa tail from the fit from lattice data.  
  \vspace{1ex}
  }
\begin{tabular}{crrrrccrrrr}
\hline \\[-0.3true cm]
             & $^{16}_{\eta_c}{\rm O}$ & $^{40}_{\eta_c}{\rm Ca}$ & $^{90}_{\eta_c}{\rm Zr}$ &  $^{290}_{\eta_c}{\rm Pb}$ 
             & & 
             & $^{16}_{J/\Psi}{\rm O}$ & $^{40}_{J/\Psi}{\rm Ca}$ & $^{90}_{J/\Psi}{\rm Zr}$ &  $^{290}_{J/\Psi}{\rm Pb}$ 
             %& 
             \\[0.15true cm]
\hline \\[-0.3true cm]
 & \multicolumn{10}{c}{$r_{\rm vdW} = 0.3~{\rm fm}$}     \\[0.0true cm]
\cline{5-8} \\[-0.3truecm]
1s           & -2.92 & -5.15 & -6.32 & -6.88 & & & -3.62 & -5.92 & -7.10 & -7.62    \\
1p           &     n & -2.06 & -4.17 & -5.55 & & &     n & -2.74 & -4.93 & -6.29    \\
2s           &     n &     n & -1.40 & -3.53 & & &     n &     n & -2.06 & -4.29    \\
2p           &     n &     n &     n & -1.50 & & &     n &     n &     n & -2.30    \\[0.2true cm]
%3s           &     n &     n &     n &     n & & &     n &     n &     n &     n    \\[0.2true cm]
%
 & \multicolumn{10}{c}{$r_{\rm vdW} = 0.5~{\rm fm}$}     \\[0.0true cm]
\cline{5-8} \\[-0.3truecm]
1s           & -3.62 & -5.99 & -7.23  & -7.79 & & & -5.23 & -7.95 & -9.24 & -9.74    \\
1p           &     n & -2.72 & -4.99 & -6.41  & & & -0.87 & -4.41 & -6.90 & -8.33    \\
2s           &     n &     n & -2.04 & -4.33  & & &     n & -0.82 & -3.71 & -6.20    \\
2p           &     n &     n &     n & -2.28  & & &     n &     n & -0.92 & -4.03    \\
%3s           &     n &     n &     n & -0.27  & & &     n &     n &     n & -1.59    \\
\hline
\end{tabular}
\end{center}
\end{table}
%%%%%%%%%%%%%%%%%%%%%%%%%%%%%%%%%%%%%%%%%%

Next, we present results from calculations based on the in-medium $J/\Psi$ 
self-energy calculated with meson Lagrangians. As mentioned previously, the 
$J/\Psi$-meson-nucleus effective potential 
is calculated in a local density approximation using the $J/\Psi$-meson mass shift 
in nuclear matter, together with the nuclear density profile $\rho_{A}(r)$ of a  
nucleus $A$ calculated in the QMC model; specifically:
\begin{equation}
\label{eqn:Vs}
V_{J/\Psi A}(r)= m^*_{J/\Psi}(\rho_A(r)) - m_{J/\Psi} = \Delta m_{J/\Psi}(\rho_{A}(r)),
\label{VJPsi-QMC}
\end{equation}
where $r$ is the distance from the center of the nucleus $A$. We solve the Schr\"{o}dinger 
equation, Eq.~(\ref{Schr-phi}), with the nuclear potential given by Eq.~(\ref{VJPsi-QMC}). 
Also, we use the reduced mass of the system instead of $m_{J/\Psi}$, which is important 
for $^4{\rm He}$.

In the situation of recoilless kinematics, it should be a 
very good approximation 
to neglect the possible energy difference 
between the longitudinal and transverse components~\cite{saitomega} of the $J/\Psi$ 
wave function ($\phi^\mu_{\Psi}$) as was assumed for the $\omega$ meson 
case~\cite{saitomega}. After imposing the Lorentz condition, 
$\partial_\mu \phi^\mu_{\Psi} = 0$, to solve the Proca equation, aside from 
a possible width, becomes equivalent to solving the Klein-Gordon equation.
 We have solved the Schr\"{o}dinger equation, and compared the 
results obtained with those obtained in solving the Klein-Gordon equation for 
some cases. We have confirmed that the two methods (without an imaginary part 
in the potential) yield nearly identical results within the accuracy of a few 
percent as we have tested also the lighter $\phi$-meson case.

%%%%%%%%%%%%%%%%%%%%%%%%%%%%%%%%%%%%%%%%%%%%%%%%%%%%%%%%%
\begin{figure}[htb]
\begin{center}
\includegraphics[height=60mm]{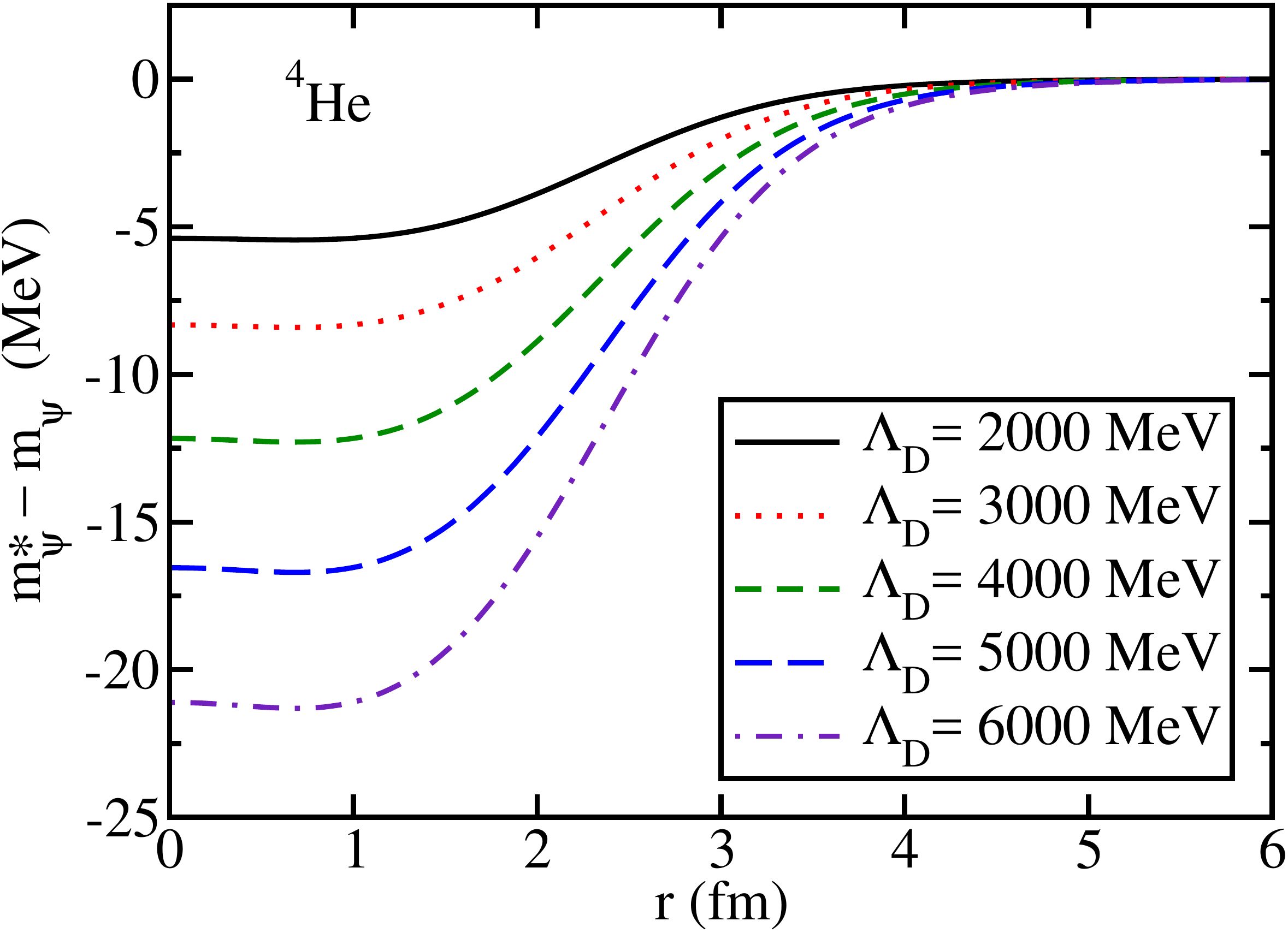}
\includegraphics[height=60mm]{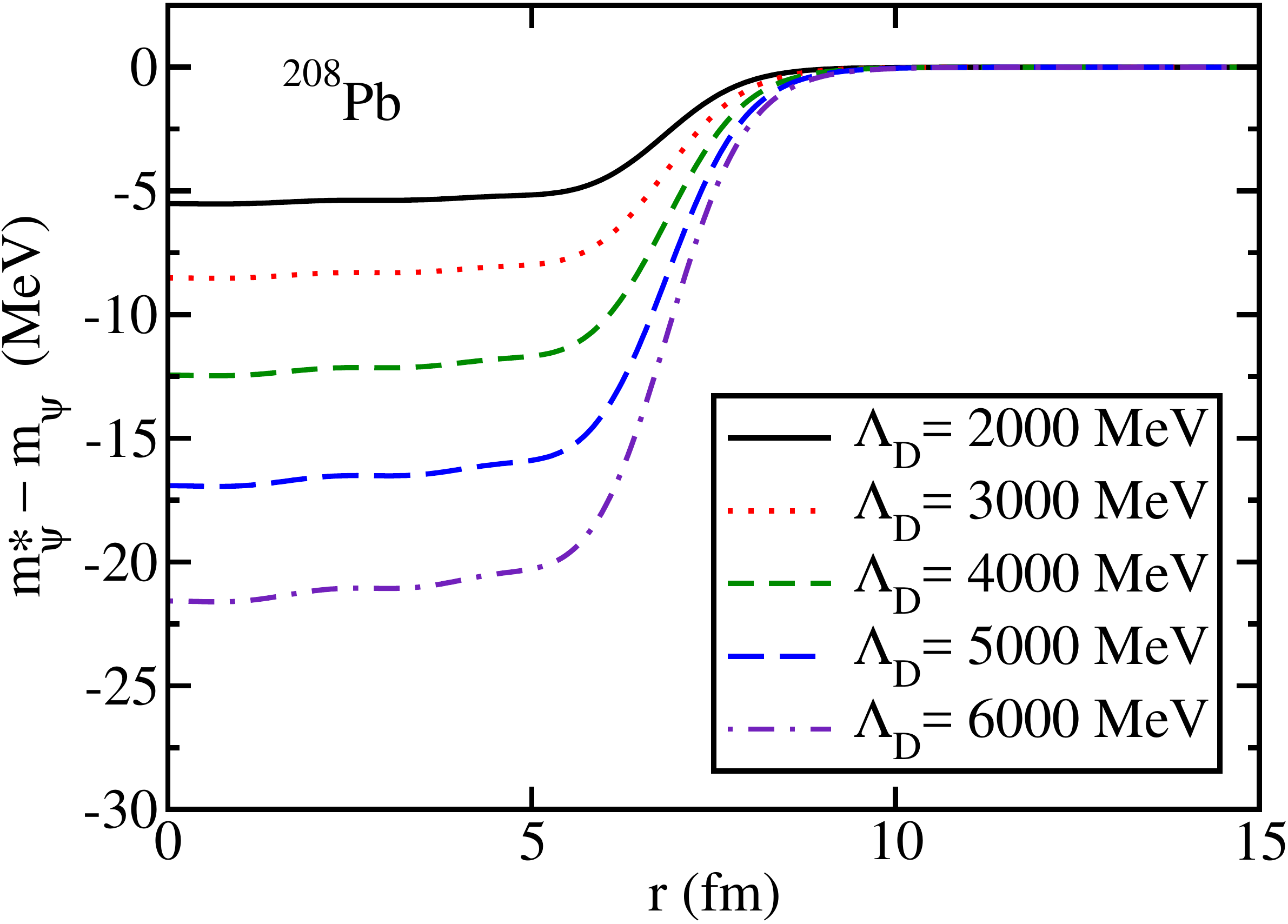}
\caption{
Potentials felt by $J/\Psi$ in a $^4$He nucleus (left panel)
and in a $^{208}$Pb nucleus (right panel) for five
values of the cutoff mass in the form factors,
$\Lambda_D = 2000, 3000, 4000, 5000$ and $6000$ MeV.
}
\label{fig:poto-nuclei}
\end{center}
\end{figure}
%%%%%%%%%%%%%%%%%%%%%%%%%%%%%%%%%%%%%%%%%%%%%%%%%%%%%%%%%%

%%%%%%%%%%%%%%%%%%%%%%%%%%%%%%%%%%%%%%%%%%%%%%%%%%%%%%%%%%%%%%%%%%%%%%%%%%%%%%%%%
%Bound state energies from the Schroedinger equation--------------------------
\begin{table}[htb]
\begin{center}
\caption{\label{tab:jpsi-nucleus-be} $J/\Psi-$nucleus bound state
  energies taking into account the change in the self-energy in medium,  calculated 
  with the Schr\"{o}dinger equation. All dimensioned quantities are given in MeV.
  \vspace{1ex}
  }
\begin{tabular}{ll|r|r|r|r|r}
  \hline 
  %\hline
  & & \multicolumn{5}{c}{Bound state energies} \\
  \hline
  & & $\Lambda_{D}=2000$ & $\Lambda_{D}=3000$ & $\Lambda_{D}= 4000$ &
$\Lambda_{D}= 5000$ & $\Lambda_{D}= 6000$ \\
\hline
$^{4}_{J/\Psi}{\rm He}$ & 1s & n & n & -0.70 & -2.70 & -5.51 \\
\hline
$^{12}_{J/\Psi}{\rm C}$ & 1s & -0.52 & -1.98 & -4.47 & -7.67 & -11.26 \\
                      & 1p & n & n & n & -1.38 & -3.84 \\
\hline
$^{16}_{J/\Psi}{\rm O}$ & 1s & -1.03 & -2.87 & -5.72 & -9.24 & -13.09 \\
                      & 1p & n & n & -0.94 & -3.48 & -6.60 \\
\hline
$^{40}_{J/\Psi}{\rm Ca}$ & 1s & -2.78 & -5.44 & -9.14 & -13.50 & -18.12 \\
                       & 1p & -0.38 & -2.32 & -5.43 & -9.32 & -13.56 \\
                       & 1d & n & n & -1.52 & -4.74 & -8.49 \\
                       & 2s & n & n & -1.27 & -4.09 & -7.60 \\
\hline
$^{48}_{J/\Psi}{\rm Ca}$ & 1s & -2.96 & -5.62 & -9.28 & -13.55 & -18.08 \\
                       & 1p & -0.73 & -2.83 & -6.03 & -9.95 & -14.18 \\
                       & 1d & n & n & -2.46 & -5.87 & -9.73 \\
                       & 2s & n & -0.07 & -1.90 & -5.00 & -8.65 \\
\hline
$^{90}_{J/\Psi}{\rm Zr}$ & 1s & -3.64 & -6.40 & -10.12 & -14.41 & -18.92 \\
                       & 1p & -1.93 & -4.42 & -7.92 & -12.03 & -16.40 \\
                       & 1d & -0.03 & -2.13 & -5.31 & -9.18 & -13.37 \\
                       & 2s & -0.02 & -1.56 & -4.51 & -8.26 & -12.37 \\
                       & 2p & n & n & -1.52 & -4.71 & -8.45 \\
\hline
$^{208}_{J/\Psi}{\rm Pb}$ & 1s & -4.25 & -7.08 & -10.82 & -15.11 & -19.60 \\
                      & 1p & -3.16 & -5.86 & -9.52 & -13.74 & -18.18 \\
                      & 1d & -1.84 & -4.38 & -7.90 & -12.01 & -16.37 \\
                      & 2s & -1.41 & -3.81 & -7.25 & -11.30 & -15.61 \\
                      & 2p & -0.07 & -1.95 & -5.10 & -8.97 & -13.14 \\
\hline 
%\hline
\end{tabular}
\end{center}
\end{table}
%---------------------------------------------------------------------------
%%%%%%%%%%%%%%%%%%%%%%%%%%%%%%%%%%%%%%%%%%%%%%%%%%%%%%%%%%%%%%%%%%%%%%%%%%%%%%%%%%%%

The results in Tab.~\ref{tab:jpsi-nucleus-be} show that
the $J/\Psi$ are expected to form $J/\Psi$-nuclear bound
states for nearly all the nuclei considered, but some cases for $^4$He.
This is insensitive to the values of cutoff mass values used in the
form factor. It will be possible to search for the bound states  
in a $^{208}$Pb nucleus at JLab after the 12 GeV upgrade.
In addition, one can expect quite rich spectra
for medium and heavy mass nuclei.
Of course, the main issue is to produce the
$J/\Psi$ meson with nearly stopped kinematics, or
nearly zero momentum relative to the nucleus.
Since the present results imply that many nuclei should
form $J/\Psi$-nuclear bound states,
it may be possible to find such kinematics by careful selection
of the beam and target nuclei.

We have been able to set the strong interaction width of the
$J/\Psi$ to be zero (and neglect its tiny natural width of
$\sim 93$ keV in free space).
Combined with the generally advocated color-octet gluon-based attraction,
or QCD color van der Waals forces, one can expect that the $J/\Psi$ meson will
form nuclear bound states, and that the signal for the formation should be
experimentally very clear, provided that the $J/\Psi$ meson is produced in recoilless
kinematics.

The $J/\Psi$-$D$ coupling constants are taken
as determined from vector meson dominance, and the cutoff masses are varied
over a large range of values. 
The QMC model predicts a $62$~MeV downward mass shift for the
$D$-meson at normal nuclear matter density. 
The $D$-meson mass shift leads to a 
corresponding in-medium $J/\Psi$ downward mass shift varying between 
$-5$~MeV and $-22$~MeV in $^4$He and $^{208}$Pb nuclei, 
for cutoff mass values in the range of $2000$~MeV and $6000$~MeV.
Such a mass shift is large enough to bind a $J/\Psi$ to a nucleus for a
$J/\Psi$ produced at low momentum in the rest frame of the nucleus.

We note that it is unclear whether the two mechanisms underlying both approaches 
we have discussed for heavy quarkonium binding are independent from each other
and their predictions should be combined. This is because a quark-gluon-based 
interaction of a quarkonium with light hadrons can, in principle, be matched 
to a hadron-based interaction. This is nicely illustrated by the derivation of 
the van der Waals force in the quarkonium-quarkonium interaction with
a chiral EFT derived from gWEFT that, in turn, is derived from pNRQCD, the latter being a 
quark-gluon based theory. In the same way, the matching of gWEFT to an EFT involving 
couplings of quarkonia to $D$ mesons would lead to an approach in which the quarkonium 
interacts with the medium via virtual $D\overline{D}$ loops. On the other hand, the 
interaction generated by the $D\overline{D}$ loop is of shorter range (of the order
of $1/2M_D$) as compared with the range of a van der Waals force and, tehrefore,
both mechanisms are expected to contribute. Such an EFT involving quarkonium and $D$ 
mesons has not been derived from pNRQCD so far and, therefore, for now one has to 
resort to a phenomenological Lagrangian approach. 

Although the present results point the the possible existence of $\eta_c$ 
and $J/\Psi$ nuclear bound states, some issues clearly require further 
investigation. 
Amongst the most important ones are the calculation of effective 
$\eta_c$ and $J/\Psi$ potentials with taking into account momentum 
dependence. In the context of the $D\overline{D}$-loop calculation, the 
width of the $D$-meson needs to be taken into account. Recent calculations~\cite{tolos} 
of in-medium $D$ and $D^*$ widths based on meson-exchange models have 
obtained somewhat contradictory results and further study is required. 
Still in this context, as emphasized in Refs.~\cite{Haidenbauer:2007jq,
Haidenbauer:2008ff,{Haidenbauer:2010ch}}, the lack of experimental information 
on the free-space interaction of $D$ mesons with nucleons is a major impediment for
constraining models and the use of symmetry principles and exploration
of the interplay between quark-gluon and baryon-meson degrees of freedom
is essential in this respect. Another issue is the dissociation of $\eta_c$ 
and $J/\Psi$ in matter by collisions with nucleons and light mesons. This 
subject has been studied vigorously in the last years using different approaches, 
like meson exchange~\cite{diss-mex} and quark models~\cite{diss-qm}, QCD sum 
rules~\cite{Navarra:2001jy}, and the NJL model~\cite{Bourque:2008es}.
Finally, as already discussed in detail in subsection~\ref{sub:HQvdw},
we stress the need for a deeper understanding of the $J/\Psi N$
interaction, including its long-range van der Waals behavior and its
relation to the $D\overline{D}$-loop mechanism within the context of the
strongly interacting many-nucleon system. The construction of an EFT 
for the quarkonium-nucleon system wherein light quark degrees of freedom 
are incorporated via chiral EFTs, possibly trailing a path similar to that in 
the derivation of a color van der Waals force with pNRQCD and gWEFT, is 
a challenge that seems worth facing in the coming years.  
%
%
%
%\input{sec_phi.tex}
%
%
%

%%%%%%%%%%%%%%%%%%%%%%%%%%%%%%%%%%%%%%%%%%%%%%%%%%%%
\section{Nuclear-bound $\phi$}
\label{phi-meson}
%%%%%%%%%%%%%%%%%%%%%%%%%%%%%%%%%%%%%%%%%%%%%%%%%%%%

Several experiments have focused 
on the light vector mesons $\rho$, $\omega$, 
and $\phi$, since their mean-free paths can be comparable with the size of 
a nucleus after being produced inside the nucleus. 
However, a unified consensus has not yet been reached among the different 
experiments{\textemdash}see Refs.~\cite{Leupold:2009kz,Hayano:2008vn,Krein:2016fqh} for 
comprehensive reviews of the current status. 

For the $\phi$-meson, although the precise values are different, 
a large in-medium broadening of the width has been reported 
by most of the experiments 
performed, while only a few of them find evidence 
for a substantial mass shift.
For example, the KEK-E325 collaboration~\cite{Muto:2005za} reported a mass 
reduction of $3.4\%$ and an in-medium decay width of $\approx 14.5$ MeV at 
normal nuclear matter density.
The latter disagrees with the SPring8~\cite{Ishikawa:2004id} result, which 
reported a large in-medium $\phi N$ cross section leading to a decay
width of 35 MeV. But this 35 MeV is in close agreement with the two JLab CLAS  
collaboration measurements reported 
in Refs.~\cite{Mibe:2007aa} and~\cite{Qian:2009ab}. 

In an attempt to clarify the situation, 
the CLAS collaboration at JLab~\cite{Wood:2010ei} 
performed new measurements of nuclear transparency ratios, 
and estimated in-medium widths in the range of 23-100 MeV. 
These values overlap with that of the SPring8 measurement~\cite{Ishikawa:2004id}.
More recently, the ANKE-COSY collaboration~\cite{Polyanskiy:2010tj} has 
measured the $\phi$-meson production from proton-induced reactions on various 
nuclear targets. 
The comparison of data with model calculations suggests an in-medium 
$\phi$ width of $\approx 50$ MeV. 
This result is consistent with that of SPring8~\cite{Ishikawa:2004id},  
as well as the one deduced from CLAS at JLab~\cite{Wood:2010ei}. 
However, the value is clearly larger than that of the KEK-E325 
collaboration~\cite{Muto:2005za}.

Thus, it is obvious that the search for evidence of 
a light vector meson mass shift is indeed complicated. 
It certainly requires further experimental 
efforts to understand better the changes of $\phi$-meson properties  
in a nuclear medium. 
For example, the J-PARC E16 collaboration~\cite{JPARCE16Proposal} intends to 
perform a more systematic study for the mass shift of vector mesons 
with higher statistics.
Furthermore, the E29 collaboration at J-PARC has recently put forward a 
proposal~\cite{JPARCE29Proposal,JPARCE29ProposalAdd} to study the 
in-medium mass modification of the $\phi$-meson via the possible formation of 
$\phi$-nucleus bound states~\cite{Aoki:2015qla}, 
using the primary reaction $\overline{p}p\rightarrow \phi\phi$. 
Finally, there is a proposal at JLab, following the 12 GeV upgrade, to study 
the binding of $\phi$ (and $\eta$) to $^4$He~\cite{JLabphi}.

On the theoretical side, various authors predict a downward shift of the 
in-medium $\phi$-meson mass and a broadening of the decay width.
The possible decrease of the light vector meson masses in a nuclear medium 
was first predicted by Brown and Rho~\cite{Brown:1991kk}.
Thereafter, many theoretical investigations have been conducted, some of 
them focused on the self-energies of the $\phi$-meson due to the kaon-antikaon loop.
Ko et al.~\cite{Ko:1992tp} used a density-dependent kaon mass determined from
chiral perturbation theory and found that at normal nuclear 
matter density, $\rho_0$,
the $\phi$-meson mass decreases very little, by at most $2\%$, and the width 
$\Gamma_\phi \approx 25$~MeV and broadens drastically for large densities.
However, their estimate has an uncertainty of factor of two for the in-medium 
kaon (and antikaon) mass used.
Hatsuda and Lee calculated the in-medium 
$\phi$-meson mass based on the QCD sum rule approach~\cite{Hatsuda:1991ez,Hatsuda:1996xt}, 
and predicted a decrease of 1.5\%-3\% at normal nuclear matter density.
Other investigations also predict a large broadening 
of the $\phi$-meson width: Ref.~\cite{Klingl:1997tm} reports a 
negative mass shift of $ < 1\%$ and a decay width of 45 MeV at $\rho_0$;   
Ref.~\cite{Oset:2000eg} predicts a decay width of 22 MeV but does not report 
a result on the mass shift; and Ref.~\cite{Cabrera:2002hc} gives a rather 
small negative mass shift of $\approx 0.81\%$ and a decay width of 30 MeV.
More recently, Ref.~\cite{Gubler:2015yna} reported a downward mass shift of
$< 2\%$ and a large broadening width of 45 MeV; 
and finally, in Ref.~\cite{Cabrera:2016rnc}, 
extending the work of Refs.~\cite{Oset:2000eg,Cabrera:2002hc}, the authors reported 
a negative mass shift of $3.4\%$ and a large decay width of 70 MeV at $\rho_0$.
The reason for these differences may lie in the different approaches 
used to estimate the kaon-antikaon loop contributions for the $\phi$-meson self-energy.

In this section we discuss possible formation of $\phi$-nucleus bound states, 
the strange quarkonium-nuclear bound states of 
$s\overline{s}$-nucleus. The results presented here are from 
Refs.~\cite{Cobos-Martinez:2017vtr,Cobos-Martinez:2017woo}.   
The method to estimate the $\phi$-meson--nuclear potential is 
similar to that for $J/\Psi$-meson, except that the $\phi$-meson  
has finite widths both in vacuum and in medium, 
due to the large decay branches of more than 80 \% 
to kaon-antikaon pairs.

%%%%%%%%%%%%%%%%%%%%%%%%%%%%%%%%%%%%%%%%%%%%%%%
\subsection{$\phi$ self-energy}
\label{phi_vacuum}
%%%%%%%%%%%%%%%%%%%%%%%%%%%%%%%%%%%%%%%%%%%%%%%
%

We use the effective Lagrangian of Refs.~\cite{{Ko:1992tp},{Klingl:1996by}}
to compute the $\phi$ self-energy; 
the interaction Lagrangian $\mathcal{L}_{int}$ involves 
$\phi K\Kbar$ and $\phi\phi K\Kbar$ couplings dictated by a 
local gauge symmetry principle:
\begin{equation}
\mathcal{L}_{int} = \mathcal{L}_{\phi K\Kbar} + \mathcal{L}_{\phi\phi K\Kbar},
\label{eqn:Lint}
\end{equation}
where
\begin{equation}
\label{eqn:phikk}
\mathcal{L}_{\phi K\Kbar} = \mi g_{\phi}\phi^{\mu}
\left[\Kbar(\partial_{\mu}K)-(\partial_{\mu}\Kbar)K\right],
\end{equation}
and
\begin{equation}
\label{eqn:phi2kk}
\mathcal{L}_{\phi\phi K\Kbar} = g^2_{\phi} \phi^\mu\phi_\mu \Kbar K.
\end{equation}
We use the convention:
\begin{equation}
\label{eqn:isospin}
K=\left(\begin{array}{c} K^{+} \\ K^{0} \end{array} \right),
\hspace{1.0cm}
\overline{K}=\left(K^{-}\;\overline{K}^{0}\;\right).
\label{KKbar}
\end{equation}

We note that the use of the effective interaction Lagrangian of Eq.~(\ref{eqn:Lint}) without the term
given in Eq.~(\ref{eqn:phi2kk}) may be considered as being motivated by the hidden gauge approach
in which there are no four-point vertices, such as (\ref{eqn:phi2kk}), that involves two
pseudoscalar mesons and two vector mesons~\cite{Lin:1999ve,Lee:1994wx}.
This is in contrast to the approach of using the minimal substitution to
introduce vector mesons as gauge particles where such four-point vertices do appear.
However, these two methods have been shown to be consistent if both the vector and
axial vector mesons are included~\cite{Yamawaki:1986zz,Meissner:1986tc,Meissner:1987ge,
Saito:1987ba}.
Therefore, we present results with and without such an interaction.

%\phi self energy loop diagram------------------------------------------------
\begin{figure}[t]
\centering
\includegraphics[scale=0.8]{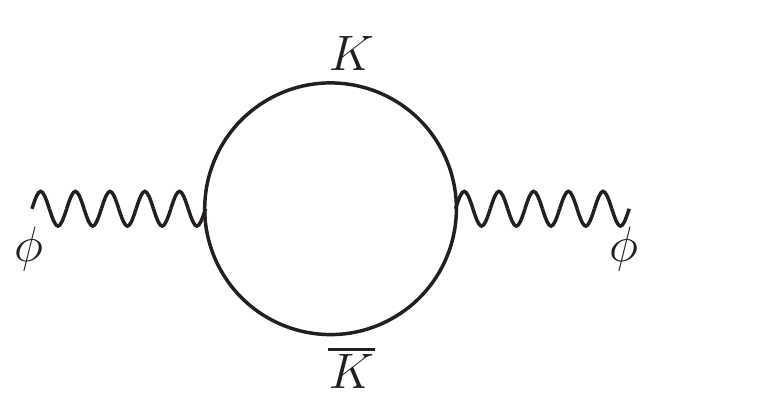}
\caption{\label{fig:phise} $K\Kbar$-loop contribution to the $\phi$-meson 
self-energy.}
\end{figure}
%-----------------------------------------------------------------------------

We first consider the contribution from the 
$\phi K\Kbar$ coupling given by Eq.~(\ref{eqn:phikk}) to 
the scalar part of the $\phi$ self-energy, 
$\Pi_{\phi}(p)$; Fig.~\ref{fig:phise} depicts this contribution. 
For a $\phi$-meson at rest, it is given
by
\begin{equation}
\label{eqn:phise}
\mi\Pi_{\phi}(p)=-\dfrac{8}{3}g_{\phi}^{2}\int\dfd{4}{4}{q}\vec{q}^{\,2}
D_{K}(q)D_{K}(q-p) \, ,
\end{equation}
where $D_{K}(q)=\left(q^{2}-m_{K}^{2}+\mi\epsilon\right)^{-1}$ is the
kaon propagator;  $p=(p^{0}=m_{\phi},\vec{0})$ is 
the $\phi$-meson four-momentum vector, 
with $m_{\phi}$ the $\phi$-meson mass; $m_{K} (=m_{\Kbar})$ is the kaon mass. 
When $m_{\phi}<2m_{K}$ the self-energy $\Pi_{\phi}(p)$ is real. However, when 
$m_{\phi}>2m_{K}$, which is the case here, $\Pi_{\phi}(p)$ acquires an imaginary
part. The mass of the $\phi$ is determined from the real part of $\Pi_{\phi}(p)$,
\begin{equation}
\label{eqn:phimassvacuum}
m_{\phi}^{2}=\left(m_{\phi}^{0}\right)^{2}+\Re\Pi_{\phi}(m_{\phi}^{2}),
\end{equation}
with $m_{\phi}^{0}$ being the bare mass of the $\phi$ and
\begin{equation}
\label{eqn:repiphi}
\Re\Pi_{\phi}=-\dfrac{2}{3}g_{\phi}^{2} \, \mathcal{P}\!\!
\int\dfd{3}{3}{q}\vec{q}^{\,2}\dfrac{1}{E_{K}(E_{K}^{2}-m_{\phi}^{2}/4)} \, .
\end{equation}
Here $\mathcal{P}$ denotes the Principal Value part of the 
integral Eq.~(\ref{eqn:phise}) 
and $E_{K}=(\vec{q}^{\,2}+m_{K}^{2})^{1/2}$.  
The decay width of $\phi$ to a $K\Kbar$ pair 
is given in terms of the imaginary part of $\Pi_{\phi}(p)$, 
\begin{equation}
\Im\Pi_{\phi} = - \dfrac{g_{\phi}^{2}}{24\pi}
m^2_\phi \left(1-\dfrac{4m_{K}^{2}}{m_{\phi}^{2}}\right)^{3/2},
\end{equation}
as
\begin{equation}
\label{eqn:phidecaywidth}
\Gamma_{\phi} = -\dfrac{1}{m_{\phi}}\Im\Pi_{\phi} = \dfrac{g_{\phi}^{2}}{24\pi} m_\phi
\left(1-\dfrac{4m_{K}^{2}}{m_{\phi}^{2}}\right)^{3/2} \, .
\end{equation}

The integral in \eqn{eqn:repiphi} is divergent and needs 
regularization; we use a phenomenological 
form factor, with a cutoff parameter $\Lambda_{K}$, 
as in Ref.~\cite{Krein:2010vp}. The coupling 
constant $g_{\phi}$ is determined by the experimental 
width of the $\phi$ in vacuum~\cite{PDG}.
{}For the $\phi$ mass, $m_{\phi}$, we use its experimental value: 
$m_{\phi}^{{\rm expt}}=1019.461$ MeV~\cite{PDG}. For 
the kaon mass $m_{K}$, there is a small 
ambiguity since $m_{K^+}\ne m_{K^0}$, as a result of charge symmetry breaking and 
electromagnetic interactions. The experimental values for 
the $K^{+}$ and $K^{0}$ meson masses in vacuum are 
$m_{K^{+}}^{{\rm expt}}=493.677$ MeV and $m_{K^{0}}^{{\rm expt}}=497.611$ MeV, 
respectively~\cite{PDG}. 
For definiteness we use the average of $m_{K^{+}}^{{\rm expt}}$ and 
$m_{K^{0}}^{{\rm expt}}$ as the value of $m_{K}$ in vacuum.
The effect of this tiny mass ambiguity on the in-medium  
kaon (antikaon) properties is negligible.
Then, we get the coupling $g_{\phi}=4.539$, and can fix  
the bare mass~$m_{\phi}^{0}$.

%%%%%%%%%%%%%%%%%%%%%%%%%%%%%%%%%%%%%%%%%%%%%%%%%%%%%%%%%%%%%%%%%
\subsection{$\phi$ nuclear bound states}
\label{phi-nuclear}
%%%%%%%%%%%%%%%%%%%%%%%%%%%%%%%%%%%%%%%%%%%%%%%%%%%%%%%%%%%%%%%%%

The in-medium $\phi$ mass is calculated by solving Eq.~(\ref{eqn:phimassvacuum}) 
by replacing $m_{K}$ by $m_{K}^{*}$ and $m_{\phi}$ by $m_{\phi}^{*}$, and
the width is obtained by using the solutions in Eq.~(\ref{eqn:phidecaywidth}).
In-medium kaon mass $m_K^*$ is calculated in the QMC model~\cite{Tsushimak} 
explained in Sec.~\ref{qmc}.
We regularize the associated loop integral with a dipole form factor using a cutoff 
mass parameter $\Lambda_{K}$. In principle, this parameter 
may be determined phenomenologically 
using, for example, a quark model---see Ref.~\cite{Krein:2010vp} for 
more details. 
However, for simplicity we keep it free and vary its value 
over a wide interval, namely
1000-4000~MeV. 

%%%%%%%%%%%%%%%%%%
\begin{table}[htb]
\begin{center}
\caption{$\phi$ mass and width at 
normal nuclear matter density, $\rho_{0}$. All quantities are given in MeV.
\vspace{2ex}
}
\label{tab:phippties} 
\begin{tabular}{l|rrrr}
\hline 
%\hline
& $\Lambda_{K}= 1000$ & $\Lambda_{K}= 2000$  & $\Lambda_{K}= 3000$ & $\Lambda_{K}= 4000$\\
\hline
$m_{\phi}^{*}$ & 1009.3 & 1000.9 & 994.9 & 990.5 \\
$\Gamma_{\phi}^{*}$ & 37.7 & 34.8 & 32.8 & 31.3  \\
\hline 
%\hline
\end{tabular}
\end{center}
\end{table}
%%%%%%%%%%%%%%%%%%%%%

In~\tab{tab:phippties}, we present the values for $m^*_{\phi}$ and 
$\Gamma_{\phi}^{*}$ at normal nuclear matter density $\rho_{0}$. A negative kaon mass 
shift of 13\% induces only maximum of $\approx$ 3\% downward mass shift of the $\phi$. 
On the other hand, $\Gamma_{\phi}^{*}$ is very sensitive 
to the change in the kaon mass; 
at $\rho_B=\rho_0$, the broadening of the $\phi$ becomes 
an order of magnitude larger than 
its vacuum value and it increases rapidly with increasing 
nuclear density, up to a factor of 
$\sim 20$ enhancement for the largest nuclear matter 
density treated, $\rho_{B}=3\rho_{0}$. 

%%%%%%%%%%%%%%%%%%%%%%%%
\begin{figure}[tb]
\centering
\includegraphics[scale=0.28]{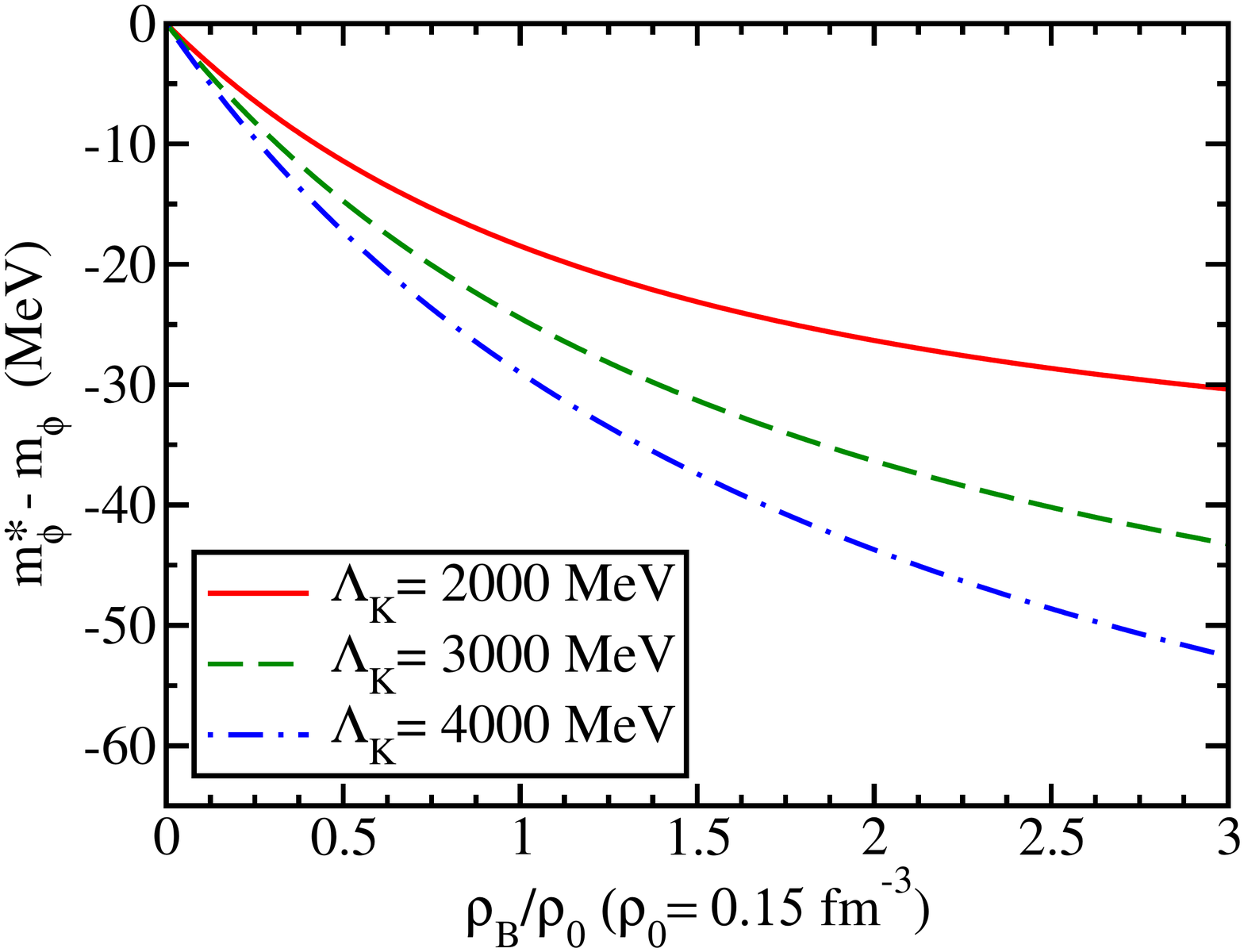}
\includegraphics[scale=0.28]{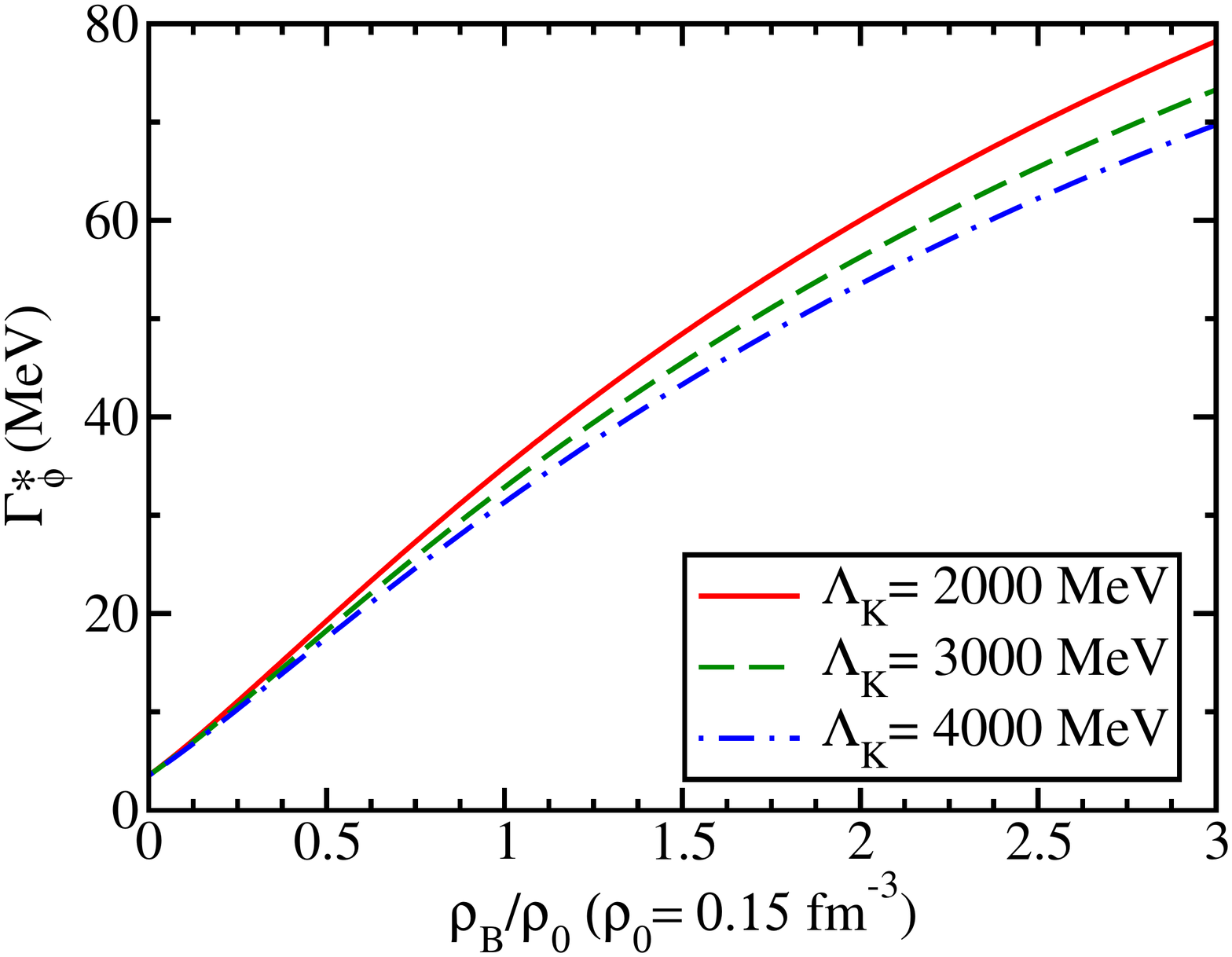}\\
\includegraphics[scale=0.28]{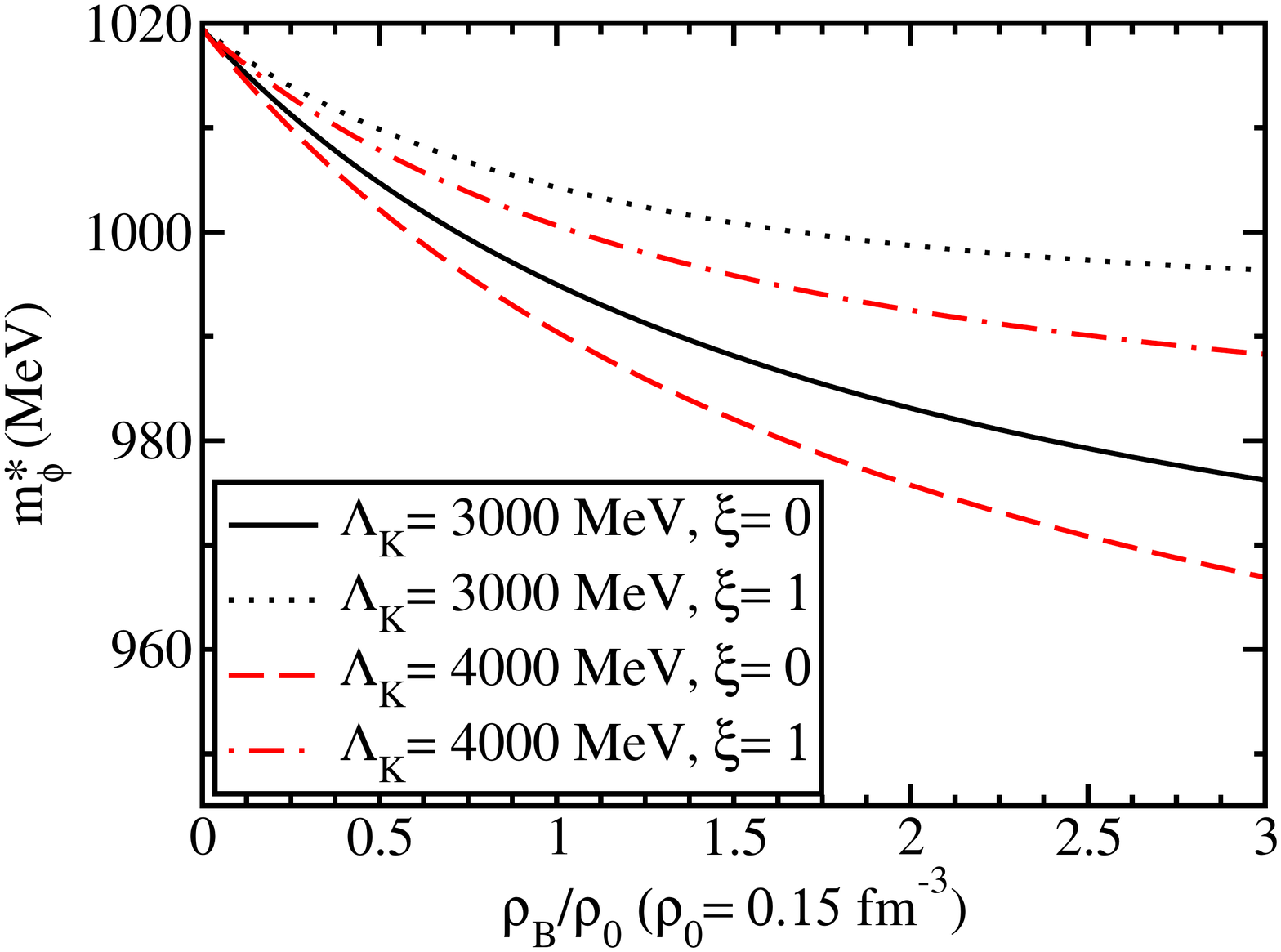}
\includegraphics[scale=0.28]{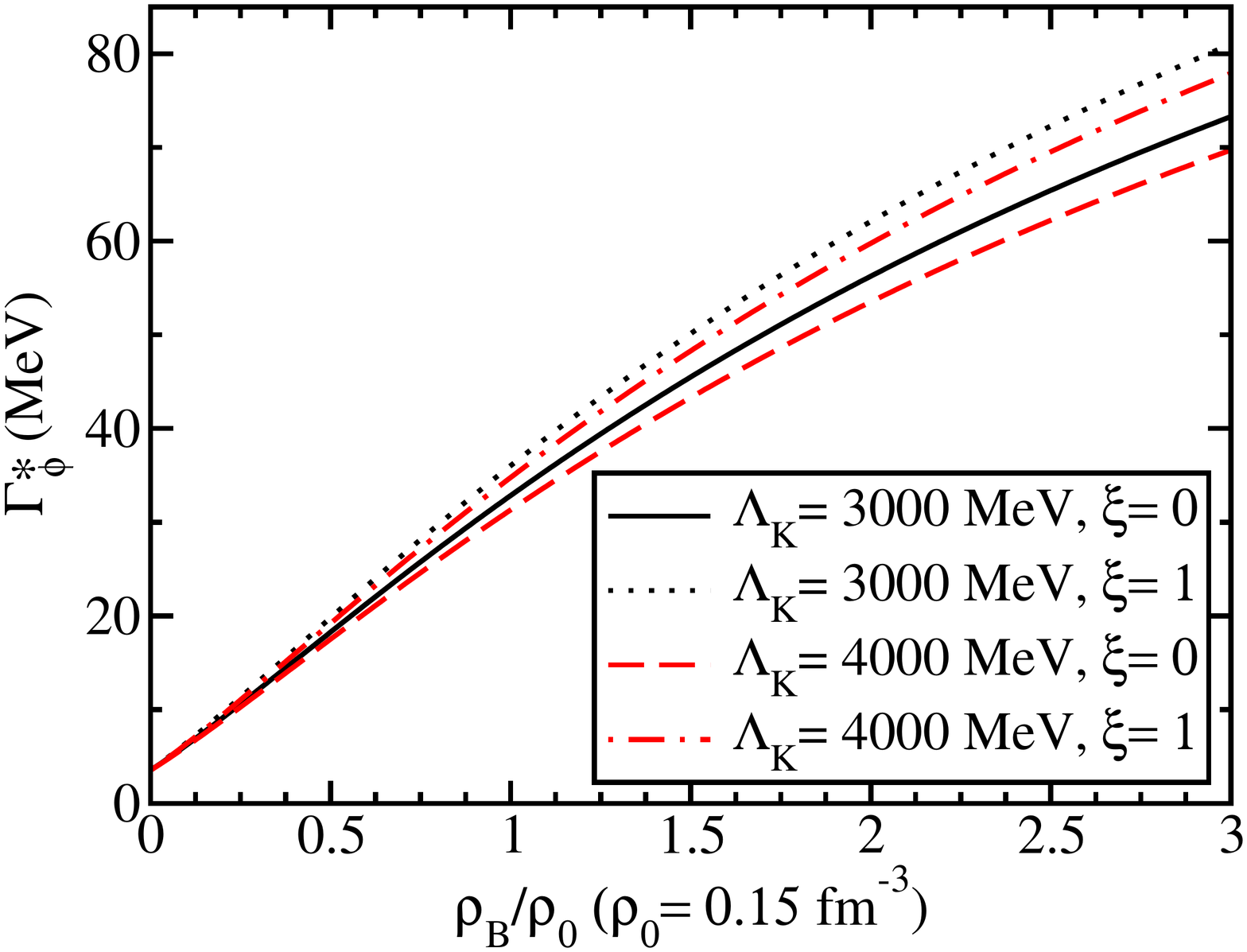}
\caption{\label{fig:mphi} In-medium $\phi$ mass shift (upper left panel) 
and width (upper right panel) without the gauge term, $\xi=0$, 
and the comparison of these with the gauge term 
($\xi=1$) (lower panels) for different values of the cutoff parameter $\Lambda_{K}$. 
Figures are taken from Refs.~\cite{Cobos-Martinez:2017vtr,Cobos-Martinez:2017woo}. }
\end{figure}
%%%%%%%%%%%%%%%%%%%%%

We present the in medium mass shift of the $\phi$ and its width 
without the gauge term, ($\xi = 0$) in the upper panel 
in Fig.~\ref{fig:mphi}. 
The effect of the in-medium kaon mass change gives a negative 
shift of the $\phi$-meson mass. However, even for 
the largest value of density treated, namely  $3 \rho_0$, 
the downward mass shift is only a few percent for 
all values of the cutoff parameter 
$\Lambda_{K}$.

To conclude and for completeness, we show the impact the $\phi\phi K \Kbar$ 
interaction of Eq.~(\ref{eqn:phi2kk}) on the 
in-medium $\phi$ mass and width. The lower panels in Fig.~\ref{fig:mphi} 
present the results. We have used the notation that $\xi = 1 (0)$ means 
that this interaction 
is (not) included in the calculation of the  $\phi$ self-energy. 
One still gets a downward shift 
of the in-medium $\phi$ mass when $\xi = 1$, although the absolute value 
is slightly different from $\xi = 0$. The in-medium width is not
very sensitive to this interaction. 

\begin{figure}[tb]
\centering
\includegraphics[scale=0.28]{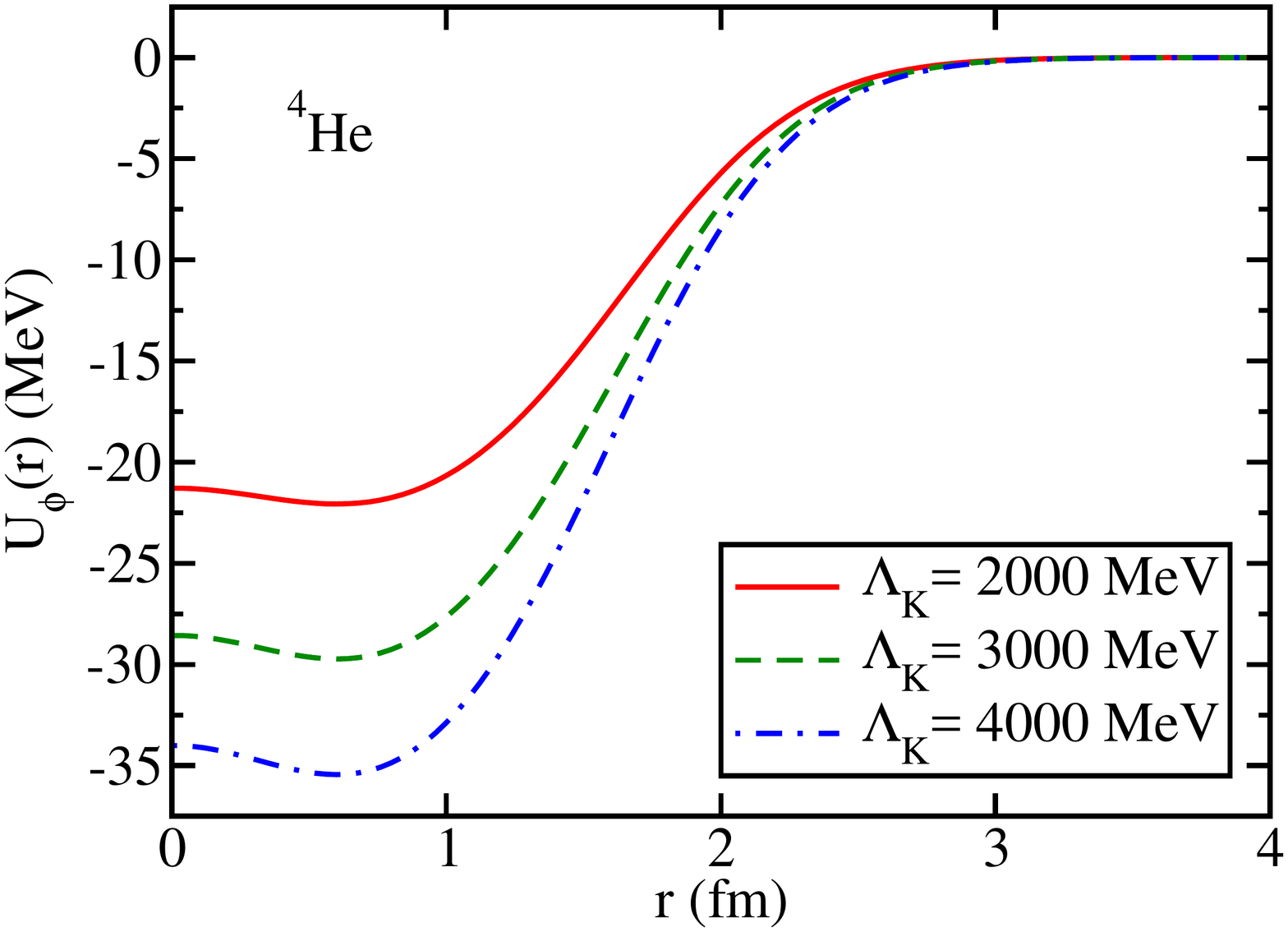}
\includegraphics[scale=0.28]{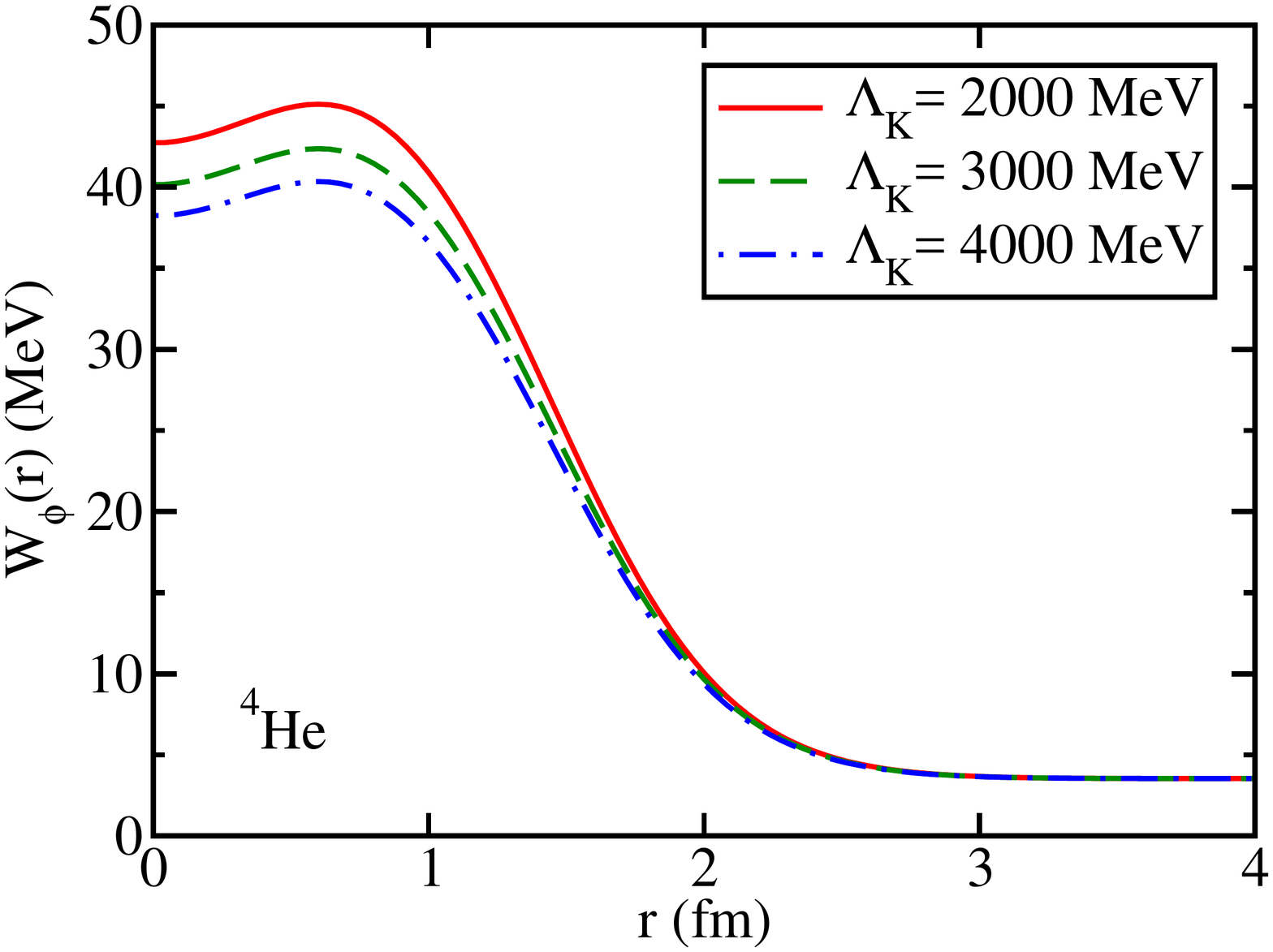}
\includegraphics[scale=0.28]{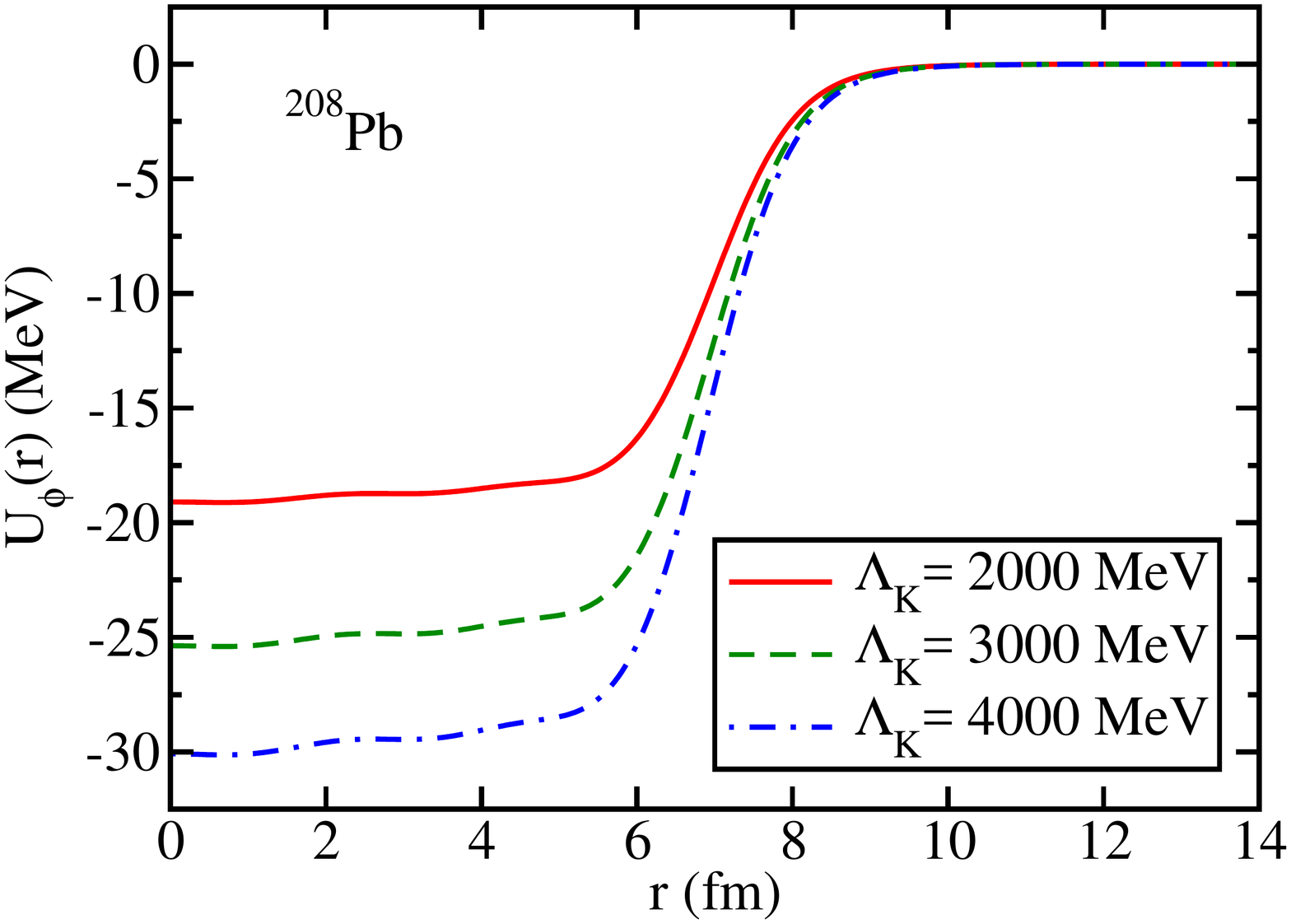}
\includegraphics[scale=0.28]{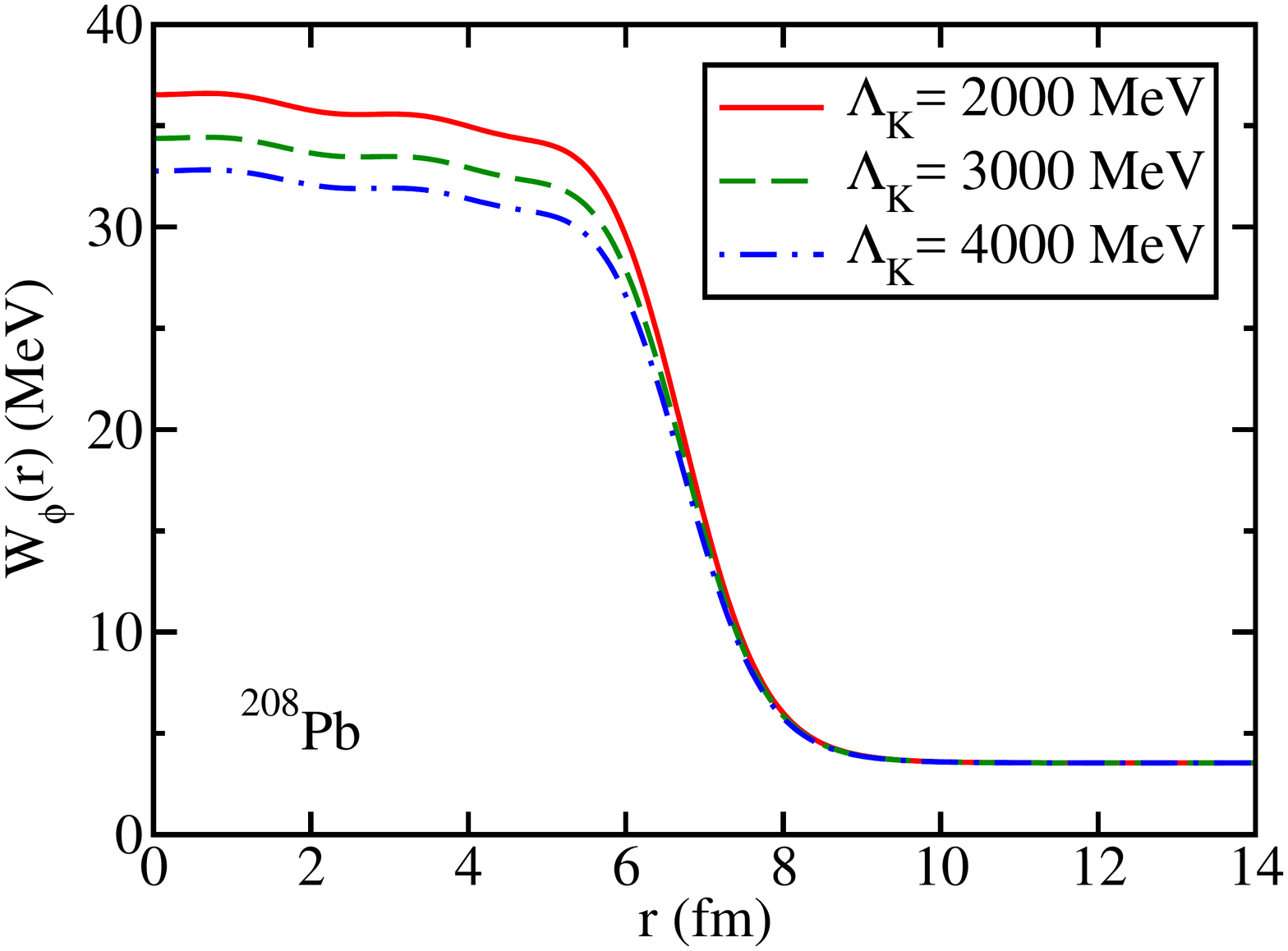}
\caption{Potential (left panel) and width (right panel) 
for $\phi$-meson in $^4$He (upper panel), and for $^{208}$Pb (lower panel), 
respectively for three values of the cutoff parameter $\Lambda_{K}$, 
from Ref.~\cite{Cobos-Martinez:2017woo}. 
\label{phipot}
}
\end{figure}

Next, we present our predictions for the single-particle energies and half widths for
$\phi$-nucleus bound states for seven nuclei selected. We solve the Klein-Gordon
equation for complex $\phi$-nucleus potentials obtained by a local-density
approximation using the nucleon density distributions calculated by the QMC model.
This leads to use the following complex $\phi$-nucleus (A) potential,  
\bg
V_{\phi A}(r) &=& \Delta m^*_{\phi}(\rho_{B}(r)) -(\mi/2)\Gamma^*_{\phi}(\rho_{B}(r)),
\\
              &\equiv& U_\phi(r) - \dfrac{i}{2} W_\phi(r),
\label{phiApot}
\en
where $U_\phi(r) \equiv \Delta m^*_{\phi}(\rho_{B}(r)) \equiv m^*_{\phi}(\rho_{B}(r)) - m_\phi$,   
$r$ is the distance from the center of the nucleus and $\rho_{B}(r)$ is 
the baryon density profile of the given nucleus calculated by the QMC model. 
As examples, we show in 
Fig.~\ref{phipot} the $\phi$-nuclear potentials $U_\phi(r)$ 
and width $W_\phi(r)$ in $^4$He and $^{208}$Pb nuclei, for three values of the 
cutoff parameter values of $\Lambda_K$.

%%%%%%%%%%%%%%%%%%%%%%%%%%
%%%%%%%%%%%%%%%%%%%%%%

%
%Bound state energies from the Klein-Gordon equation--------------------------
\begin{table}[ht]
\begin{center}
%\scalebox{0.85}{
\caption{$\phi$-nucleus single-particle energy
$E$ and half width $\Gamma/2$ obtained, with and without the imaginary 
part of the potential $W_\phi(r)$, 
for three values of the cutoff parameter $\Lambda_K$.
When only the real part is included, where the corresponding single-particle 
energy $E$ is given inside the brackets, $\Gamma=0$ for all nuclei.
``n'' denotes that no bound state is found.
All quantities are given in MeV.
\vspace{1ex}
}
\label{tab:phibse} 
\begin{tabular}{ll|rr|rr|rr} 
\hline 
%\hline
& & \multicolumn{2}{c|}{$\Lambda_{K}=2000$} &
\multicolumn{2}{c}{$\Lambda_{K}=3000$} & 
\multicolumn{2}{c}{$\Lambda_{K}=4000$}  \\
\hline
 & & $E$ & $\Gamma/2$ & $E$ & $\Gamma/2$ & $E$ & $\Gamma/2$ \\
\hline
$^{4}_{\phi}{\rm He}$ & 1s & n (-0.8) & n & n (-1.4) & n & -1.0 (-3.2) & 8.3 \\
\hline
$^{12}_{\phi}{\rm C}$ & 1s & -2.1 (-4.2) & 10.6 & -6.4 (-7.7) & 11.1 & -9.8 (-10.7) & 11.2 \\
\hline
$^{16}_{\phi}{\rm O}$ & 1s & -4.0 (-5.9) & 12.3 & -8.9 (-10.0) & 12.5 & -12.6 (-13.4) & 12.4 \\
& 1p & n (n) & n & n (n) & n & n (-1.5) & n \\
\hline
$^{40}_{\phi}{\rm Ca}$ & 1s & -9.7 (-11.1) & 16.5 & -15.9 (-16.7) & 16.2 & -20.5 (-21.2) & 15.8 \\
& 1p & -1.0 (-3.5) & 12.9 & -6.3 (-7.8) & 13.3 & -10.4 (-11.4) & 13.3 \\
& 1d & n (n) & n & n (n) & n & n (-1.4) & n \\
\hline
$^{48}_{\phi}{\rm Ca}$ & 1s & -10.5 (-11.6) & 16.5 & -16.5 (-17.2) & 16.0 & -21.1 (-21.6) & 15.6 \\
& 1p & -2.5 (-4.6) & 13.6 & -7.9 (-9.2) & 13.7 & -12.0 (-12.9) & 13.6 \\
& 1d & n (n) & n & n (-0.8) & n & -2.1 (-3.6) & 11.1 \\
\hline
$^{90}_{\phi}{\rm Zr}$ & 1s & -12.9 (-13.6) & 17.1 & -19.0 (-19.5) & 16.4 & -23.6 (-24.0) & 15.8 \\
& 1p & -7.1 (-8.4) & 15.5 & -12.8 (-13.6) & 15.2 & -17.2 (-17.8) & 14.8 \\
& 1d & -0.2 (-2.5) & 13.4 & -5.6 (-6.9) & 13.5 & -9.7 (-10.6) & 13.4 \\
& 2s & n (-1.4) & n & -3.4 (-5.1) & 12.6 & -7.4 (-8.5) & 12.7 \\
& 2p & n (n) & n & n (n) & n & n (-1.1) & n \\
\hline
$^{208}_{\phi}{\rm Pb}$ & 1s & -15.0 (-15.5) & 17.4 & -21.1 (-21.4) & 16.6 & -25.8 (-26.0) & 16.0 \\
& 1p & -11.4 (-12.1) & 16.7 & -17.4 (-17.8) & 16.0 & -21.9  (-22.2) & 15.5 \\
& 1d & -6.9 (-8.1) & 15.7 & -12.7 (-13.4) & 15.2 & -17.1 (-17.6) & 14.8 \\
& 2s & -5.2 (-6.6) & 15.1 & -10.9 (-11.7) & 14.8 & -15.2 (-15.8) & 14.5 \\
& 2p & n (-1.9) & n & -4.8 (-6.1) & 13.5 & -8.9 (-9.8) & 13.4 \\
& 2d & n (n) & n & n (-0.7) & n & -2.2 (-3.7) & 11.9 \\
\hline 
%\hline
\end{tabular}
%}
\end{center}
\end{table}
%---------------------------------------------------------------------------

Using the $\phi$-meson potentials obtained in this manner, we calculate
the $\phi$-meson--nuclear bound state energies and absorption widths for
the seven nuclei selected.
Before proceeding, a few comments on the use of \eqn{eqn:kg} (below) 
are in order. 
In this study we consider the situation where the $\phi$-meson is
produced nearly at rest; therefore, the same remarks made in the previous
section on neglecting a possible energy difference between the longitudinal 
and transverse components of the $J/\Psi$ apply to the present case
of the $\phi$. We solve the Klein-Gordon
equation ($\phi$-meson wave function is represented by $\phi(\vec{r})$ here), 
\begin{equation}
\label{eqn:kg}
\left(-\nabla^{2} + \mu^{2} + 2\mu V(\vec{r})\right)\phi(\vec{r})
= \mathcal{E}^{2}\phi(\vec{r}),
\end{equation}
where $\mu=m_{\phi}m_{A}/(m_{\phi}+m_{A})$ is the reduced mass of
the $\phi$-meson-nucleus system with $m_{\phi}$ $(m_{A}$) the mass of the
$\phi$-meson (nucleus $A$) in vacuum, and $V(\vec{r})$ is the complex
$\phi$-meson-nucleus potential of Eq.~(\ref{phiApot}).
This equation is solved by using the momentum space
methods developed in Ref.~\cite{Kwan:1978zh}.
Here, \eqn{eqn:kg} is first converted to momentum space representation
via a Fourier transform, followed by a partial wave-decomposition of the
Fourier-transformed potential. Then, for a given value of angular momentum, the
eigenvalues of the resulting equation are found by the inverse iteration
eigenvalue algorithm.
The calculated bound state energies ($E$) and widths ($\Gamma$), which are
related to the complex energy eigenvalue $\mathcal{E}$ by $E=\Re\mathcal{E}-\mu$
and $\Gamma=-2\Im\mathcal{E}$, are listed in \tab{tab:phibse} for three values of
the cutoff parameter $\Lambda_{K}$, with and without the imaginary part of the
potential, $W_{\phi A}(r)$.

Table~\ref{tab:phibse} shows the results for the real and imaginary
parts of the single-particle energies $\mathcal{E}=E-(\mi/2)\Gamma$ in 
seven nuclei selected.
We present results with and without the imaginary 
(absorptive) part of the $\phi$-nucleus potential $V_{\phi A}(r)$. One sees that
$\phi$ is not bound to $^{4}{\rm He}$ when the imaginary part of the potential is
included. For larger nuclei, the $\phi$ does bind but 
while the binding is substantial the 
energy levels are quite broad; the half widths being roughly the same size as the 
central values of the real parts. 

We first discuss the case in which the imaginary part of the $\phi$-nucleus  
potential $W_\phi(r)$ is set to zero. 
The results are listed in the brackets in \tab{tab:phibse}. 
From the values shown in brackets, we see that
the $\phi$-meson is expected to form bound states with all the seven nuclei
selected, for all values of the cutoff parameter $\Lambda_{K} = 2000, 3000$
and $4000$ MeV.
(For the variation in the potential depths due to the $\Lambda_K$ values, see Fig.~\ref{phipot}.)
However, the bound state energy is obviously dependent on $\Lambda_{K}$, as expected.

Next, we discuss the results obtained with the full potential, 
including the imaginary part $W_\phi(r)$. 
Adding the absorptive part of the potential, the situation changes appreciably.
From the results presented in \tab{tab:phibse} 
we note that, for the largest value of the cutoff parameter $\Lambda_K = 4000$ MeV 
which yields the deepest attractive potentials, 
the $\phi$-meson is expected to form bound states in all the seven nuclei selected, 
including the lightest $^4$He nucleus.
However, whether or not the bound states can be observed experimentally,  
is sensitive to the value of the cutoff parameter $\Lambda_K$.
One can also observe that the width of the bound
state is insensitive to the values of $\Lambda_{K}$ for all nuclei.
Furthermore, since the imaginary part induces a repulsive potential,
the bound states disappear completely in some cases, 
even though the bound states are found when the absorptive part is set to zero. 
This feature is obvious for the $^4$He nucleus.
Thus, it would be very interesting for the future experiments    
planned at J-PARC and JLab, using light and medium-heavy 
nuclei~\cite{Aoki:2015qla,JLabphi,Buhler:2010zz,Ohnishi:2014xla}.

Here we comment that we have also solved the Schr\"{o}edinger equation 
with the potential \eqn{phiApot} with and without the imaginary part, 
and obtained the single-particle energies and widths, 
and compared with those given in \tab{tab:phibse}. 
The results found for both, with and without the imaginary part, 
are essentially the same.

To summarize, essential to the calculation of the $\phi$-nucleus bound states 
is the in-medium kaon mass which is calculated in the QMC model, 
where the scalar and vector meson mean fields couple directly to the 
light $u$ and $d$ quarks (antiquarks) in the $K$ ($\Kbar$) meson.

At normal nuclear matter density, allowing for a very large variation of the 
cutoff parameter $\Lambda_{K}$, although one finds a sizable 
negative mass shift of 13\% in the kaon mass, this induces only a few percent 
downward shift of the $\phi$-meson mass. 
On the other hand, it induces an order-of-magnitude broadening of the decay width.

Given the nuclear matter results, a local density approximation was used  
to infer the position dependent $\phi$-nucleus (A) complex scalar potential, 
$V_{\phi A}(\rho_B(r)) = U_\phi(r) - (i/2) W_\phi(r)$, 
in a finite nucleus. 
This allowed us to study the binding and absorption of a number of 
$\phi$-nuclear systems, given the nuclear density profiles, $\rho_B(r)$, 
calculated using the QMC model.

The Klein-Gordon equation has been solved to obtain the bound state 
single-particle energies for seven nuclei selected.
While the results found in here show that one should expect the 
$\phi$-meson to bound in all but the lightest nuclei, 
the broadening of these energy levels, which is comparable to the amount 
of binding, suggests that it may be challenging to observe such states experimentally.
%
%
%
%\input{sec_mesic_other.tex}
% 
%
%
%%%%%%%%%%%%%%%%%%%%%%%%%%%%%%%%%%%%%%%%%%%%%%%%%%%%%%%%%%%%%%%%%%%%%%%%%%%%%%%%%%%%%
\section{Nuclear-bound of $\omega$, $\eta$ and $\eta'$  mesons}
\label{etaod}
%%%%%%%%%%%%%%%%%%%%%%%%%%%%%%%%%%%%%%%%%%%%%%%%%%%%%%%%%%%%%%%%%%%%%%%%%%%%%%%%%%%%%

As mentioned already, to study the medium modification of the
light vector ($\rho$, $\omega$ and $\phi$) meson masses is very interesting
since it can provide us with information on
partial restoration of chiral symmetry in a nuclear medium.
Such experiments carried out by the CERES and HELIOS 
collaborations at the CERN/SPS~\cite{ceres}, 
and those at JLab~\cite{jlab1,Wood:2008ee} and GSI~\cite{gsi,Friese:1999qm}, 
are closely related to this issue.
An alternative approach to study meson mass shifts in nuclei
was suggested by Hayano {\it et al.}~\cite{hayano0} to produce
$\eta$ and $\omega$ mesons with nearly zero recoil, 
which inspired the theoretical investigations 
of $\eta$- and $\omega$-mesic nuclei~\cite{hayano2,etao}. 
We review here the results for $\omega$-, $\eta$- and $\eta'$-mesic nuclei, 
those studied using the quark-meson coupling (QMC) model~\cite{Tsushimad,etao}.

In this section we discuss the nuclear binding of  
the $\omega, \eta$ and  $\eta'$ mesons.
These mesons contain  
the {\it hidden light-quark} ($q\overline{q}$) components, 
while the $D$ and $\overline{D}$ mesons, which will be discussed in next section, 
are respectively represented by $c\overline{q}$ and $\overline{c}q$ mesons. 
Since the interactions of the mesons with the nuclear medium, 
or nuclear many-body systems are mainly 
mediated by the scalar-isoscalar $\sigma$ 
and vector-isoscalar $\omega$ mean fields which directly 
couple to the light quarks in the QMC model, the treatment 
for these mesons is different from that for 
the hidden strange meson $\phi$, and hidden charm meson $J/\Psi$.
Thus, at first order, they directly interact with the 
surrounding nuclear environment, not necessary through  
the higher order process such as appeared in 
the self-energy diagrams for $\phi$ and $J/\Psi$.

%%%%%%%%%%%%%%%%%%%%%%%%%%%%%%%%%
\subsection{In-medium masses of $\omega, \eta$ and $\eta'$ mesons}
\label{massetaod}
%%%%%%%%%%%%%%%%%%%%%%%%%%%%%%%%%

First, we discus the $\omega, \eta$, and $\eta'$ meson masses
in symmetric nuclear matter, since the treatments need some  
more explanations in addition to those given in subsection~\ref{hadronmass}.
They are calculated in Ref.~\cite{etao}
(see also Eq.~(\ref{hmass})):
\bg
m_{\eta,\omega}^*(\vr) &=& \dfrac{2 [a_{P,V}^2\Omega_q^*(\vr)
+ b_{P,V}^2\Omega_s(\vr)] - z_{\eta,\omega}}{R_{\eta,\omega}^*}
+ \dfrac{4}{3}\pi R_{\eta,\omega}^{* 3} B,
\qquad ({\rm for}\, \eta',\, a_P \leftrightarrow b_P),
\label{metao}\\
& &\left.\dfrac{\partial m_j^*(\vr)}
{\partial R_j}\right|_{R_j = R_j^*} = 0, \quad\quad
(j = \omega,\eta,\eta'),
\label{equil}\\
%
%{\rm with}\qquad \hfill\nn \\
%
a_{P,V} &\equiv& \sqrt{1/3} \cos\theta_{P,V}
- \sqrt{2/3} \sin\theta_{P,V},\quad
b_{P,V} \equiv \sqrt{2/3} \cos\theta_{P,V}
+ \sqrt{1/3} \sin\theta_{P,V}, 
\label{abpv}
\en
where $\theta_{P,V}$ are the octet and singlet mixing angles 
for the pseudoscalar and vector mesons, respectively.

\begin{figure}[htb]
\begin{center}
\epsfig{file=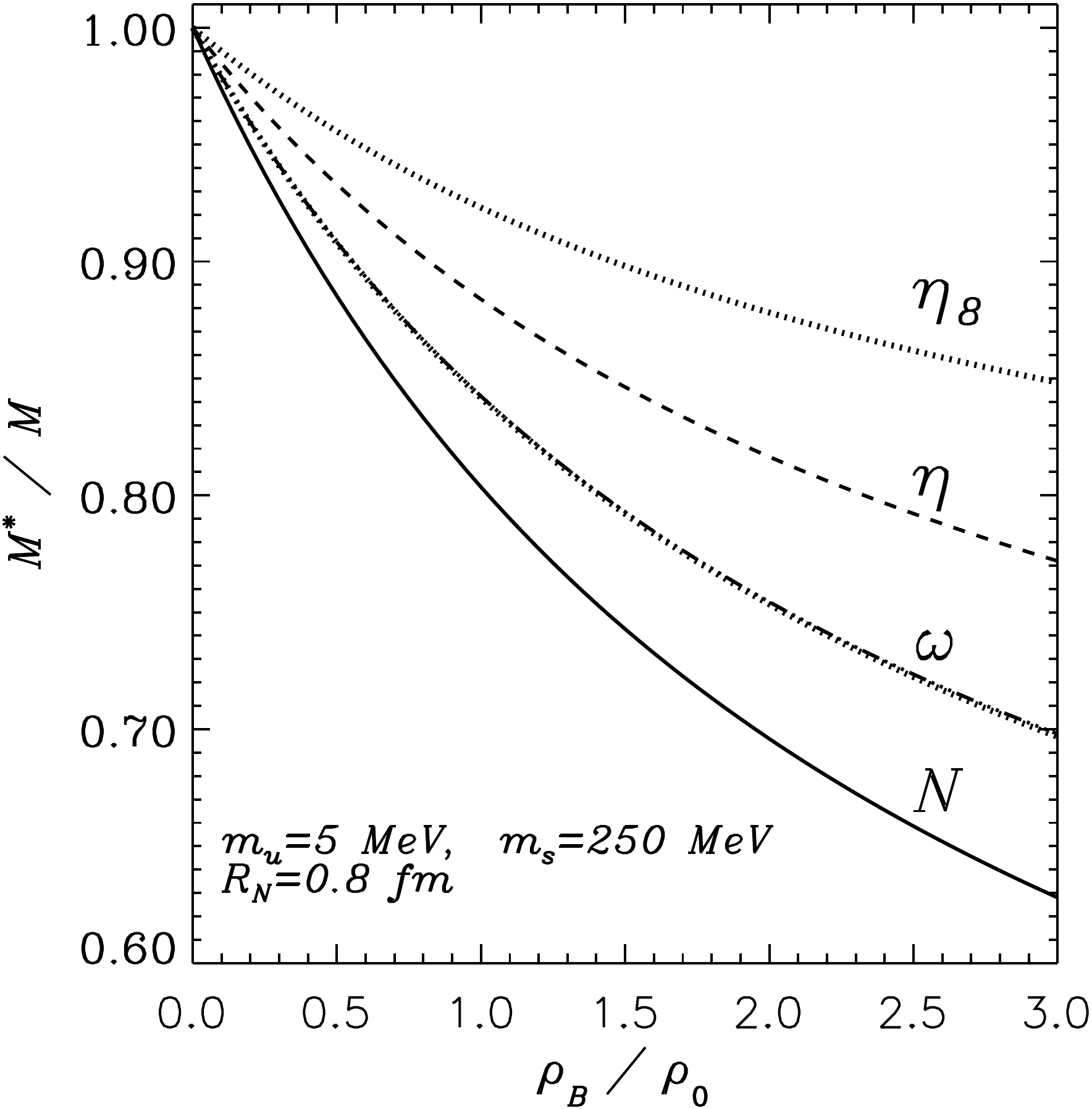,height=7cm}
\hspace{5ex}
\epsfig{file=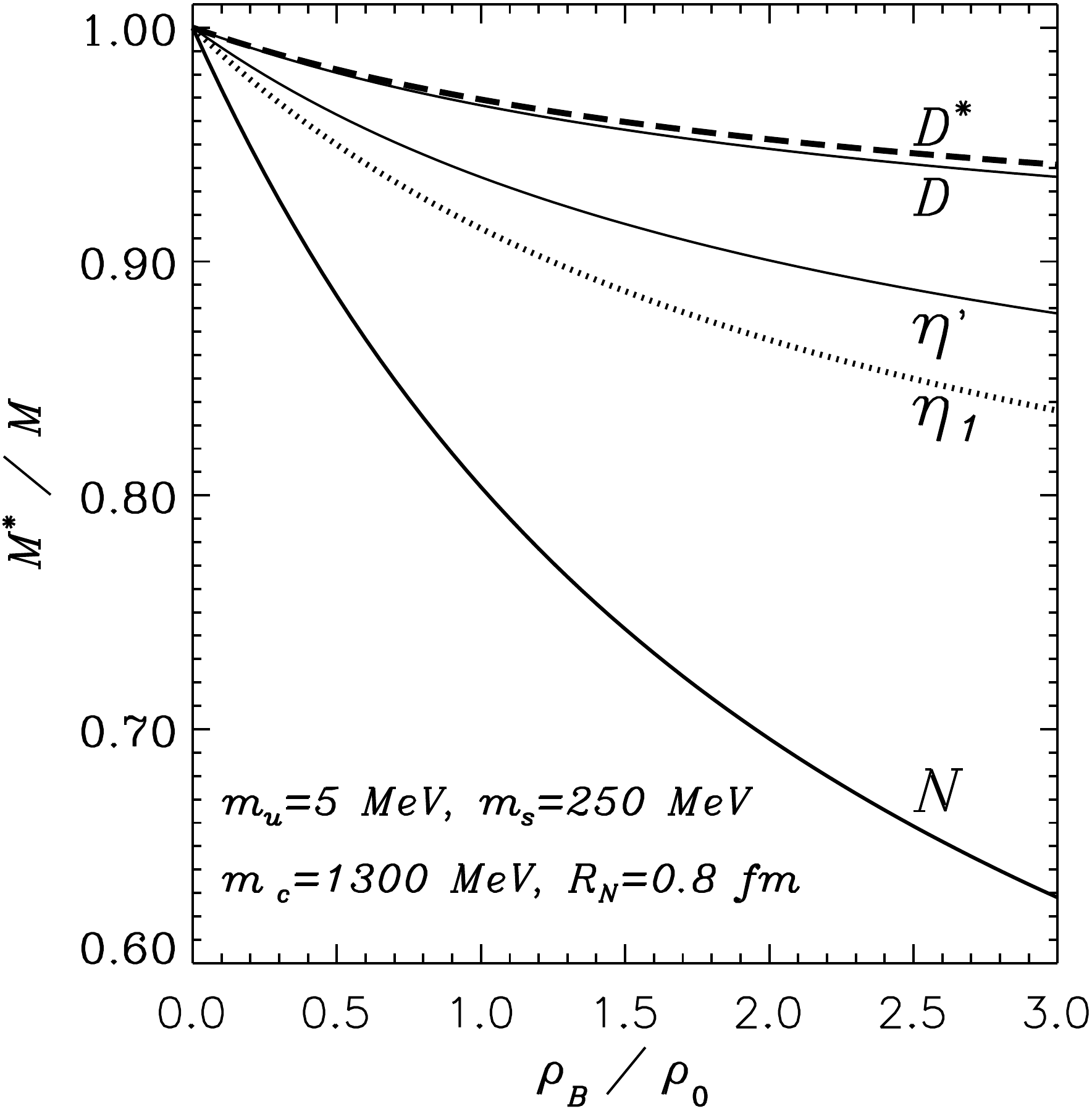,height=7cm}
\caption{Effective mass ratios in medium to those in vacuum 
in symmetric nuclear matter ($\rho_0$ = 0.15 fm$^{-3}$).
$\eta_8$ and $\eta_1$ are respectively the octet and singlet states 
without the effect of the mixing angle $\theta_P$.
}
\label{etaomassmatter}
%\end{minipage}
\end{center}
\end{figure}

In Fig.~\ref{etaomassmatter}
we show the calculated in-medium to vacuum mass ratios of the mesons  
in symmetric nuclear matter~\cite{Tsushimad,etao}.
The masses for the physical $\omega$, $\eta$ and $\eta'$ are calculated  
using the corresponding octet ($\eta_8$) and 
singlet ($\eta_1$) mixing angles,
$\theta_P=-10^\circ$ for ($\eta,\eta'$), and
$\theta_V=39^\circ$ for ($\phi,\omega$)~\cite{PDG}. 
Note that for the $\omega$ and $\phi$ mesons, they are 
known to be nearly {\it ideal-mixing}, so the mixing effect for their 
in-medium masses are expected to be negligible, and in fact they have been treated 
as $q\overline{q}$ and $s\overline{s}$ mesons, respectively.
The masses for the octet and singlet states without the 
mixing effect for calculating the $\eta$ and $\eta'$ system 
are also shown (the dotted lines indicated by $\eta_8$ and $\eta_1$).

%%%%%%%%%%%%%%%%%%%%%%%%%%%%%%%%%%%%%%%%%%%%%%%%%%%%%%%%%%%%%%%%%%%%%%%%%%%%%
%%%%%%%%%%%%%%%%%%%%%%%%%%%%%%%%%%%%%%%%%%%%%%%%%%%%%%%%%%%%%%%%%%%%%%%%%%%%%
%

%%%%%%%%%%%%%%%%%%%%%%%%%%%%%%%%%%%%%%%%%%%%%%%%%%%%%%%%%%%%%%%%%%%%%%%%%
\subsection{$\omega$-, $\eta$- and $\eta'$-nuclear bound states}
\label{etaoetapresult}
%%%%%%%%%%%%%%%%%%%%%%%%%%%%%%%%%%%%%%%%%%%%%%%%%%%%%%%%%%%%%%%%%%%%%%%%%

To calculate the bound state energies for the
mesons with the situation of almost zero momenta, we 
solve the Klein-Gordon equation~\cite{etap,Tsushimad,etao}:

\bg
[ \nabla^2 &+& (E^*_j - V^j_v(r))^2- \tilde{m}^{*2}_j(r) ]\,
\phi_j(\vr) = 0,\qquad (j=\omega,\eta,\eta',D,\overline{D}),
\label{kgequation1}
\\
\tilde{m}^*_j(r) &\equiv& m^*_j(r) - \dfrac{i}{2}
\left[ (m_j - m^*_j(r)) \gamma_j + \Gamma_j^0 \right]\,
\equiv\, m^*_j(r) - \dfrac{i}{2} \Gamma^*_j (r),
\label{width}
\en
where $E^*_j$ is the total energy of the meson,
$V^j_v(r)$, $m_j$ and $\Gamma_j^0$
are respectively the sum of the vector and Coulomb potentials,
the corresponding masses and widths in free space, and
$\gamma_j$ are treated as phenomenological
parameters to describe the in-medium meson widths,
$\Gamma^*_j(r)$~\cite{etao}.
Note that, for the $\eta'$-meson, 
any effects of in-medium width has been ignored. 
Thus, for the $\eta'$-meson, the results may have larger uncertainties 
due to the lack of including the possible widths in nucleus.
We discuss this issue below based on recent experimental developments.

We show the nuclear bound state energies calculated for $\eta$, $\omega$ and $\eta'$ 
mesons respectively for $\gamma_\eta = 0.5$, $\gamma_\omega = 0.2$ and $\gamma_{\eta'} = 0$.
This order seems to be experimentally promising. 
For the $\eta$ and $\omega$ cases, they are expected to correspond best with experiment
according to the estimates in Refs.~\cite{hayano0,fri}, although for $\eta'$, 
no clue existed when it was studied~\cite{etap}.
The bound state single-particle energies obtained  
are listed in Tab.~\ref{etaoenergy}. 

%%%%%%%%%%%%%%%%%%%%%%%%%%%%%%%%%%%%%%%%%%%%%%%%%%%%%%%%%%%%%%%%%%%%%%%%%
%Omega
\begin{table}[htb]
%\vspace{-2em}
\begin{center}
\caption{
$\eta$, $\omega$ and $\eta'$ bound state single-particle energies (in MeV),
$E_j = Re (E^*_j - m_j)\,(j=\eta,\omega,\eta')$,
where all widths for the $\eta'$ are set to zero.
The eigenenergies are given by,
$E^*_j = E_j + m_j - i \Gamma_j/2$.
\vspace{1ex}
}
\label{etaoenergy}
\begin{tabular}[t]{lcccccl}
\hline
%\hline
& &$\gamma_\eta=0.5$ & &$\gamma_\omega$=0.2
& &$\gamma_{\eta'}=0$ \\
\hline 
& &$E_\eta$ &$\Gamma_\eta$ &$E_\omega$ &$\Gamma_\omega$
&$E_{\eta'}$ \\
\hline
$^{6}_j$He &1s &-10.7&14.5 &-55.6&24.7 &* (not calculated)\\
%
%\hline
$^{11}_j$B &1s &-24.5&22.8 &-80.8&28.8 &* \\
%
%\hline
$^{26}_j$Mg &1s &-38.8&28.5 &-99.7&31.1 &* \\
            &1p &-17.8&23.1 &-78.5&29.4 &* \\
            &2s & --- & --- &-42.8&24.8 &* \\
%
%\hline \hline
$^{16}_j$O &1s &-32.6&26.7 &-93.4&30.6 &-41.3 \\
           &1p &-7.72&18.3 &-64.7&27.8 &-22.8 \\
%
%\hline
$^{40}_j$Ca &1s &-46.0&31.7 &-111&33.1  &-51.8 \\
            &1p &-26.8&26.8 &-90.8&31.0 &-38.5 \\
            &2s &-4.61&17.7 &-65.5&28.9 &-21.9 \\
%
%\hline
$^{90}_j$Zr &1s &-52.9&33.2 &-117&33.4  &-56.0 \\
            &1p &-40.0&30.5 &-105&32.3  &-47.7 \\
            &2s &-21.7&26.1 &-86.4&30.7 &-35.4 \\
%
%\hline
$^{208}_j$Pb &1s &-56.3&33.2 &-118&33.1 &-57.5 \\
             &1p &-48.3&31.8 &-111&32.5 &-52.6 \\
             &2s &-35.9&29.6 &-100&31.7 &-44.9 \\
%
%\hline 
\hline
\end{tabular}
\vspace{1ex}
\end{center}
\end{table}
%
%%%%%%%%%%%%%%%%%%%%%%%%%%%%%%%%%%%%%%%%%%%%%%%%%%%%%%%%%%%%%%%%%%%%%%%%%%%%
%

We first discuss the $\eta$-nuclear bound states.
Studies of the $\eta$-nuclear bound states have a relatively long history following the studies
in Refs.~\cite{Haider:1986sa,Liu:1986rd}.  
Later in Ref.~\cite{hayano2} experimental feasibility of observing the $\eta$-nuclear 
bound states was studied.
The results for the $\eta$ in Tab.~\ref{etaoenergy} show 
that the $\eta$-meson will be bound in all nuclei considered. 
However, the half widths ($\Gamma_\eta/2$) obtained are relatively large compared to the 
corresponding binding energies. A similar conclusion was also derived 
in Ref.~\cite{hayano2}. They argued that, despite the relatively large widths 
of the bound states, it would be feasible to experimentally observe the 
$\eta$ mesic nuclei in the excitation energy spectrum.
$\eta$ bound states in nuclei was also studied focusing on the 
$\eta$-$\eta'$ mixing and flavor-singlet dynamics in Refs.~\cite{Bass:2010kr,Bass:2005hn}.

As for the $\omega$-nuclear bound states, the results in Tab.~\ref{etaoenergy} 
indicate that one may expect deeper bound state energies and smaller widths 
than those for the $\eta$. A similar trend was also found in the study 
made based on QHD~\cite{Saito:1998ev}.
However, as already discussed in section~\ref{phi-meson}, the present situation 
for the vector meson properties in nuclear medium, is still  
controversial~\cite{Leupold:2009kz,Hayano:2008vn,Krein:2016fqh}.

As already mentioned, we have shown the result for the $\eta'$-nuclear 
bound state energies calculated for some nuclei, by setting the possible 
width to be zero. 
(Based on the QCD symmetry, the $\eta'$ (and $\eta$) mesic nuclei 
was further studied in Ref.~\cite{Bass:2013nya}.)  
However, a recent experimental study of the 
$\eta'$-nucleus optical potential by the photoproduction of $\eta'$ (and $\omega$) 
off carbon and niobium nuclei~\cite{Friedrich:2016cms}, found the imaginary 
part of the $\eta'$-nuclear potential to be about three times smaller than 
the real part. Thus, one can extract the main features 
of the $\eta'$-nuclear bound states from the results 
shown in Tab.~\ref{etaoenergy}, the results obtained by setting the 
width to be zero. 

Theoretically, the $\eta'$-nucleus bound states were studied 
in Refs.~\cite{etap,Nagahiro:2004qz,Nagahiro:2011fi,Nagahiro:2012aq}, 
where some of them focused on the reactions involving a $^{12}$C nucleus.
However, recent measurement of excitation spectra in the 
$^{12}$C($p,d$) reaction near the $\eta'$ emission threshold, 
observed no distinct structure associated with the formation of 
$\eta'$-nucleus bound states~\cite{Tanaka:2016bcp,Tanaka:2017cme}. 
Thus, we need to wait for further experimental efforts, ideally in low 
recoil momentum conditions, to draw a definite conclusion, possibly 
using heavier nuclear targets.

%%%%%%%%%%%%%%%%%%%%%%%%%%%%%%%%%%%%%%%%%%%%%%%%%%%%%%%%%%%%%%%%%%%%%%%%%

%%%%%%%%%%%%%%%%%%%%%%%%%%%%%%%%%%%%%%%%%%%%%%%%%%%%%%
\section{Nuclear-bound heavy-flavor hadrons}
\label{sec:Hflavor}
%%%%%%%%%%%%%%%%%%%%%%%%%%%%%%%%%%%%%%%%%%%%%%%%%%%%%

Very little is presently known about the strength of the interaction of charmed hadrons 
with ordinary baryons and mesons. For the feasibility of experimental studies of formation
of nuclear bound states, the production rate for charmed hadrons is a key factor. For the
specific case of experiments of antiproton annihilation on the nucleon and nuclei, as those
planned to be conducted by the
$\overline{\rm P}$ANDA collaboration~\cite{Wiedner:2011} at the FAIR
facility, several predictions have been made in recent years. These include predictions for 
charmed mesons and baryons in free space~\cite{{Kroll:1988cd},{Kerbikov:1994xx},{Kaidalov:1994mda},
{Titov:2008yf},{Goritschnig:2009sq},{Haidenbauer:2009ad},{Haidenbauer:2010nx},{Titov:2011vc},
{Khodjamirian:2011sp},{Goritschnig:2012vs},{Haidenbauer:2014rva},{Shyam:2014dia},{Wang:2016fhj},
{Haidenbauer:2016pva}} and $D\bar{D}$ production in nuclei~\cite{Shyam:2016bzq} and formation of 
$\Lambda_c^+$-hypernuclei~\cite{Shyam:2016uxa}. 
Systematic studies were made on the property changes of heavy hadrons 
which contain a charm or a bottom quark 
in nuclear matter~\cite{QMCbc} (see subsection~\ref{hadronmass}), 
and on $\Lambda_c^+$- and $\Lambda_b$-hypernuclei~\cite{QMChypbc}. 
Furthermore, production of $D\overline{D}$ meson pair 
in nucleus~\cite{Shyam:2016bzq}, These results suggest that it is quite likely 
to form the charmed and bottom hypernuclei, 
which were predicted first in mid 70's~\cite{Tyapkin,Dover}.
The experimental possibilities were also studied later~\cite{bcexp}.
For the strange hypernuclei including recent development, 
see Refs.~\cite{QMChyp,Guichon:2008zz,Tsushima:2009zh,Tsushima:2010ew}.
In addition, the $B^-$-nuclear (atomic) 
bound states were also predicted, based on the analogy with the 
kaonic atoms~\cite{Katom}, and a study made for 
the $D$- and $\overline{D}$-nuclear bound states~\cite{Tsushimad} (subsection~\ref{DDbar_result})  
using the QMC model~\cite{QMCGuichon,Guichonfinite,QMCreview,Saitofinite} 
(section~\ref{qmc}). 

The issue of charmed mesic nuclei~\cite{Tsushimad} is in some ways even more 
exciting, in that it promises more specific information on the
relativistic mean fields in nuclei and the nature of dynamical chiral
symmetry breaking. We focus on systems containing an anti-charm quark
and a light quark, $\bar{c}q$ $(q=u,d)$, 
which have no strong decay channels if bound.
If we assume that dynamical chiral symmetry breaking is the
same for the light quark in the charmed meson as in purely light-quark
systems, we expect the same coupling constant 
$g^q_\sigma$, in the QMC model.
Thus, the $D$ and $\overline{D}$ mesons   
can provide very nice systems to study the dynamical symmetry breaking 
due to their light-quark components. 
That is, whether or not the light quarks in the $D$ and $\overline{D}$ mesons 
reveal the same dynamical symmetry breaking  
in a nuclear medium, as those in nucleons.

We note that, in the absence of any strong interaction, the $D^-$ will form
atomic states, bound by the Coulomb potential.
The resulting binding for, say, the $1s$ level in $^{208}$Pb
is between 10 and 30 MeV and should provide a very clear
experimental signature. On the other hand,
although we expect the D-meson (systems of $\bar{q}c$) will
form deeply bound $D$-nucleus states, they will also couple
strongly to open channels such as $D N \rightarrow B_c (\pi's)$, 
with $B_c$ a charmed baryon.
Unfortunately, because our present knowledge does not permit
an accurate calculation of the $D$-meson widths in a nucleus,
results for the $D$-mesic nuclei may not give useful information
for experimenters.

In the following we discuss the nuclear binding of 
$D$, $\overline{D}$, $\Lambda^+_c$ and $\Lambda_b$ heavy hadrons. 
One of the interests to study such systems is 
dynamically symmetry breaking of light quarks inside the heavy flavored hadrons,
whether or not the light quarks inside the heavy hadrons feel the same forces 
with those in light hadrons (nucleons and light mesons).
The role of the light quarks in heavy hadrons in connection with 
the partial restoration of chiral symmetry in a nuclear medium, 
is one of the very interesting issues in this 
section{\textemdash}we refer the reader to Ref.~\cite{Hosaka:2016ypm} for 
an extended overview on this topic.

%%%%%%%%%%%%%%%%%%%%%%%%%%%%%%%%%%%%%%%%%%%%%%%%%%%%%%%%%%%%%%%%%%%%%%%%%%%%
\subsection{Predictions for $D$- and $\overline{D}$-mesic nuclei in $^{208}$Pb}
\label{DDbar_result}
%%%%%%%%%%%%%%%%%%%%%%%%%%%%%%%%%%%%%%%%%%%%%%%%%%%%%%%%%%%%%%%%%%%%%%%%%%%%

In this subsection we discuss the possible formation of 
$D$- and $\overline{D}$-nuclear bound states 
in a $^{208}$Pb nucleus. 
The reason we consider the $^{208}$Pb nucleus is that, it induces sufficient 
attractive Coulomb potential for the negatively charged $D^-$ meson, 
and this can assist to form the $D^-$-$^{208}$Pb bound state, 
namely, a Coulomb-assisted mesic nuclei (nuclear bound states). 
Theoretically, possible nuclear bound states involving the 
$D$ and $\overline{D}$ mesons were studied 
in Refs.~\cite{Tsushimad,GarciaRecio:2010vt,GarciaRecio:2011xt}.

Because the role of the Coulomb potential for the $D^-$-meson is important, 
we explicitly give potentials for the $D$ and $\overline{D}$ mesons also including 
the Coulomb potential. But we do not show the potential for $D^+$, 
since the Coulomb repulsion for $D^+$ would not allow the $D^+$-$^{208}$Pb bound states.
The potentials are given by:
\bg
V^D_s(r)
&=& m^*_D(r) - m_D,
\label{spotj}
\\
V^{D^-}_v(r) &=&
 V^q_\omega(r) - \dfrac{1}{2}V^q_\rho(r) - A(r),
\label{vdpot1}
\\
V^{\bar{D^0}}_v(r) &=&
V^q_\omega(r) + \dfrac{1}{2}V^q_\rho(r),
\label{vdpot2}
\\
V^{D^0}_v(r) &=&
- V^q_\omega(r) - \dfrac{1}{2}V^q_\rho(r) ,
\label{vdpot3}
\en
where $A(r)$ is the Coulomb potential associated with the $D^-$ and $^{208}$Pb.
Note that the $\rho$-meson mean field potential, $V^q_\rho(r)$, is negative
in a nucleus with a neutron excess, such as e.g., $^{208}$Pb, the present case.
To discuss the effects of the larger $\omega$-meson coupling 
suggested by $K^+$-nucleus scattering, $V^q_\omega(r)$ is replaced by 
$\tilde{V}^q_\omega(r) = 1.4^2 V^q_\omega(r)$~\cite{Tsushimak,Tsushimad}.

Before showing the calculated potentials for the $D^-$ in $^{208}$Pb,
which are particularly interesting in view of the strong Coulomb field,
recall that we have already shown in Figs.~\ref{fig:DDsmass} and~\ref{etaomassmatter} 
(right panel) the mass shift of $D (\bar{D})$ calculated in symmetric
nuclear matter.

\begin{figure}[hbt]
\begin{center}
\epsfig{file=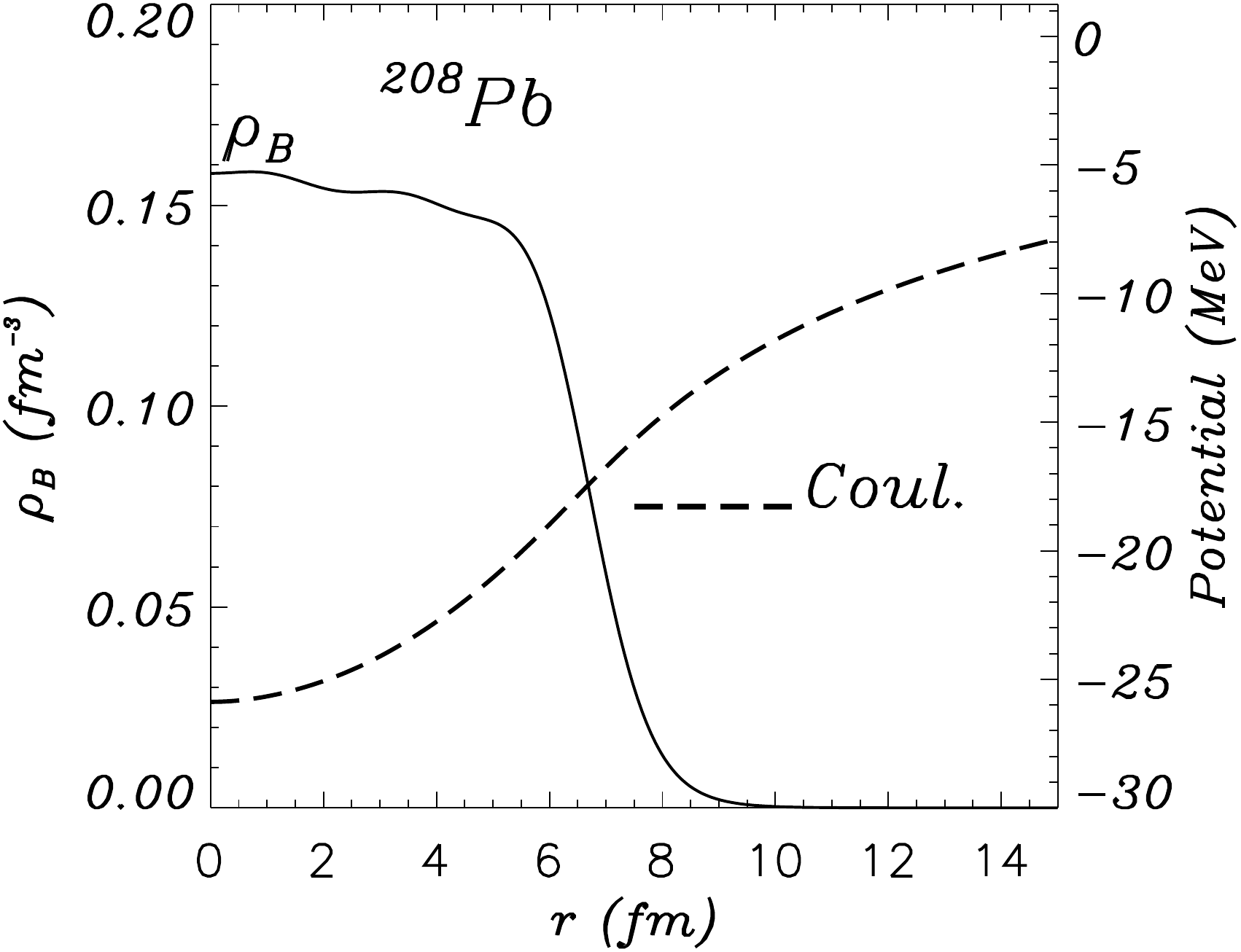,height=6.5cm}
\hspace{0.5cm}
\epsfig{file=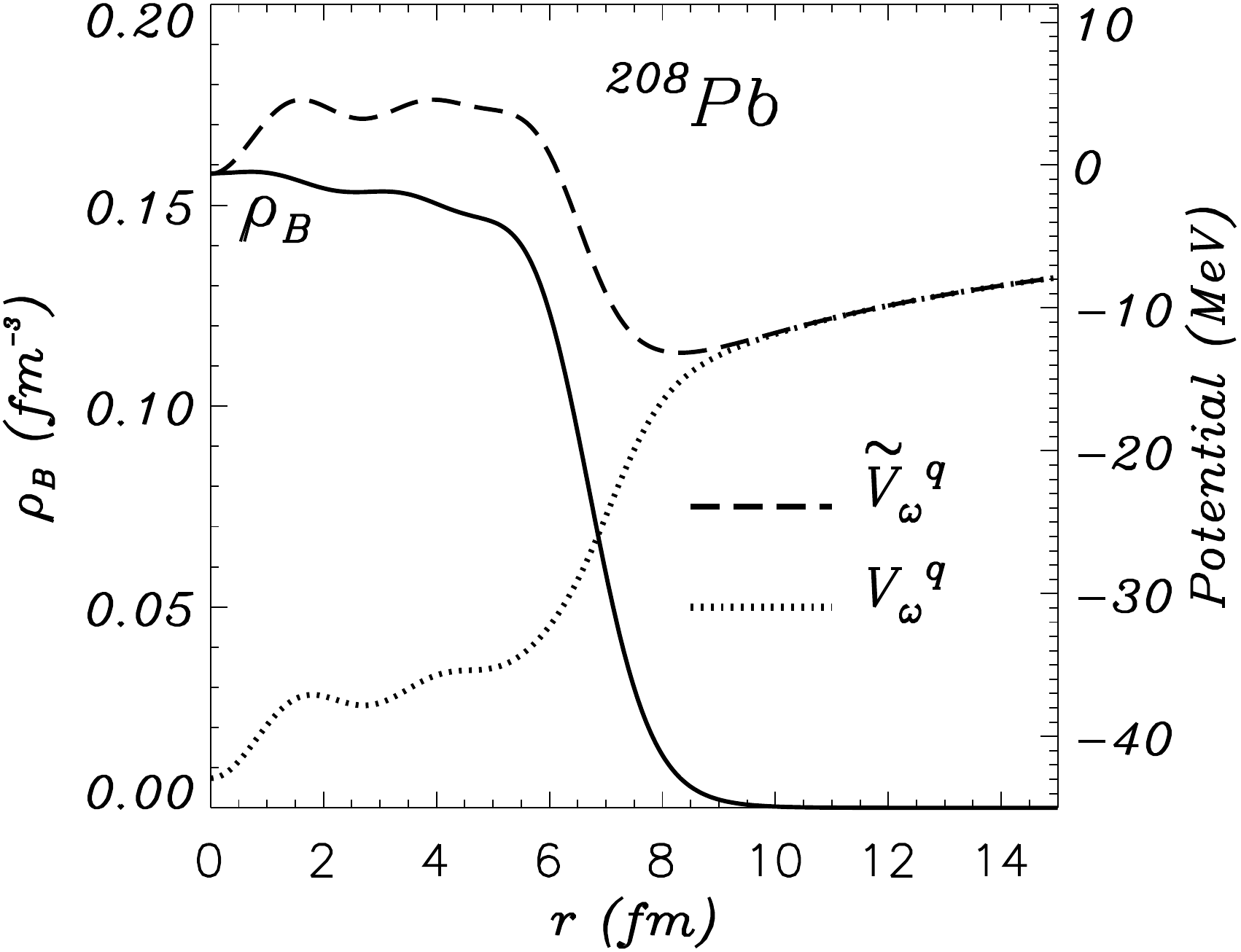,height=6.5cm}
\caption{Coulomb potential in a $^{208}$Pb (left panel) and 
sum of the scalar, vector and Coulomb potentials for the $D^-$
meson in $^{208}$Pb (right panel) for two cases,
$(m^*_{D^-}(r) - m_{D^-}) + \tilde{V}^q_\omega(r)
- {1}/{2} V^q_\rho(r) - A(r)$ (the dashed line) and
$(m^*_{D^-}(r) - m_{D^-}) + V^q_\omega(r)
- {1}/{2} V^q_\rho(r) - A(r)$ (the dotted line), 
where $\tilde{V}^q_\omega(r) = 1.4^2 V^q_\omega(r)$.
}
\label{dmespot}
\end{center}
\end{figure}

In Fig.~\ref{dmespot} we show respectively 
the Coulomb potential felt by $D^-$ in $^{208}$Pb 
calculated by the QMC model (left panel), 
and the total potential, including the 
Coulomb and $\rho$-mean field potentials (right panel),
the sum of the potentials for the two choices of
$V^{D^-}_s(r) + V^{D^-}_v(r)$ (the dashed line corresponds to
$\tilde{V}^q_\omega(r) = 1.4^2 {V}^q_\omega(r)$, 
and the dotted line to the usual ${V}^q_\omega(r)$).
Because the $D^-$ meson is heavy and may be described well
in the (nonrelativistic) Schr\"{o}dinger equation, one
expects the existence of the $D^-$-$^{208}$Pb bound states
just from inspection of the naive sum of the potentials,
in a way which does not distinguish the Lorentz vector or
scalar character.

%
%%%%%%%%%%%%%%%%%%%%%%%%%%%%%%%%%%%%%%%%%%%%%%%%%%%%%%%%%%%%%%%%%%%%%%%%%%%%%
%
%
%\newpage
%%%%%%%%%%%%%%%%%%%%%%%%%%%%%%%%%%%%%%%%%%%%%%%%%%%%%%%%%%%%%%%%%%%%%%%%%%%%%

Now we are in a position to calculate the bound state energies for the
$D$ and $\bar{D}$ in $^{208}$Pb,
using the potentials calculated in the QMC model.
For the $D^-$ and $\overline{D^0}$, the widths are set to zero which is 
(nearly) exact. 
There are several variants of the dynamical equation for a bound meson-nucleus
system. Consistent with the mean field picture of QMC, we 
solve the Klein-Gordon equation,
%%%%%%
\bg
[ \nabla^2 &+& (E_j - V^j_v(r))^2- m^{*2}_j(r) ]\,
\phi_j(\vr) = 0 , 
\label{kgequation}
\en
%%%%%
where $E_j$ is the total energy of the meson $j$ 
(the binding energy is $E_j-m_j$).
To deal with the long range Coulomb potential, we first expand the quadratic
term (the zeroth component of Lorentz vector) as ,
$(E_j - V^j_v(r))^2 = E_j^2 + A^2(r) + V^2_{\omega\rho}(r) +
2 A(r)V_{\omega\rho}(r) - 2 E_j [A(r)+V_{\omega\rho}(r)]$,
where $V_{\omega\rho}(r)$ is the combined potential due to
$\omega$ and $\rho$ meson mean fields 
($V_{\omega\rho}(r) = V_\omega^q(r)- {1}/{2}V_\rho^q(r)$ for $D^-$).
Then Eq.~(\ref{kgequation}) can be rewritten as an effective
Schr\"odinger-like equation,
%%%%%
\begin{equation}
\left[ - {\nabla^2\over 2m_j} + V_j(E_j,r)\right] \Phi_j(r)
= {E_j^2-m_j^2 \over 2m_j}\Phi_j(r),
\label{schrodinger}
\end{equation}
%%%%%
where $\Phi_j(r) = 2m_j\phi_j(r)$ and
$V_j(E_j,r)$ is an effective energy-dependent
potential which can be split
into three pieces (Coulomb, vector and scalar parts),
%%%%%
\begin{equation}
V_j(E_j,r) = {E_j\over m_j} A(r) + {2E_jV^j_{\omega\rho}(r)
- (A(r)+V^j_{\omega\rho}(r))^2\over 2m_j} + {{m^*_j}^2(r)-m^2_j\over 2m_j}.
\end{equation}
%%%%%
Note that, only the first term in this equation is a long-range
interaction and thus needs special treatment, while the second and third
terms are  short range interactions.
In practice, Eq.~(\ref{schrodinger}) is first converted into
momentum space representation via a Fourier transformation and is then
solved  using the Kwan-Tabakin-Land\'e technique~\cite{Kwan:1978zh}.
It should be emphasized that no reduction has been made to derive
the Schr\"odinger-like equation, so that all relativistic corrections
are included in the calculation.
The calculated  meson-nucleus bound state energies for $^{208}$Pb,
are listed in Tab.~\ref{denergy}.

%%%%%%%%%%%%%%%%%%%%%%%%%%%%%%%%%%%%%%%%%%%%%%%%%%%%%%%%%%%%%%%%%%%%%%%%%%%%%
\begin{table}[htb]
\begin{center}
\caption{
$D^-$, $\overline{D^0}$ and $D^0$ bound state energies (in MeV).
The widths are all set to zero. }
\label{denergy}
\begin{tabular}[t]{lcccccc}
\hline
%\hline
State  &$D^- (\tilde{V}^q_\omega)$ &$D^- (V^q_\omega)$
&$D^- (V^q_\omega$, no Coulomb) &$\d0bar (\tilde{V}^q_\omega)$
&$\d0bar (V^q_\omega)$ &$D^0 (V^q_\omega)$ \\
\hline
%$^{208}_{\bar{D},D^0}$Pb &
                         1s &-10.6 &-35.2 &-11.2 &unbound &-25.4 &-96.2\\
                         1p &-10.2 &-32.1 &-10.0 &unbound &-23.1 &-93.0\\
                         2s & -7.7 &-30.0 & -6.6 &unbound &-19.7 &-88.5\\
\hline
%\hline
%
\end{tabular}
\end{center}
%\vspace{-2em}
\end{table}
%%%%%%%%%%%%%%%%%%%%%%%%%%%%%%%%%%%%%%%%%%%%%%%%%%%%%%%%%%%%%%%%%%%%%%%%%%%
%

The results in Tab.~\ref{denergy} show that both the $D^-$ and $\overline{D^0}$ are bound in $^{208}$Pb
with the usual QMC $\omega$ coupling constant, $g^q_\omega$.
For $D^-$ the Coulomb force 
provides roughly 24 MeV of attractive potential for the $1s$ state, and is strong enough
to bind the system even with the much more repulsive $\omega$ coupling $1.4^2 g^q_\omega$ 
(viz., $\tilde{V^q_\omega} = 1.4^2V_\omega^q$).
The  $\overline{D^0}$ with the stronger $\omega$ coupling $1.4^2 g^q_\omega$ is not bound.
Note that the difference between $\overline{D^0}$ and $D^-$ 
without the Coulomb force is due to the interaction with the
$\rho$-meson mean field $\dfrac{1}{2}V^q_\rho$, 
which is attractive for the $\overline{D^0}$ but repulsive for
the $D^-$. For completeness, we have also calculated the binding energies
for the $D^0$, which is deeply bound since the $\omega$ interaction
with the light antiquarks is attractive. However, the expected
large width associated with strong absorption may render it
experimentally inaccessible. It is an extremely important experimental
challenge to see whether it can be detected.

%%%%%%%%%%%%%%%%%%%%%%%%%%%%%%%%%%%%%%%%%%%%%%%%%%%%%%%%%%%%%%%%%%%%%%%%%%%%%
%
\begin{figure}[hbt]
\begin{center}
\epsfig{file=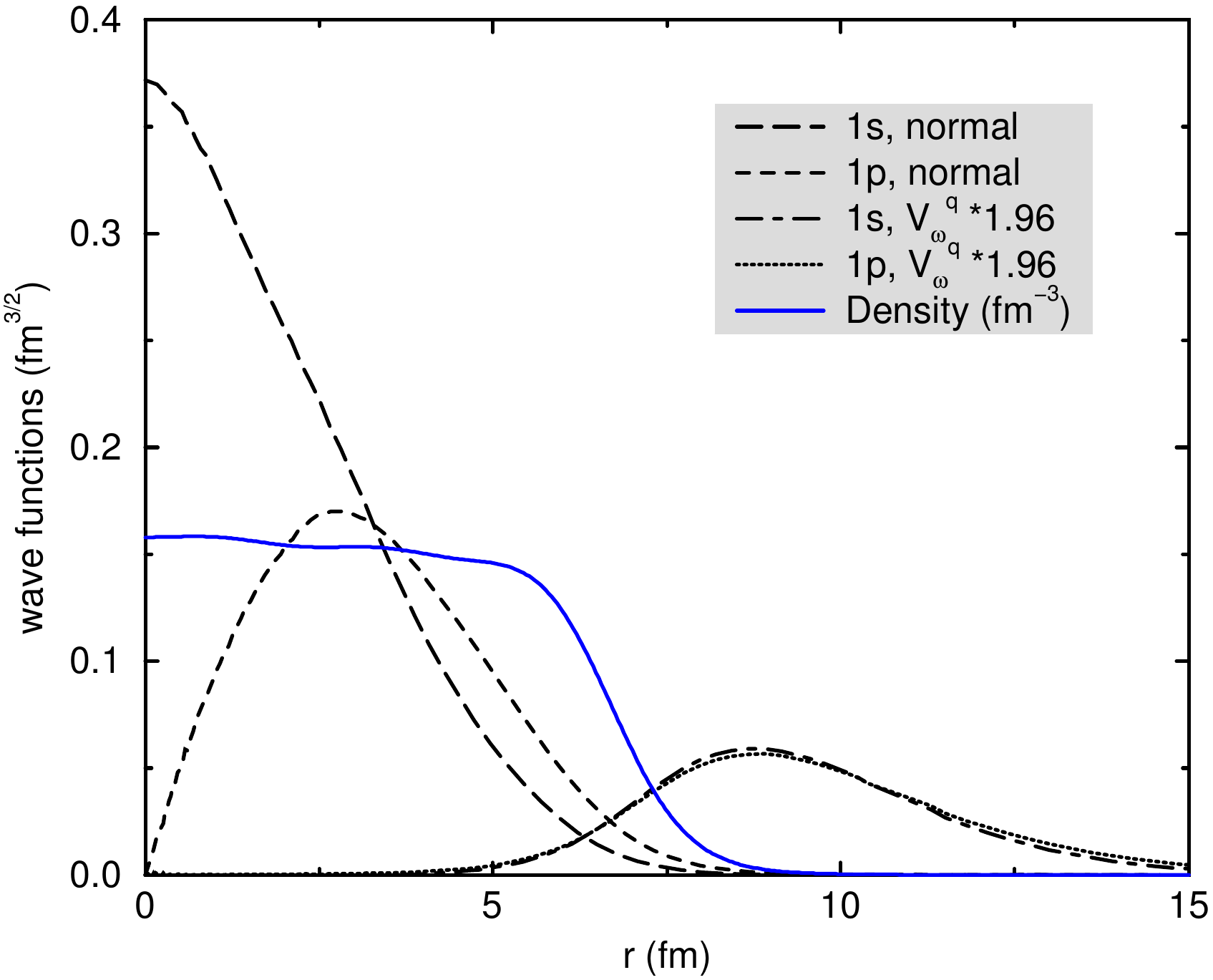,height=6cm}
\caption{The $D^-$-meson bound state wave functions in $^{208}$Pb 
obtained by solving the Schr\"odinger-like equation Eq.~(\ref{schrodinger}) 
for two different $\omega$ meson coupling strengths, $g^q_\omega$ 
and $1.4^2 g^q_\omega = 1.96 g^q_\omega$.
See also the caption of Fig.~\ref{dmespot}.
The wavefunction is normalized as,
$\int_0^{\infty}\!dr\,4\pi r^2 |\Phi(r)|^2 = 1$.
}
\label{dmeswf}
\end{center}
\end{figure}
%
%%%%%%%%%%%%%%%%%%%%%%%%%%%%%%%%%%%%%%%%%%%%%%%%%%%%%%%%%%%%%%%%%%%%%%%%%%%%%

The $D^-$ bound state wave functions obtained by solving the Schr\"odinger-like equation 
Eq.~(\ref{schrodinger}) are shown
in  Fig.~\ref{dmeswf}, together with the baryon density distribution
in $^{208}$Pb.
For the usual $\omega$ coupling $g^q_\omega$, the eigenstates ($1s$ and $1p$)
are well within the nucleus, and behave as expected at the origin.
For the stronger $\omega$ coupling $1.4^2 g^q_\omega$, however, the $D^-$ meson
is considerably pushed out of the nucleus. In this case,
the bound state (an atomic state) is formed
solely due to the Coulomb force.
An experimental determination of whether this is a nuclear state or an atomic
state would give a strong constraint on the $\omega$ coupling.
We note, however, that
because it is very difficult to produce $D$-mesic nuclei with
small momentum transfer, and also the $D$-meson production
cross subsection is small compared with the background from other channels,
it will be a challenging task to detect such bound states
experimentally.

It should be emphasized again that, whether or not the $\overline{D^0}$-$^{208}$Pb 
bound states exist, would give new information as to whether the interactions of
light quarks in a heavy meson are the same as those in a nucleon in nucleus. 
The enormous difference 
between the binding energies of the $D^0$ ($\sim
100$ MeV) and the $\overline{D^0}$ ($\sim 10$ MeV) 
is a simple consequence of the
presence of a strong Lorentz vector mean-field, while the existence of
any binding at all would give us important information concerning the
role of the Lorentz scalar $\sigma$ field (and hence dynamical symmetry
breaking) in heavy-flavor hadron systems. In spite of the perceived experimental
difficulties, we feel that the search for these bound systems should have a very high
priority.

To summarize, the $D^-$ meson should be bound in $^{208}$Pb, due to two 
different mechanisms, namely, the scalar and attractive
$\sigma$ mean field for the case of $V^q_\omega(r)$ even without the Coulomb
force, or solely due to the Coulomb force for the case of
$\tilde{V}^q_\omega(r) = 1.4^2 V^q_\omega(r)$.
The existence of any bound states at all would give us important
information concerning the role of the Lorentz scalar $\sigma$ field,
and hence dynamical symmetry breaking.

%%%%%%%%%%%%%%%%%%%%%%%%%%%%%%%%%%%%%%%%%%%%%%%%%%%%%%%%%%%%%%%%
\subsection{$\Lambda_c^+$- and $\Lambda_b$-hypernuclei}
\label{bchyp}
%%%%%%%%%%%%%%%%%%%%%%%%%%%%%%%%%%%%%%%%%%%%%%%%%%%%%%%%%%%%%%%%

Systematic studies were made on the property changes of heavy hadrons 
which contain a charm or a bottom quark 
in nuclear matter~\cite{QMCbc} (subsection~\ref{hadronmass}), 
and on $\Lambda_c^+$- and $\Lambda_b$-hypernuclei~\cite{QMChypbc}. 
Furthermore, production of $D\overline{D}$ meson pair 
in nucleus~\cite{Shyam:2016bzq}, and formation reaction of 
$\Lambda_c^+$-hypernuclei~\cite{Shyam:2016uxa} were 
studied. These results suggest that it is quite likely 
to form the charmed and bottom hypernuclei, 
which were predicted first in mid 70's~\cite{Tyapkin,Dover}.
The experimental possibilities were also studied later~\cite{bcexp}.
(For the strange hypernuclei including recent development, 
see Refs.~\cite{QMChyp,Guichon:2008zz,Tsushima:2009zh,Tsushima:2010ew}.)
In addition, the $B^-$-nuclear (atomic) 
bound states were also predicted, based on analogy with 
kaonic atom~\cite{Katom}, and a study made for 
the $D$- and $\Dbar$-nuclear bound states~\cite{Tsushimad} 
(subsection~\ref{DDbar_result})  
using the QMC model~\cite{QMCGuichon,Guichonfinite,Saitofinite,QMCreview} 
(section~\ref{qmc}). 

We discuss in the following subsections some results for $\Lambda^+_c$- and $\Lambda_b$-hypernuclei. 
The results are obtained by solving a system of coupled differential 
equations for finite nuclei, embedding a $\Lambda^+_c$ or a $\Lambda_b$ 
to a closed-shell nucleus in Hartree mean-field approximation.
(See section~\ref{qmc}.)
The results are compared with those for the 
$\Lambda$-hypernuclei~\cite{QMChyp}, which were also studied in the QMC model. 
It is shown that, although the scalar and vector potentials 
felt by the $\Lambda$, $\Lambda^+_c$ and $\Lambda_b$ 
in the corresponding hypernuclear multiplet with the same baryon numbers 
are quite similar, the wave functions obtained, e.g., for  
$1s_{1/2}$ state are very different. 
Namely, the $\Lambda^+_c$ baryon density distribution in $^{209}_{\Lambda^+_c}$Pb 
is much more pushed away from the center than that
for the $\Lambda$ in $^{209}_\Lambda$Pb due to the repulsive Coulomb force.
On the contrary, the $\Lambda_b$ baryon density distributions 
in $\Lambda_b$-hypernuclei are much more centralized 
than those for the $\Lambda$ in the corresponding 
$\Lambda$-hypernuclei due to its heavy mass.
Furthermore, the level spacing for the 
$\Lambda_b$ single-particle energies is much smaller than that for 
the $\Lambda$ and $\Lambda^+_c$, which may imply many interesting 
new phenomena, that will possibly be discovered in experiments. 
The studies for such heavy-flavor-hypernuclei open a new possibility 
of experiments, for the facilities such as JLab, J-PARC, and FAIR.

%%%%%%%%%%%%%%%%%%%%%%%%%%%%%%%%%%%%%%%%%%%%%%%%%%%%%%%%%%%%%%%%%%%%%%%
\subsection{Description of $\Lambda_c^+$- and $\Lambda_b$-hypernuclei}
\label{bchypdetail}
%%%%%%%%%%%%%%%%%%%%%%%%%%%%%%%%%%%%%%%%%%%%%%%%%%%%%%%%%%%%%%%%%%%%%%%

Let us start to consider static, (approximately) spherically symmetric
charmed and bottom hypernuclei (closed shell plus one heavy baryon
configuration) ignoring small non-spherical 
effects due to the embedded heavy baryon in 
Hartree mean-field approximation.
In this approximation, $\rho NN$ tensor coupling gives
a spin-orbit force for a nucleon bound 
in a static spherical nucleus, although
in Hartree-Fock it can give a central force which contributes to
the bulk symmetry energy~\cite{Guichonfinite,Saitofinite}. 
Furthermore, it gives no contribution for nuclear
matter, since the meson fields are independent of position
and time. Thus, we ignore the $\rho NN$ tensor coupling in this
study as usually adopted in the Hartree treatment of
quantum hadrodynamics (QHD)~\cite{QHD1,QHD2}.

Here we recall the effective Pauli potential discussed in 
subsection~\ref{qmc_finite}. (See  Eq.~(\ref{Pauli}).)
This potential is regard to reflect the Pauli blocking effects originating 
from the underling quark structure of hyperons. As explained already, 
for the strange hyperon sector, it includes the effects of 
the $\Sigma N - \Lambda N$ channel coupling, as well as those to 
reproduce the observed $_\Lambda^{208}$Pb $1s$ state single-particle 
energy of $\cong -27$ MeV~\cite{QMChyp}. 
For the $\Lambda_c^+$- and $\Lambda_b$-hypernuclei, 
the same effective Pauli potential is included.

Next, we briefly discuss the spin-orbit force in QMC~\cite{Guichonfinite}. 
The origin of the spin orbit force for a composite nucleon moving
through scalar and vector fields which vary with position, 
was explained in detail in Ref.~\cite{Guichonfinite}. The situation for the
$\Lambda$ and also for other hyperons are discussed in detail  
in Ref.~\cite{QMChyp,QMChypbc}. 
In order to include the spin-orbit potential (approximately) 
properly, for example, for the $\Lambda^+_c$-hypernuclei, 
it is added perturbatively the correction due to the vector potential,
\be
-\dfrac{2}{2 M^{\star 2}_{\Lambda^+_c} (\vec{r}) r}
\, \left( \dfrac{d}{dr} g^{\Lambda^+_c}_\omega \omega(\vec{r}) \right)
\vec{l}\cdot\vec{s},
\ee
to the single-particle energies obtained by solving the Dirac
equation, in the same way as that added in Ref.~\cite{QMChyp}.
This may correspond to a correct spin-orbit force which 
is calculated by the underlying quark 
model~\cite{Guichonfinite,QMChyp}:
%%%%%
\begin{equation}
V^{\Lambda^+_c}_{S.O.}(\vec{r}) \vec{l}\cdot\vec{s}
= - \dfrac{1}{2 M^{\star 2}_{\Lambda^+_c} (\vec{r}) r}
\, \left( \dfrac{d}{dr} [ M^\star_{\Lambda^+_c} (\vec{r})
+ g^{\Lambda^+_c}_\omega \omega(\vec{r}) ] \right) \vec{l}\cdot\vec{s},
\label{soQMC}
\end{equation}
%%%%%
since the Dirac equation at the hadronic level solved in usual QHD-type 
models leads to, 
%%%%%
\begin{equation}
V^{\Lambda^+_c}_{S.O.}(\vec{r}) \vec{l}\cdot\vec{s}
= - \dfrac{1}{2 M^{\star 2}_{\Lambda^+_c} (\vec{r}) r}
\, \left( \dfrac{d}{dr} [ M^\star_{\Lambda^+_c} (\vec{r})
- g^{\Lambda^+_c}_\omega \omega(\vec{r}) ] \right) \vec{l}\cdot\vec{s},
\label{soQHD}
\end{equation}
%%%%%
which has the opposite sign for the vector potential,  
$g^{\Lambda^+_c}_\omega \omega(\vec{r})$.
The correction to the spin-orbit force, which appears naturally in the
QMC model, may also be modeled at the hadronic level of the Dirac equation by
adding a tensor interaction, motivated by the quark 
model~\cite{Jennings2,Cohen}.
Here, we should make a comment that, as was discussed 
in detail Ref.~\cite{Gal}, a one boson exchange model with underlying (approximate) 
SU(3) symmetry in the strong interaction, also leads to weaker 
spin-orbit forces for the (strange) hyperon-nucleon ($YN$) 
than that for the nucleon-nucleon ($NN$).

However, in practice, because of its heavy mass 
$M^\star_{\Lambda^+_c}$, the contribution to the single-particle energies 
from the spin-orbit potential with or without 
including the correction term, turned 
out to be even smaller than that for the 
$\Lambda$-hypernuclei, and further smaller for the $\Lambda_b$-hypernuclei.
Contribution from the spin-orbit potential with the correction term 
is typically of order $0.01$ MeV,   
and even the largest case is $\cong 0.1$ MeV. 
This can be understood when one considers the limit, 
$M^\star_{\Lambda^+_c} \to \infty$
in Eq.~(\ref{soQMC}), where the quantity inside the square brackets varies   
smoothly from an order of hundred MeV to zero near the surface of 
the hypernucleus, and the derivative with respect to $r$ is finite.    
(See also Fig.~\ref{CaPbpot} presented below.)

%%%%%%%%%%%%%%%%%%%%%%%%%%%%%%%%%%%%%%%%%%%%%%%%%%%%%%%%%%%%%%%%%%%
\subsection{Predictions for single-particle energies in $\Lambda_c^+$- and $\Lambda_b$-hypernuclei}
\label{hypbcresults}
%%%%%%%%%%%%%%%%%%%%%%%%%%%%%%%%%%%%%%%%%%%%%%%%%%%%%%%%%%%%%%%%%%

\begin{figure}[htb]
\begin{center}
\epsfig{file=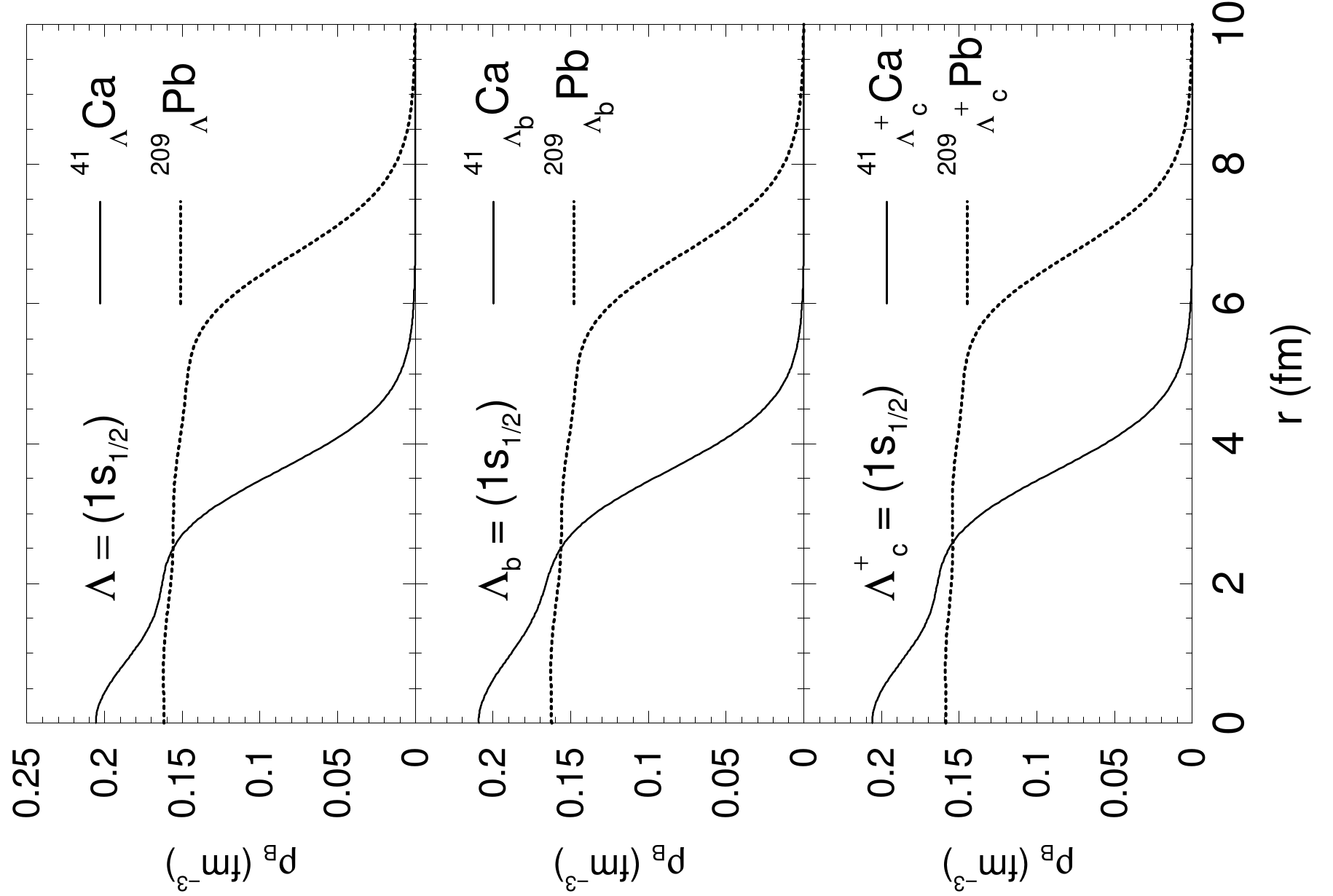,height=6cm,angle=-90}
\caption{
Total baryon density distributions in  
$^{41}_j$Ca and $^{209}_j$Pb ($j = \Lambda,\Lambda^+_c,\Lambda_b$),  
for $1s_{1/2}$ configuration for the $\Lambda, \Lambda_c^+$ and $\Lambda_b$.
\label{CaPbden}
}
\end{center}
\end{figure}

Now we discuss the results.
First, in Fig.~\ref{CaPbden} we show the total baryon density distributions 
in $^{41}_j$Ca and $^{209}_j$Pb ($j=\Lambda,\Lambda^+_c,\Lambda_b$), 
for $1s_{1/2}$ configuration in each hypernucleus. 
Note that because of the self-consistency, the total baryon density 
distributions are dependent on the configurations of 
the embedded hyperons. The total baryon density distributions are quite 
similar for the $\Lambda$-, $\Lambda^+_c$- and $\Lambda_b$-hypernuclear 
multiplet with the same baryon numbers, A, since the effect of 
$\Lambda,\Lambda^+_c$ and $\Lambda_b$ is $\cong 1/A$ for 
each hypernucleus.
Nevertheless, one can notice that the 
$\Lambda_b$-hypernuclear density near the center 
is slightly higher than the corresponding 
$\Lambda$- and $\Lambda^+_c$-hypernucleus. 
This is because the $\Lambda_b$ is 
heavy and localized nearer the center, and 
contributes to the total baryon density there.
The baryon (probability) density distributions for the 
$\Lambda$, $\Lambda^+_c$ and $\Lambda_b$ in the corresponding 
hypernuclei will be shown later.

%%%%%%%%%%%%%%%%%%%%%%%%%%%%%%%%%%%%%%%%%%%%%%%%%%%%
%%%%%%%%%%%%%%%%%%%%%%%%%%%%%%%%%%%%%%%%%%%%%%%%%%%%

\begin{figure}[htb]
\begin{center}
\epsfig{file=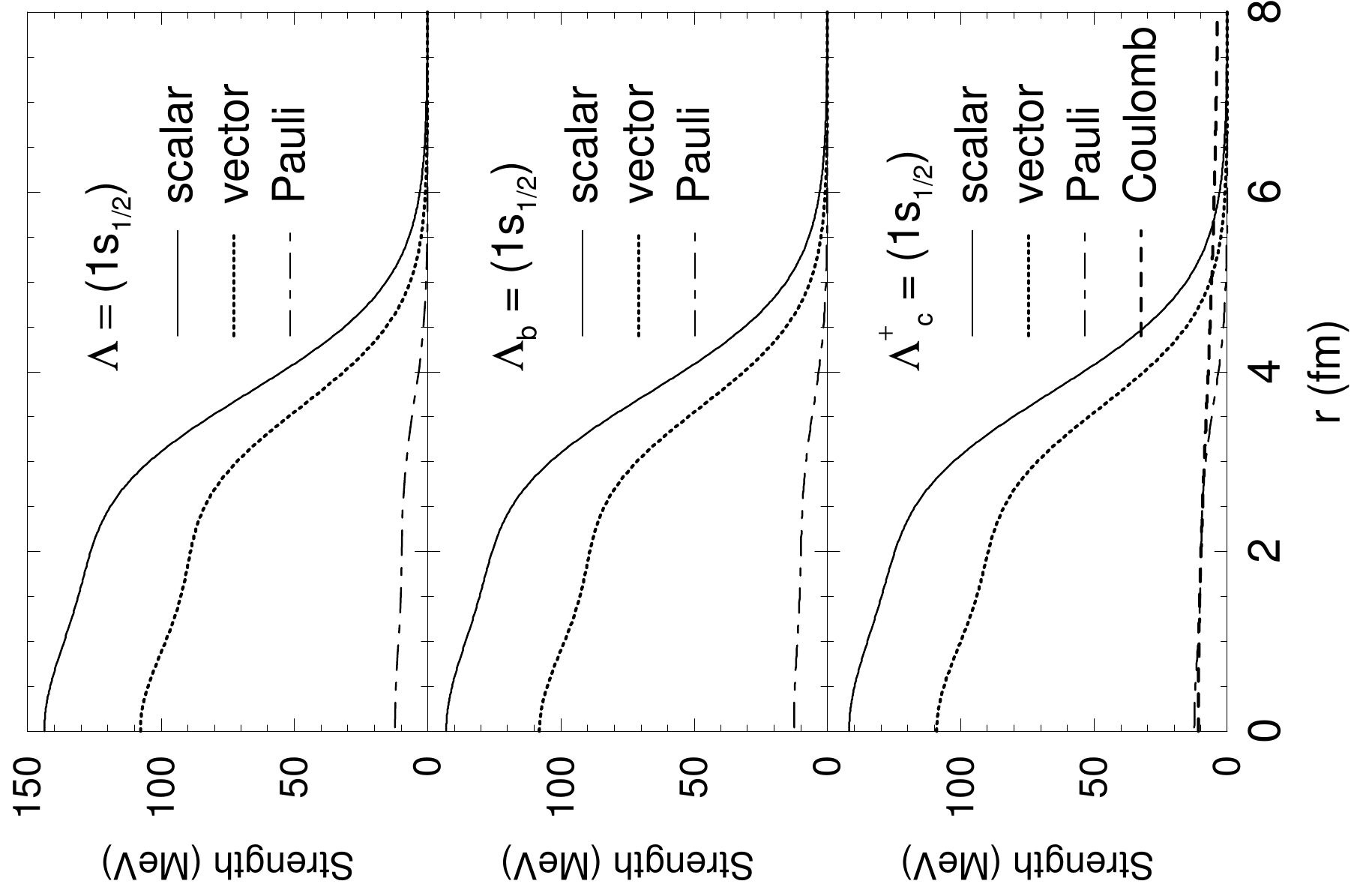,height=6cm,angle=-90}
\hspace{0.5cm}
\epsfig{file=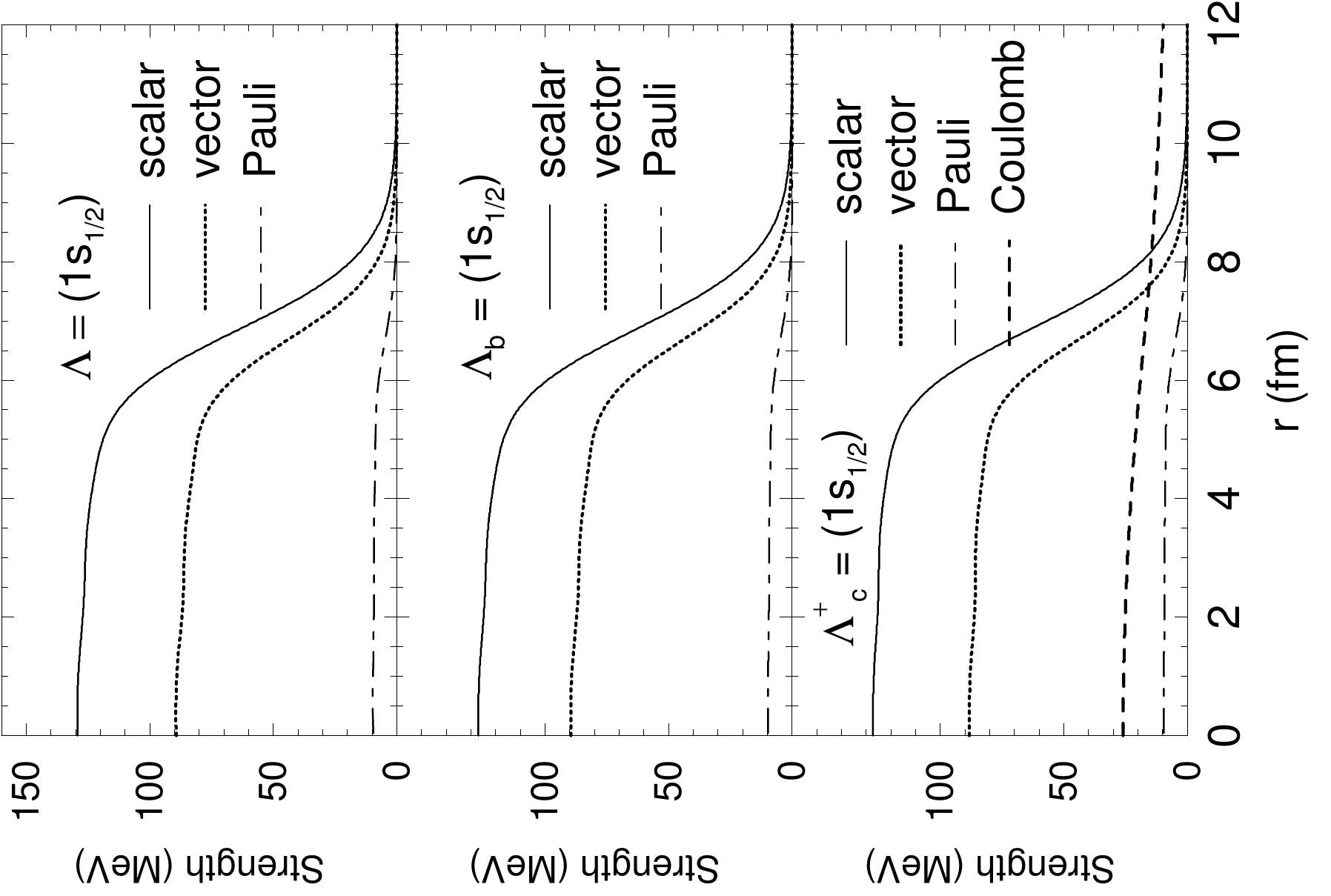,height=6cm,angle=-90}
\caption{
Potential strengths for $1s_{1/2}$ state felt by 
the $\Lambda,\Lambda^+_c$ and $\Lambda_b$ 
in $^{41}_j$Ca  (left panel) 
and those in $^{209}_j$Pb ($j = \Lambda,\Lambda^+_c,\Lambda_b$).
"Pauli" stands for the effective, repulsive, potential representing 
the Pauli blocking at the quark level plus
the $\Sigma_{c,b} N - \Lambda_{c,b} N$ channel coupling,
introduced at the baryon level
phenomenologically~\protect\cite{QMChyp}.
\label{CaPbpot}
}
\end{center}
\end{figure}

\begin{figure}[htbp]
\begin{center}
\epsfig{file=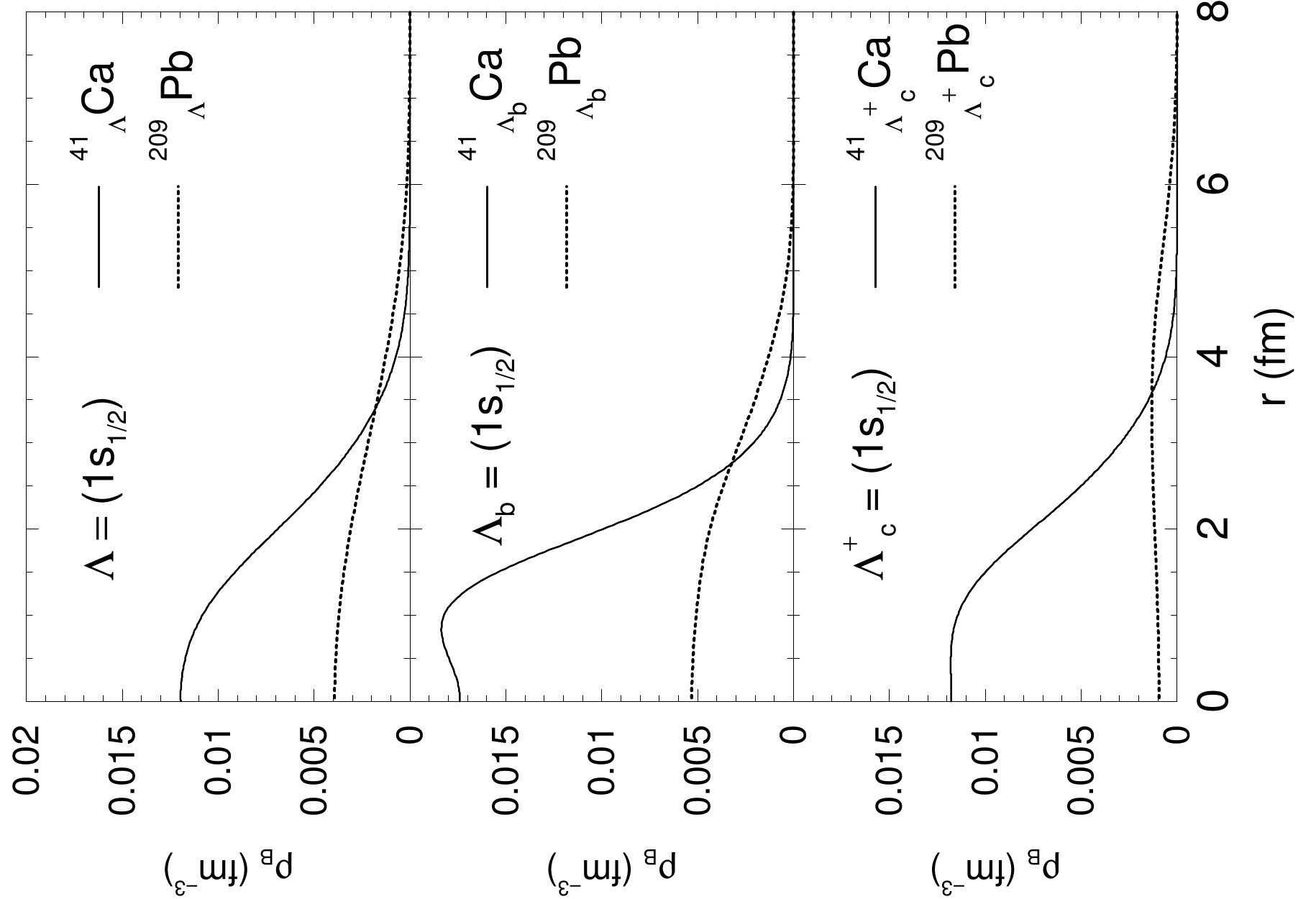,height=6cm,angle=-90}
\caption{$\Lambda,\Lambda^+_c$ and $\Lambda_b$ 
baryon (probability) density distributions for $1s_{1/2}$ state 
in $^{41}_j$Ca and 
$^{209}_j$Pb ($j = \Lambda,\Lambda^+_c,\Lambda_b$).
\label{HBdensity}
}
\end{center}
\end{figure}

Next, in Fig.~\ref{CaPbpot}, it is shown the scalar, vector,  
and effective Pauli-blocking potentials (denoted by Pauli) 
felt by the $\Lambda$, $\Lambda^+_c$ and $\Lambda_b$ 
for $1s_{1/2}$ state in $^{41}_j$Ca (left panel) 
and $^{209}_j$Pb (right panel) [$j=\Lambda,\Lambda^+_c,\Lambda_b$].
For the $\Lambda^+_c$-hypernuclei, the Coulomb potentials are also shown.
The corresponding probability density distributions 
are shown in Fig.~\ref{HBdensity}.

As for the case of the nuclear matter~\cite{QMCbc}, the scalar and vector 
potentials felt by these particles in hypernuclear multiplet 
with the same baryon numbers are also quite similar.
(See subsection~\ref{hadronmass}.)
Thus, as far as the total baryon density distributions and 
the scalar and vector potentials are concerned, 
$\Lambda$-, $\Lambda^+_c$- and $\Lambda_b$-hypernuclei show quite similar 
features within the multiplet. 
However, as shown in Fig.~\ref{HBdensity}, 
the wave functions obtained for $1s_{1/2}$ state are very different.
The $\Lambda^+_c$ baryon density distribution in $^{209}_{\Lambda^+_c}$Pb
is much more pushed away from the center
than that for the $\Lambda$ in $^{209}_\Lambda$Pb due to the Coulomb force.
On the contrary, the $\Lambda_b$ baryon density distributions
in $\Lambda_b$-hypernuclei are much larger near the origin than those for
the $\Lambda$ in the corresponding $\Lambda$-hypernuclei
due to its heavy mass.

Having obtained reasonable ideas about the potentials felt by  
$\Lambda,\Lambda^+_c$ and $\Lambda_b$, we show 
the calculated single-particle energies in 
Tabs.~\ref{table1} and~\ref{table2}~\cite{QMChypbc}.
Results for the $\Lambda$-hypernuclei are from Ref.~\cite{QMChyp}.
In these calculations, effective Pauli blocking,  
the effect of the $\Sigma_{c,b} N - \Lambda_{c,b} N$ 
channel coupling, and correction to the
spin-orbit force based on the underlying quark structure,    
are all included in the same way as adopted in Ref.~\cite{QMChyp}.
However, the correction term for the spin-orbit 
force, as well as the contribution from the spin-orbit force itself 
are small, their effects on the final result are very small.
Note that since the mass difference of the $\Lambda_c^+$ and $\Sigma_c$ 
is larger than that of the $\Lambda$ and $\Sigma$,   
and it is also true for the $\Lambda_b$ and $\Sigma_b$, 
we expect the effect of the channel coupling for the charmed and bottom 
hypernuclei is smaller than those for the 
strange hypernuclei, although the same parameters were used.
In addition, the single-particle states were searched up to 
the same highest state as that of the core neutrons in each hypernucleus, 
since the deeper levels are usually easier to observe in experiment.

%\newpage
%%%%%%%%%%%%%%%%%%%%%%%%%%%%%%%%%%%%%%%%%%%%%%%%%%
%%%%%%%%%%%%%%%%%%%%%%%%%%%%%%%%%%%%%%%%%%%%%%%%%%

From the results listed in Tabs.~\ref{table1} and~\ref{table2}, 
first, it is clear that the $\Lambda^+_c$ 
single-particle energy levels are higher 
than the corresponding levels for the $\Lambda$ and $\Lambda_b$. 
This is a consequence of the 
Coulomb force. This feature becomes stronger as 
the proton number in the core nucleus increases. 

Second, the level spacing for the $\Lambda_b$ single-particle energies 
is much smaller than that for the $\Lambda$ and $\Lambda^+_c$.
This is ascribed to its heavy effective mass, $M^\star_b$. 
In the Dirac equation for the $\Lambda_b$, 
the mass term dominates more than that of the term  
proportional to Dirac's $\kappa$, which classifies the states,  
or single-particle 
wave functions. (See Refs.~\cite{Saitofinite,QMChyp} for detail.)
This small level spacing would make it very difficult to  
distinguish the states in experiment, 
or to achieve such high resolution.
On the other hand, this may imply also many new phenomena. 
It will have a large probability to trap a $\Lambda_b$ among 
one of those many states, especially in a heavy nucleus such as lead (Pb).
What are the possible consequences? 
Maybe the $\Lambda_b$ weak decay happens inside a heavy nucleus with a very low 
probability? Does it emit many photons when the $\Lambda_b$ gradually 
makes transitions from a deeper state to a shallower state?
All these questions may raise a lot of interesting issues.

%%%%%%%%%%%%%%%%%%%%%%%%%%%%%%%%%%%%%%%%%%%%%%%%%%
%%%%%%%%%%%%%%%%%%%%%%%%%%%%%%%%%%%%%%%%%%%%%%%%%%

%\samepage
%
%%%%%%%%%%%%%%%%%%%%%%%%%%%%%%%%%%%%%%%%%%%%%%%%%%%%%%%%%%%%%%%%%%%%%%%%%%%%%
%^17_CO
\begin{table}[htbp]
\begin{center}
\caption{Single-particle energies (in MeV)
for $^{17}_j$O, $^{41}_j$Ca and $^{49}_j$Ca 
($j=\Lambda,\Lambda^+_c,\Lambda_b$). 
Single-particle energy levels are calculated up to the same highest states 
as that of the core neutrons. 
Results for the hypernuclei are taken 
from Ref.~\cite{QMChyp}.
Experimental data for $\Lambda$-hypernuclei 
are taken from Ref.~\cite{chr}, where 
spin-orbit splittings for $\Lambda$-hypernuclei 
are not well determined by the experiments.}
\label{table1}
\begin{tabular}[t]{c|cccc|cccc|ccc}
%\hline
\hline
& & & & & & & & & & &\\
&$^{16}_\Lambda$O  &$^{17}_\Lambda$O 
&$^{17}_{\Lambda^+_c}$O  &$^{17}_{\Lambda_b}$O
&$^{40}_\Lambda$Ca &$^{41}_\Lambda$Ca
&$^{41}_{\Lambda^+_c}$Ca &$^{41}_{\Lambda_b}$Ca
&$^{49}_\Lambda$Ca    &$^{49}_{\Lambda^+_c}$Ca &$^{49}_{\Lambda_b}$Ca\\
&(Exp.)& & & &(Exp.)& & & & & & \\
\hline
$1s_{1/2}$&-12.5&-14.1&-12.8&-19.6&-20.0&-19.5&-12.8&-23.0&-21.0&-14.3&-24.4\\
$1p_{3/2}$&-2.5 &-5.1 &-7.3 &-16.5&-12.0&-12.3&-9.2 &-20.9&-13.9&-10.6&-22.2\\
$1p_{1/2}$&($1p_{3/2}$)&-5.0&-7.3 &-16.5
&($1p_{3/2}$)&-12.3&-9.1&-20.9&-13.8&-10.6&-22.2\\ 
$1d_{5/2}$&     &     &     &     &     &-4.7 &-4.8 &-18.4&-6.5 &-6.5 &-19.5\\
$2s_{1/2}$&     &     &     &     &     &-3.5 &-3.4 &-17.4&-5.4 &-5.3 &-18.8\\
$1d_{3/2}$&     &     &     &     &     &-4.6 &-4.8 &-18.4&-6.4 &-6.4 &-19.5\\
$1f_{7/2}$&     &     &     &     &     &     &     &     &---  &-2.0 &-16.8\\
%\hline
\hline
\end{tabular}
\end{center}
\end{table}
%
%%%%%%%%%%%%%%%%%%%%%%%%%%%%%%%%%%%%%%%%%%%%%%%%%%%%%%%%%%%%%%%%%%%%%%%
%\newpage
%
%
%\samepage
%%%%%%%%%%%%%%%%%%%%%%%%%%%%%%%%%%%%%%%%%%%%%%%%%%%%%%%%%%%%%%%%%%%%%%%%%%%%%
\begin{table}[htbp]
\begin{center}
\caption{Single-particle energies (in MeV)
for $^{91}_j$Zr and $^{208}_j$Pb 
($j=\Lambda,\Lambda^+_c,\Lambda_b$).
Experimental data are taken from Ref.~\protect\cite{aji}.
See caption of Table~\protect\ref{table1} for other explanations.
}
\label{table2}
\begin{tabular}[t]{c|cccc|cccc}
\hline 
%\hline
 & & & & & & & & \\
&$^{89}_\Lambda$Yb          &$^{91}_{\Lambda}$Zr
&$^{91}_{\Lambda^+_c}$Zr    &$^{91}_{\Lambda_b}$Zr
&$^{208}_\Lambda$Pb         &$^{209}_{\Lambda}$Pb
&$^{209}_{\Lambda^+_c}$Pb   &$^{209}_{\Lambda_b}$Pb\\
&(Exp.)& & & &(Exp.)& & & \\
\hline 
$1s_{1/2}$&-22.5&-23.9&-10.8&-25.7&-27.0&-27.0&-5.2 &-27.4\\
$1p_{3/2}$&-16.0&-18.4&-8.7 &-24.2&-22.0&-23.4&-4.1 &-26.6\\
$1p_{1/2}$&($1p_{3/2}$)&-18.4&-8.7 &-24.2
&($1p_{3/2}$)&-23.4&-4.0 &-26.6\\
$1d_{5/2}$&-9.0 &-12.3&-5.8 &-22.4&-17.0&-19.1&-2.4 &-25.4\\
$2s_{1/2}$&---  &-10.8&-3.9 &-21.6&---  &-17.6&---  &-24.7\\
$1d_{3/2}$&($1d_{5/2}$)&-12.3&-5.8 &-22.4
&($1d_{5/2}$)&-19.1&-2.4 &-25.4\\
$1f_{7/2}$&-2.0 &-5.9 &-2.4 &-20.4&-12.0&-14.4&---  &-24.1\\
$2p_{3/2}$&---  &-4.2 &---  &-19.5&---  &-12.4&---  &-23.2\\
$1f_{5/2}$&($1f_{7/2}$)&-5.8 &-2.4 &-20.4
&($1f_{7/2}$)&-14.3&---  &-24.1\\
$2p_{1/2}$&     &-4.1 &---  &-19.5&---  &-12.4&---  &-23.2\\
$1g_{9/2}$&     &---  &---  &-18.1&-7.0 &-9.3 &---  &-22.6\\
$1g_{7/2}$&     &     &     &     
&($1g_{9/2}$)&-9.2 &---  &-22.6\\
$1h_{11/2}$&    &     &     &     &     &-3.9 &---  &-21.0\\
$2d_{5/2}$&     &     &     &     &     &-7.0 &---  &-21.7\\
$2d_{3/2}$&     &     &     &     &     &-7.0 &---  &-21.7\\
$1h_{9/2}$&     &     &     &     &     &-3.8 &---  &-21.0\\
$3s_{1/2}$&     &     &     &     &     &-6.1 &---  &-21.3\\
$2f_{7/2}$&     &     &     &     &     &-1.7 &---  &-20.1\\
$3p_{3/2}$&     &     &     &     &     &-1.0 &---  &-19.6\\
$2f_{5/2}$&     &     &     &     &     &-1.7 &---  &-20.1\\
$3p_{1/2}$&     &     &     &     &     &-1.0 &---  &-19.6\\
$1i_{13/2}$&    &     &     &     &     &---  &---  &-19.3\\
%\hline
\hline
\end{tabular}
\end{center}
\end{table}
%
%%%%%%%%%%%%%%%%%%%%%%%%%%%%%%%%%%%%%%%%%%%%%%%%%%%%%%%%%%%%%%%%%%%%%%%
%

To summarize, a quantitative study for $\Lambda^+_c$- and $\Lambda_b$-hypernuclei 
in the QMC model has shown that, although the scalar and 
vector potentials felt by the $\Lambda$, $\Lambda^+_c$ and $\Lambda_b$ 
are quite similar in the corresponding hypernuclear multiplet with the 
same baryon numbers, the single-particle wave functions, 
and single-particle energy level spacings are quite different. 
For the $\Lambda^+_c$-hypernuclei, the Coulomb force  
plays a crucial role, and so does the heavy $\Lambda_b$ mass 
for the $\Lambda_b$-hypernuclei. 
It should be emphasized that the values used   
for the coupling constants of $\sigma$ 
(or $\sigma$-field dependent strength), $\omega$ and $\rho$ 
to the $\Lambda, \Lambda^+_c$ and $\Lambda_b$, 
are determined automatically based on the underlying quark model, 
the same as for the nucleon and other baryons.
(Recall that the values for the vector $\omega$ fields to 
any baryons can be obtained by the number of light quarks in a baryon, 
but those for the $\sigma$ are different as 
shown in Eq.~(\ref{Ssigma}).)
Phenomenology would determine ultimately whether or not the coupling constants 
(strengths) determined by the underlying quark model 
actually work for $\Lambda^+_c$ and $\Lambda_b$.
This also provides us with the information on dynamical 
chiral symmetry breaking for the light quarks inside 
the different heavy-flavor baryons.
Although one may speculate about some aspects of the present results,  
they clearly show that the $\Lambda^+_c$- and $\Lambda_b$-hypernuclei 
should exist in realistic experimental conditions.  
Experiments at facilities like JLab, J-PARC and FAIR would provide 
further inputs to gain a better understanding of the interaction of $\Lambda^+_c$ 
and $\Lambda_b$ with the nuclear matter. 
The study of the presence of $\Lambda^+_c$ and $\Lambda_b$ in finite nuclei 
experimentally is expected to be realized in the near future.
Careful investigations, both theoretical and experimental, would lead to 
a much better understanding of the role of heavy quarks in 
a nuclear medium (finite nuclei).
%
%
%
%\input{sec_concl.tex}
%
%
%

% % % % % % % % % % % % % % % % % % % % % 
\section{Conclusions and Perspectives}
\label{sec:CONCL}

We have presented a review of the theoretical ideas underlying the study of quarkonia in 
atomic nuclei, along with the latest experimental information. This is an exciting field 
which promises to provide a far deeper understanding of how QCD works. It is an field 
where experimental interest is high and important new results may be expected in the 
near future.

In the context of understanding whether or not one can expect to find $J/\Psi$ mesons 
bound to atomic nuclei, we reviewed the predictions of both heavy quark effective field theory 
and lattice QCD for the attraction they might be expected to feel. The current expectation 
is for an attractive nuclear potential from a few MeV up to as much as 20 MeV, certainly 
sufficient to lead to bound states. Further support for this suggestion comes when one 
also calculates the reduction in the $J/\Psi$ effective mass which arises through the 
modification of self-energy loops in-medium, calculated within the QMC model. These effects 
combined lead one to anticipate substantial binding energies in medium and large nuclei. 
We stress that these states should also be expected to be rather narrow, 
with widths below 1 MeV. That should improve the chances of finding them if one can 
produce the $J/\Psi$ mesons in essentially recoilless conditions.

Even though there have been rather more experiments aimed at examining the interaction 
of $\phi$ mesons with nuclei, the situation is currently quite uncertain. While it is clear that 
there is substantial broadening of the resonance, there is as yet no consensus as to 
whether or not the real part of the interaction is attractive. Our review of recent theoretical 
progress leads us to expect that one should find $\phi$-nucleus bound states for almost all 
nuclei. A number of experiments aimed at finding such states were described. 
It was also pointed out that these experiments are made even more challenging because the 
widths are expected to be comparable with the amount of binding.

For mesons which are not strictly quarkonia, like the $\omega , \eta$ and $\eta^\prime$, there has 
also been a great deal of experimental work at laboratories such as CERN/SPS, GSI, JLab and Mami. 
There are hints of binding for both the $\omega$ and $\eta$ that strongly suggest more work 
would be valuable. There is compelling theoretical evidence that the real part of their interaction 
with nuclei  is attractive but the possibility of strong absorption clouds those predictions, making 
the need for new expeirmental work even greater. 

Other systems involving naked, rather than hidden, heavy quarks such as the $D$ and 
$\overline{D}$ mesons, as well as $\Lambda^+_c$ and $\Lambda_b$ baryons have also 
been reviewed. In some cases there are predictions of remarkably deep binding. In any case,  
there is considerable interest in the study of the binding of such systems because of the 
information which it may provide concerning dynamical symmetry breaking 
and the partial restoration of chiral symmetry in a nuclear medium.

The various theoretical predictions discussed here are of direct relevance for the 
existing modern experimental facilities such as FAIR, Jefferson Lab at 12 GeV and J-PARC, 
and we await the future experimental results with great anticipation.
On the theoretical side, we look forward to the new information 
that one may expect lattice QCD simulations will be able to provide concerning the binding
of heavy-flavor-quarkonia and (heavy-flavor) baryons to atomic nuclei. 

%
%
%\input{sec_acknow.tex}
%
%

% % % % % % % % % % % % % % % % % % % % % 
\section{Acknowledgments}
\label{sec:ACKNOW}
The authors would like to thank J.J. Cobos-Mart\'{\i}nez for the updated $J/\Psi$ results
presented in this article. G.K. thanks Jaume Tarr\'us-Castell\'a for discussions.
This work was partially supported by Conselho Nacional 
de Desenvolvimento Cient\'{\i}fico e Tecnol\'ogico - CNPq, 
Grant Nos. 305894/2009-9 (G.K.), 313800/2014-6 (A.W.T), 
400826/2014-3 and 308088/2015-8 (K.T.),  and
Funda\c{c}\~ao de Amparo \`a Pesquisa do Estado de S\~ao Paulo - FAPESP, 
Grant Nos. 2013/0197-0 (G.K.) and 2015/17234 (K.T.).
This research was also supported by the Australian Research Council through the
ARC Centre of Excellence for Particle Physics at the Terascale (CE110001104),
and through Grant No. DP151103101 (A.W.T.).
%
%

%
%
%\input{references.tex}
%
%
%%%
%%%%%%%%%%%%%%%%%%%%%%%%%%%%%%%%%%%%%%%%%%%%%%%%%%%%%%%%%%%%%%%%%%%%%%%%%%%%%%%%%%%%%%%%%
%%%%%%% Reference format (European style)
%%%\def\Journal#1#2#3#4{{#1} {#2} (#4) #3 }
%%%\def\ANNRNPS{{\em Annu. Rev. Nucl. Part. Sci.}}
%%%\def\NPB{{\em Nucl. Phys.} B}
%%%\def\PLA{{\em Phys. Lett.} A}
%%%\def\PLB{{\em Phys. Lett.} B}
%%%\def\PRL{\em Phys. Rev. Lett.}
%%%\def\PREV{\em Phys. Rev.}
%%%\def\PREP{\em Phys. Rep.}
%%%\def\PRD{{\em Phys. Rev.} D}
%%%\def\PRC{{\em Phys. Rev.} C}
%%%\def\ZPC{{\em Z. Phys.} C}
%%%\def\ZPA{{\em Z. Phys.} A}
%%%%%%%%%%%%%%%%%%%%%%%%%%%%%%%%%%%%%%%%%%%%%%%%%%%%%%%%%%%%%%%%%%%%%%%%%%%%%%%%%%%%%%%%%

% % % % % % % % % % % % % % % % % % % % % 

\end{document}